\documentclass{JFM-FLM_Au}

\usepackage{tcolorbox}
\usepackage{bm}
\usepackage{ulem}
\def\aaa{{\it a}}
\def\bbb{{\it b}}
\def\ccc{{\it c}}
\def\ddd{{\it d}}
\newcommand{\mylab}[3]{\raisebox{#2}[0mm][0mm]{%
\makebox[0mm][l]{\hspace*{#1}\textbf{#3}}}}

\lefttitle{G. Boga and A. Cimarelli}
\righttitle{Journal of Fluid Mechanics}

\title{Boundary effects on turbulence dissipation and energy cascade: an enstrophy point of view}

\author{G. Boga\aff{1} \and A. Cimarelli\aff{1}}

\affiliation{\aff{1}Department of  Engineering "Enzo Ferrari", University of Modena and Reggio Emilia, 41125 Modena, Italy}

\corresau{Gabriele Boga, \email{gabriele.boga@unimore.it}}

\begin{document}
\maketitle

\begin{abstract}
In the present work, we investigate the single- and two-point enstrophy dynamics in a temporal boundary layer. The analysis focuses on the confinement effects induced by the solid wall and the turbulent/non-turbulent interface, where interscale energy transfer and viscous dissipation depart substantially from the classical picture of turbulence. Since enstrophy is related to the energy dissipation and contributes to the energy cascade process through the vortex-stretching mechanism, its analysis provides a complementary perspective on regions where these two processes are modified. Enstrophy is generated at the wall, with the vorticity field initially comprising only two non-zero components, primarily through the footprint of structures that are not sustained locally but instead populate the buffer and outer layers. This initially two-dimensional vorticity field diffuses into the flow and is transported away from the wall. Part of the enstrophy, generated at $r_y^+ \lesssim 55$, remains confined to the viscous sublayer, where it undergoes a local dynamics characterized by a direct interscale transfer towards smaller scales, where it is dissipated. Conversely, enstrophy generated at $r_y^+ \gtrsim 55$ crosses the buffer layer and feeds the outer region. Within the buffer layer, where enstrophy amplification by vortex stretching is strongest, vorticity is also redistributed among its three components, particularly towards the wall-normal component, mainly through the vertical tilting of mean spanwise vorticity, consistently with hairpin-vortex dynamics. Finally, in the outer region, enstrophy exhibits an almost isotropic structure, with a weak split interscale transfer feeding both larger and smaller scales, reflecting the imprint of the attached-eddy dynamics.
\end{abstract}


\section{Introduction}
\label{sec:intro}

Many flows of engineering and environmental relevance develop in the presence of solid boundaries. Even in the simplest configuration, such as the flow over a flat, infinitely extended plate, the presence of a solid boundary poses significant challenges to our understanding and prediction of the turbulent dynamics. Wall-bounded turbulence has therefore been the subject of extensive theoretical and experimental investigation over several decades (see \citet{Townsend1976,Smits2011,MarusicMonty2019} for background and reviews), revealing substantial departures from the classical picture of turbulence. In this classical picture, energy is injected at the largest scales of motion and subsequently transferred towards progressively smaller scales through a direct energy cascade, up to the smallest dissipative scales \citep{Richardson1922,Kolmogorov1941,Frisch1995}. In this process of energy cascade, the influence of the large-scale anisotropic motions  is expected to progressively weaken towards the small scales, which tend to exhibit isotropic and universal behaviour. However, the presence of a solid boundary profoundly modifies this dynamics. In fact, wall-turbulent flows exhibit energy injection at intermediate scales in the near-wall region, associated with the characteristic structures of the near-wall cycle, while the scale separation between these energy-containing scales and the Kolmogorov scale does not increase with the Reynolds number. Indeed, in wall-turbulent flows, the energy-injection scale and the Kolmogorov scale, hereafter denoted by $L_S$ and $\eta$, respectively, scale as $L_S / \delta_\nu = \kappa z / \delta_\nu$ and $\eta / \delta_\nu = (\kappa z / \delta_\nu)^{1/4}$, where $z$ is the wall-normal coordinate, $\delta_\nu$ is the viscous length scale, and $\kappa$ is the von K\'arm\'an constant. Hence, for a fixed wall distance in wall units, their ratio reads
\begin{equation}
    \lim_{\substack{\nu \to 0 \\ z/\delta_\nu = \mathrm{const}}} \frac{L_S}{\eta} = \left( \frac{\kappa z}{\delta_\nu} \right)^{3/4} = \mathrm{const}
\end{equation}
In particular, since the location of maximum energy production is at $z / \delta_\nu = 11$, and therefore moves closer to the wall as the Reynolds number increases, we obtain
\begin{equation}
     \frac{L_S}{\eta} = \left( 11 \kappa \right)^{3/4} \approx 3,
\end{equation}
showing that the energy-injection and Kolmogorov scales decrease in size with increasing Reynolds number at the same rate (see \citet{Marati2004,cimarelli2024spatially} for more details). As shown in previous studies \citep{Piomelli1991,Domaradzki1994,Hartel1994,DunnMorrison2005, cimarelli2013paths}, this results in a transfer of energy towards larger wall-parallel scales, thus representing a substantial departure from the classical picture. Moreover, the transfer of energy towards such large scales also requires a dissipation mechanism capable of removing energy from them. In the near-wall region, this is primarily achieved through sharp wall-normal velocity gradients, corresponding to small wall-normal scales, which dissipate energy associated with motions extending over much larger wall-parallel scales \citep{cimarelli2024spatially}. The presence of both this inverse interscale energy transfer and the modified dissipative dynamics can be interpreted in terms of a confinement effect induced by the solid wall.

Some features of the modified turbulent dynamics observed near solid boundaries have also been reported at the turbulent/non-turbulent interface (TNTI), which separates the rotational turbulent region from the surrounding irrotational flow \citep{DaSilva2014}. In particular, the TNTI is characterized by an inverse energy flux towards larger horizontal scales, together with a direct transfer of energy towards smaller vertical scales, accompanied by the development of strong vertical gradients associated with energy dissipation (see \citet{cimarelli2015spectral}, \citet{zhou2020energycascade} and \citet{cimarelli2024spatially}). The TNTI may therefore also be regarded as inducing a form of confinement, although weaker than that associated with a solid wall.

Further insight into the physical phenomena described above, namely the inverse interscale energy transfer and the dissipation associated with large wall-parallel scales, can be gained by investigating the enstrophy dynamics. Indeed, enstrophy is closely related both to energy dissipation, through an exact relationship, and to the energy cascade through the vortex-stretching mechanism. Single-point enstrophy statistics have previously been investigated in both wall-bounded turbulence and at the TNTI. In wall-bounded flows, the analysis of enstrophy statistics has provided a perspective at least as informative as that offered by the energy dynamics. In particular, the no-slip condition gives rise to the generation of vorticity at the wall and its subsequent diffusion into the fluid, a process commonly described in terms of boundary-vorticity flux \citep{Lighthill1963,Eyink2008,Bechlars_Sandberg_2017}. More recently, this concept has been extended to enstrophy through the introduction of the boundary-enstrophy flux \citet{Terrington2023}. More broadly, previous studies have employed enstrophy budgets to investigate the mechanisms governing enstrophy production and transport in wall-bounded flows, highlighting the roles of strain-vorticity alignment, different contributions to enstrophy production and wall-normal vorticity transport (see \citet{Gorski1994,Bechlars_Sandberg_2017,Keylock2025,kumar2023vorticity} for a partial review). Enstrophy has also proven to be a useful quantity for investigating the TNTI, providing insight into the dynamics of the interface between turbulent and non-turbulent regions \citep{Bisset2002,Westerweel2005,Holzner2007,DaSilvaPereira2008,DaSilva2011, cimarelli2015spectral}.

However, the multiscale nature of the phenomena discussed above, makes it necessary to adopt a statistical framework capable of capturing their scale-dependent features. To this end, we employ the two-point formalism introduced by \citet{Hill2002exact}, extended to the two-point turbulent enstrophy. The two-point enstrophy framework has previously been applied to homogeneous isotropic turbulence \citep{Baj2022simultaneous,boga2026two} and polymeric turbulence \citep{Chiarini_2025}, with interesting results. The analysis is carried out using direct numerical simulations of a temporally evolving boundary layer. Given its statistical homogeneity in the streamwise direction, this flow configuration allows for simpler budget equations and a reduction in the dimensionality of the two-point budgets compared with the more traditional spatially evolving boundary layer.  At the same time, despite its simpler configuration, the temporally evolving boundary layer, first introduced by \citet{kozul2016}, qualitatively exhibits the characteristic features of both wall-bounded turbulence \citep{cimarelli2024energy, cimarelli2024spatially} and the TNTI \citep{ZhangTBL2018}.

The analysis is organized as follows. Section \ref{sec:DNS} describes the flow configuration and the DNS settings. The discussion of the results begins in section \ref{sec:SP} with an analysis of single-point enstrophy statistics, including the enstrophy budget terms and the streamwise, spanwise and wall-normal enstrophy contributions, thereby providing an initial, global picture of the enstrophy dynamics. The subsequent sections focus on two-point enstrophy statistics. Section \ref{sec:TP} introduces the two-point framework, while section \ref{sec:TP_enstr_contr} examines the second-order structure functions of the streamwise, spanwise and wall-normal vorticity components. The analysis then proceeds to the two-point enstrophy budget, which is discussed in sections \ref{sec:TP_enstr_3D} and \ref{sec:TP_enstr_1D}. Finally, concluding remarks are provided in section \ref{sec:concl}.

\section{Numerical setup and flow configuration}\label{sec:DNS}

The present section describes the flow configuration and numerical settings used to generate the dataset analysed throughout this work. The flow configuration studied is a temporally evolving boundary layer, which consists of an initially quiescent fluid in a rectangular domain with a solid wall at the bottom moving at a constant velocity $U_w$. The flow configuration is the one adopted in \citet{kozul2016}. Contrary to the more traditional spatially evolving boundary layer, the temporal boundary layer is statistically homogeneous in both the streamwise and spanwise directions, and evolves in time rather than in the streamwise direction (see figure \ref{fig:Vol}(b)). This additional homogeneity considerably simplifies the statistical equations describing it, also thanks to its null average wall-normal velocity. Despite these simplifications, previous studies have shown that the temporal boundary layer reproduces all the main physical mechanisms observed in the spatially evolving boundary layer \citep{kozul2016, ZhangTBL2018, cimarelli2024energy, cimarelli2024spatially}.

The computational domain dimensions are denoted by $(L_x,L_y,L_z)$ in the streamwise $x$, spanwise $y$ and wall-normal $z$ directions, respectively. Periodic boundary conditions are imposed for both velocity and pressure in the homogeneous $x$ and $y$ directions. In the wall-normal direction, Dirichlet boundary conditions are imposed at the bottom of the domain for the velocity, $z=0$, imposing the streamwise $\upsilon^*$, spanwise $v^*$ and wall-normal $w^*$ velocity components to $(\upsilon^*,v^*,w^*)=(U_w,0,0)$. At the top of the domain, $z=L_z$, impermeable free-slip boundary conditions are imposed, $(\partial \upsilon^* / \partial z, \partial v^* / \partial z, w^*) = (0, 0, 0)$. Neumann boundary conditions are used for the pressure both at the bottom, $z=0$, and at the top, $z=L_z$ of the domain, $\partial p^* / \partial z = 0$.

The flow is initialized with the streamwise velocity profile
\begin{equation}
    \upsilon^*(x,y,z,t=0) = (1+c) \frac{U_w}{2} \left[ 1 + \tanh{\left( \frac{D - z}{ 2h} \right)} \right],
\end{equation}
where $D$ is the initial profile thickness and $h = 54 \nu / U_w$ is the momentum thickness of the imposed shear layer with $\nu$ the kinematic viscosity. To accelerate the transition to turbulence, a uniformly distributed white-noise perturbation $c(x,y,z) \in [-0.05,0.05]$ with zero mean is superimposed onto the velocity profile. The spanwise and wall-normal velocity components are initially set to zero. The Reynolds number based on the initial profile thickness is set to $Re = U_w D / \nu = 500$ and is chosen to obtain a rapid loss of memory of the initial conditions (see \citet{kozul2016} for further details). Unless otherwise specified, all quantities are nondimensionalized using the characteristic scales $D$, $U_w$, and $D/U_w$ for length, velocity, and time, respectively.

As the boundary layer develops, the friction Reynolds number $Re_{\tau} = u_{\tau} \delta / \nu$, where $\delta$ is the boundary layer thickness based on the $99\%$ of the mean velocity and $u_{\tau}$ is the friction velocity, increases with time. The analyses presented in this work are performed at the time instant corresponding to $Re_{\tau} = 1501$. The domain dimensions have been designed to properly describe this final state and are reported in table \ref{tab:tab1}. The domain is discretized using $(N_x,N_y,N_z)=(3072,3072,768)$ grid points. Uniform spacing is used in the streamwise and spanwise directions, whereas an error-function stretching (Chapter 2 in \citet{Orlandi2000}) is employed in the wall-normal direction with clustering parameter $\alpha_z=2.25$. The resulting grid resolution in viscous units is summarized in table \ref{tab:tab1}, where the superscript $\cdot^+$ indicates normalization using the viscous length scale $\delta_\nu = \nu / u_\tau$ and the friction velocity $u_\tau$.  The reader is referred to \citet{cimarelli2024energy} for further details.

The simulations have been performed using the open-source code CaNS \citep{costa2018fft}, which employs a standard pressure-projection method and a staggered second-order finite-difference scheme for the spatial discretization. Time integration is carried out using a mixed approach. In particular, viscous terms in the wall-normal direction are implicitly integrated through the use of a Crank–Nicholson scheme, while all other terms are explicitly integrated by using a three-step Runge–Kutta method with $\mathit{CFL} = 0.95$.

\begin{table}
\centering
\caption{Domain size and spatial resolution at the friction Reynolds number analysed. The subscripts $\cdot_w$ and $\cdot_\delta$ indicate quantities evaluated at the wall, $z=0$, and at the boundary layer edge, $z=\delta$, respectively, and $\eta$ is the Kolmogorov length scale.}
\label{tab:tab1}
\begin{tabular}{lcccccc}
\hline
$Re_\tau$ & $(L_x,L_y,L_z)/\delta$ & $(\Delta x^+,\Delta y^+)$ & $(\Delta z_w^+,\Delta z_\delta^+)$ & $\Delta z_\delta / \eta$ \\
\hline
1501 & (11.9, 5.9, 2.8) & (5.8, 2.9) & (0.09, 9.2) & 1.13 & \\
\hline
\end{tabular}
\end{table}

In order to stick with the customary view of wall-bounded flows, in the rest of the paper we will consider a reference frame moving at a constant speed with the wall. The resulting streamwise velocity $u^* = \upsilon^* - U_w$ is zero at the wall and $U_w$ outside the boundary layer. Throughout this paper, the standard Reynolds decomposition is used $u_i^* = U_i + u_i$, where the superscript $\cdot^*$ indicates total quantities, capital letters indicate average quantities $U = \langle u^* \rangle$ and small case letters indicate fluctuating quantities. The average operator is indicated by $\langle \cdot \rangle$ and, it is computed by averaging in the statistically homogeneous streamwise and spanwise directions, $x$ and $y$, and between four independent ensembles obtained by performing four independent simulations with a different seed of the random noise in the initial condition $c$.

\section{Turbulent enstrophy}\label{sec:SP}

Turbulent enstrophy, defined as $\zeta = \omega_i \omega_i$, where $\boldsymbol{\omega} = \boldsymbol{\nabla} \times \boldsymbol{u}$ is the fluctuating vorticity, is an observable closely related to the very nature of turbulence. In fact, in addition to its rotational nature, enstrophy is closely related to both energy dissipation and the energy cascade through the vortex stretching mechanism. For these reasons, it may give an alternative view in cases where the energy cascade and dissipation mechanisms are altered, such as in the vicinity of solid walls or near a turbulent/non-turbulent interface (TNTI).

\begin{figure}[t!]
\centering
\includegraphics[width=0.4\linewidth]{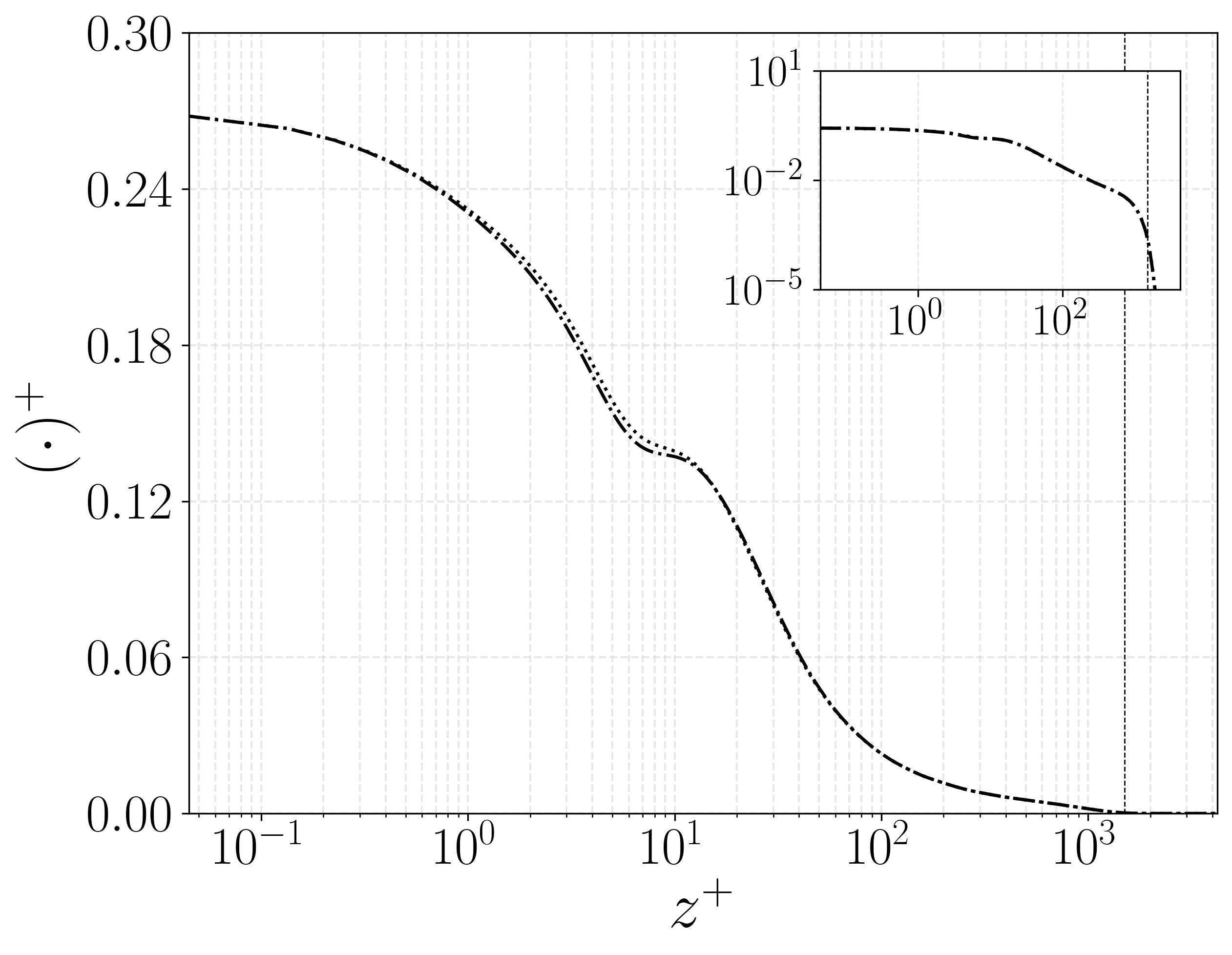}\mylab{-55mm}{44mm}{(\aaa)}
\includegraphics[trim=6cm 1cm 4.5cm 3cm, clip=true, width=0.59\linewidth]{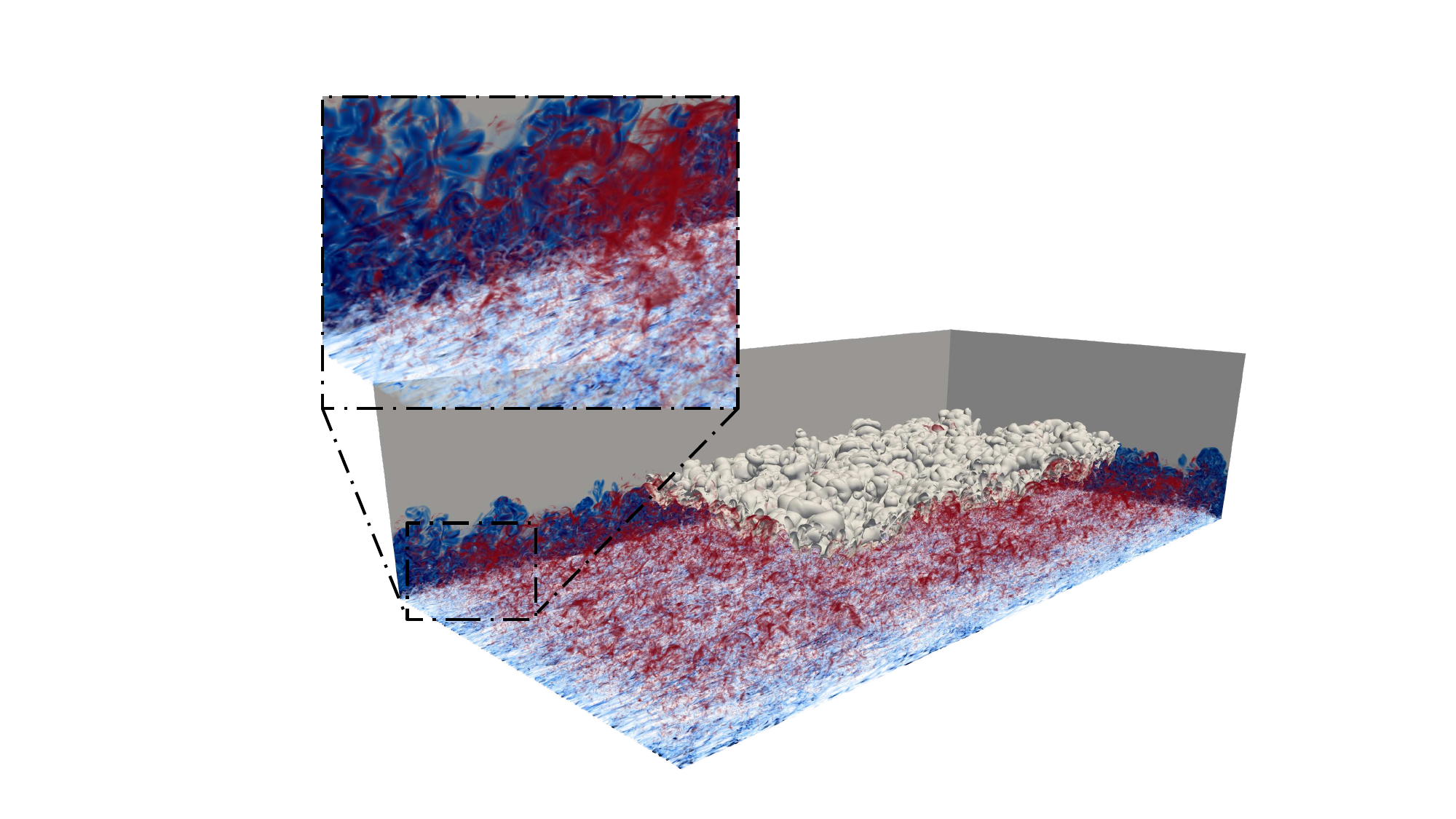}\mylab{-10mm}{44mm}{(\bbb)}
        \caption{(a) Turbulent enstrophy multiplied by viscosity $(\nu \langle \zeta \rangle)^+$ (dash-dotted lines) compared to turbulent pseudo-dissipation $\langle \epsilon^+ \rangle$ (dotted lines). The two terms are reported in logarithmic scale in the inset panel in order to better visualise the outer region. The vertical dotted line indicates the mean interface location. (b) Instantaneous snapshot of the enstrophy field. The red volume rendering represents the relative difference between the squared strain rate and the  enstrophy $|\frac{\epsilon / \nu - \zeta}{\epsilon / \nu}|$, where high values are shown as opaque red and low values are transparent. The enstrophy field is also volume-rendered, with high values shown as opaque blue and low values fading to transparent white. Two slices of the enstrophy field are reported using a logarithmic colormap, ranging from blue (high values) to white (low values). A portion of the enstrophy isosurface $\zeta = 1.4 \cdot 10^{-6} \langle \zeta \rangle|_w$ is shown in opaque grey. Both figures refer to a friction Reynolds number $Re_{\tau} = 1501$.
\label{fig:Vol}}
\end{figure}

The mean turbulent enstrophy and the mean turbulent dissipation are related through the following equation,
\begin{equation}
\langle \epsilon \rangle = \nu \langle \zeta \rangle + \nu \langle \frac{\partial u_i}{\partial x_j} \frac{\partial u_j}{\partial x_i} \rangle,
\end{equation}
where $\epsilon = \nu (\partial u_i/\partial x_j)(\partial u_i/\partial x_j)$ denotes the pseudo-dissipation. For the symmetries of the temporal boundary layer, this relation simplifies to
\begin{equation}
\langle \epsilon \rangle = \nu \langle \zeta \rangle + \nu \frac{\partial^2 \langle w w \rangle}{\partial z^2}.
\label{eq_ens_diss}
\end{equation}
Hence, the difference between the mean pseudo-dissipation and $\nu\langle\zeta\rangle$ is determined by the curvature of the vertical profile of wall-normal velocity variance $\langle w w \rangle$. For the present flow configuration, this difference is found to be negligible and mostly confined to the region between the viscous sublayer and the buffer layer, as shown in figure \ref{fig:Vol}(a), where the mean enstrophy and pseudo-dissipation profiles are compared. This suggests that the considerations discussed below in relation to enstrophy may also be relevant, to some extent, to the energy dissipation.

Although enstrophy and turbulent dissipation are closely related in terms of their mean values, their instantaneous topology and spatial distributions differ substantially \citep{CarboneBragg2020JFM}. In wall-bounded turbulence, however, this difference becomes much less pronounced in the near-wall region than in the outer region because vortex sheets are the dominant structure types close to the wall. Under these conditions, regions of high enstrophy and high dissipation mostly coincide \citep{Bechlars_Sandberg_2017}, as also illustrated in figure~\ref{fig:Vol}(b) for the present simulation.



\subsection{Single-point enstrophy budget}

In this subsection, we first examine the single-point enstrophy budget to identify the dominant mechanisms governing the evolution of turbulent enstrophy. This analysis provides the basis for the following sections, where the enstrophy dynamics is investigated in greater detail by examining how enstrophy is divided among its streamwise $\langle \omega_x^2 \rangle$, spanwise $\langle \omega_y^2 \rangle$ and wall-normal $\langle \omega_z^2 \rangle$ contributions, and subsequently extending the analysis to two-point statistics.

Under the statistical symmetries of the temporal boundary layer, the evolution of turbulent enstrophy $\langle \zeta \rangle$ is described by the following transport equation,
\begin{equation}
\begin{gathered}
        \frac{\partial \langle \zeta \rangle}{\partial t} = \underbrace{- \frac{\partial \langle \zeta w \rangle}{\partial z}}_{T_{\zeta}} \underbrace{+ \nu \frac{\partial^2 \langle \zeta \rangle}{\partial z^2}}_{D_{\zeta}} \underbrace{-2 \langle \omega_y w \rangle \frac{\partial \Omega_y}{\partial z}}_{P_{\zeta}} \\
        \underbrace{+ 2 \langle \omega_x \omega_z \rangle \frac{\partial U}{\partial z} + 2 \Omega_y \langle \omega_i \frac{\partial u_i}{\partial y}\rangle + 2 \langle \omega_i \omega_j \frac{\partial u_i}{\partial x_j} \rangle}_{\xi} - \underbrace{2 \nu \langle \frac{\partial \omega_i}{\partial x_j} \frac{\partial \omega_i}{\partial x_j} \rangle}_{\chi}.
\label{eq:enstr_SP_TBL}
\end{gathered}
\end{equation}
where $\Omega_y = \langle \omega_y \rangle$ is the average spanwise vorticity. Here, $T_{\zeta}$ and $D_{\zeta}$ denote the inertial and viscous transport terms, respectively, $P_{\zeta}$ represents enstrophy production due to the mean vorticity gradient, $\xi$ is the vortex stretching and tilting term, and $\chi$ is the viscous destruction of enstrophy associated with vorticity gradients (referred to as enstrophy dissipation in the following).

\begin{figure}[t!]
\centering
\includegraphics[width=0.477\linewidth]{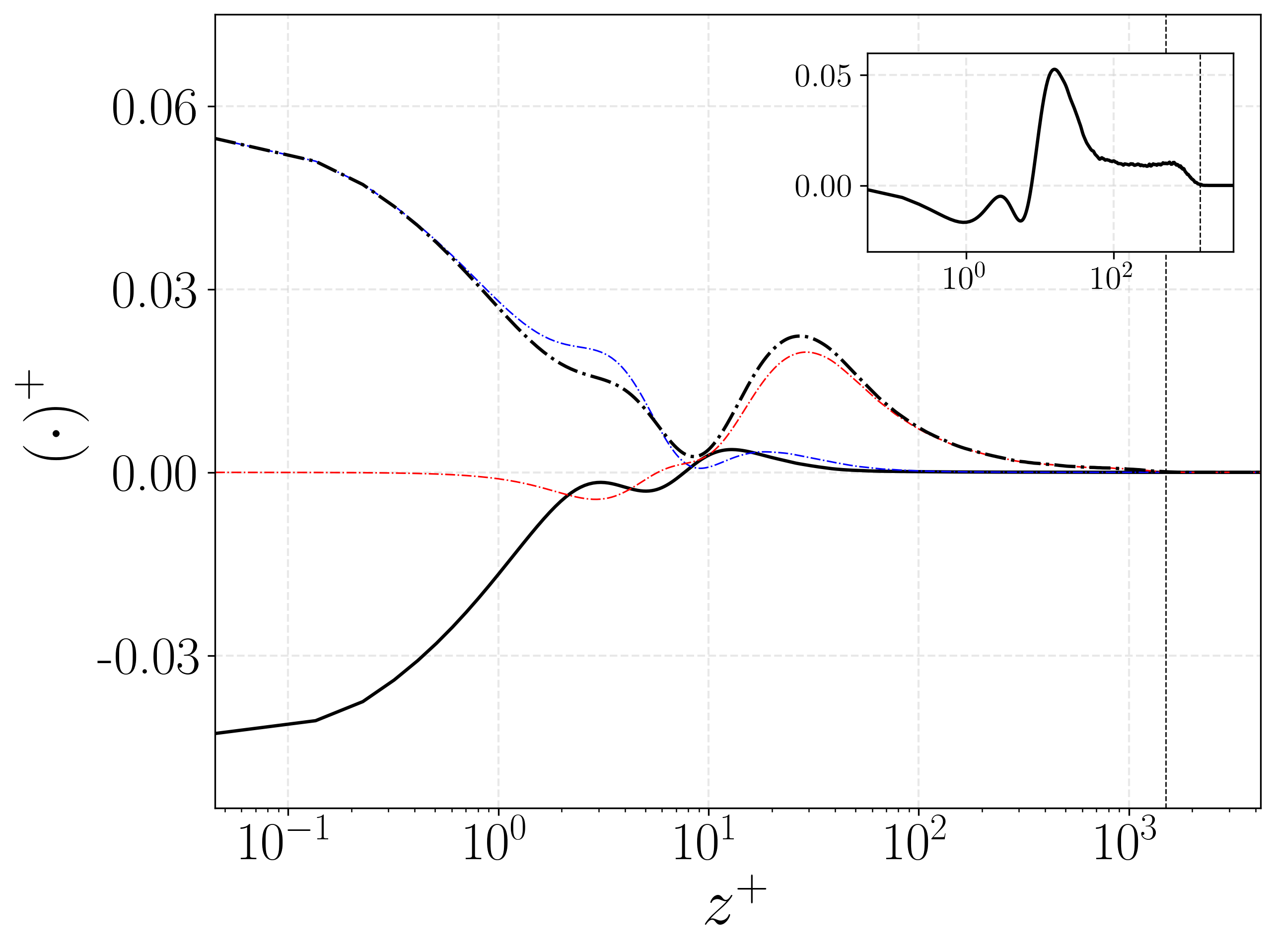}\mylab{-66mm}{44mm}{(\aaa)}
\includegraphics[width=0.503\linewidth]{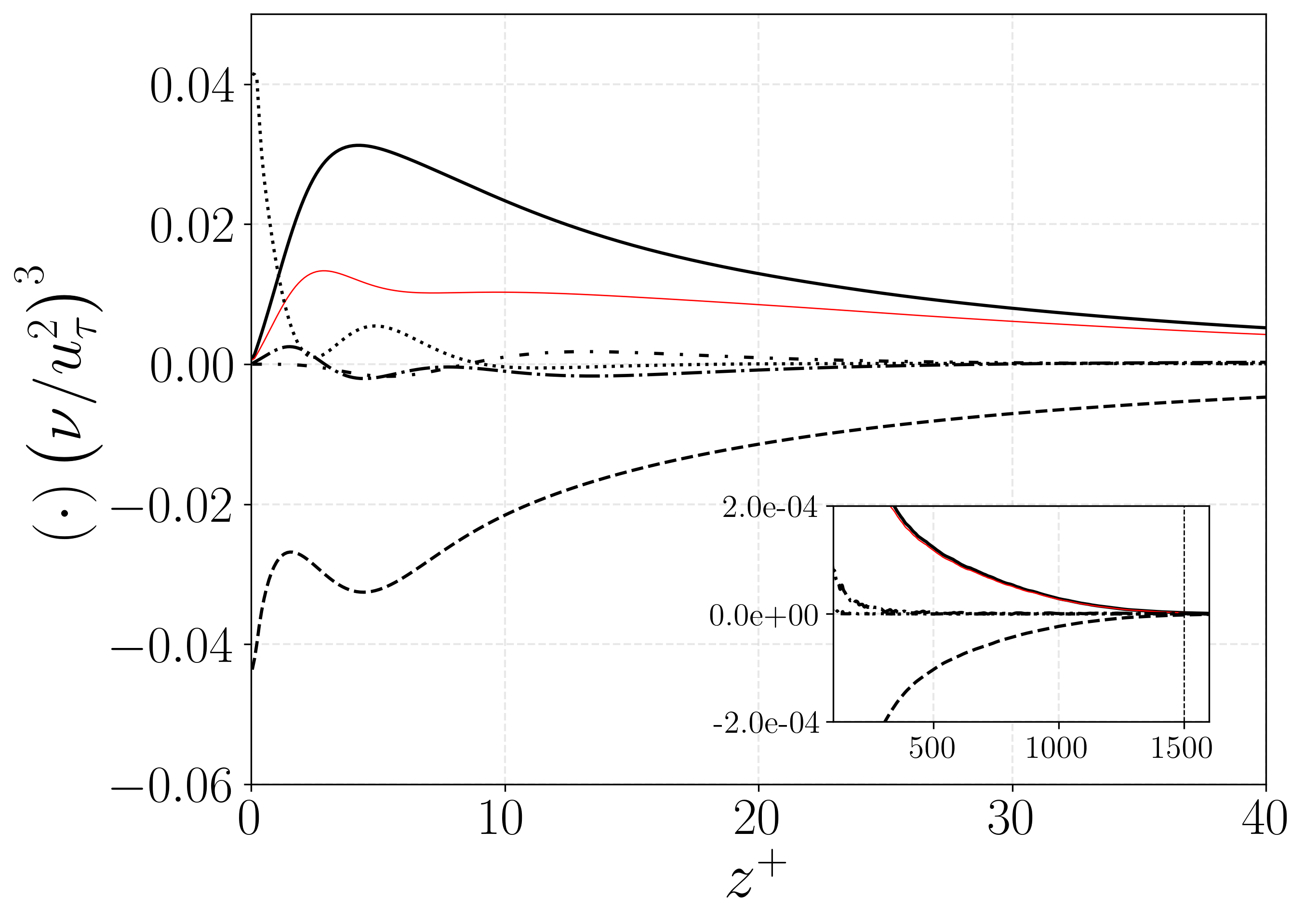}\mylab{1mm}{44mm}{(\bbb)}
        \caption{(a) Source and fluxes of turbulent enstrophy. The source term $S^+$ is reported as a black solid line, the total flux $\phi^+$ as a black dash-dotted line, the viscous flux $- (\nu \partial \langle \zeta \rangle / \partial z)^+$ as a blue dash-dotted line and the turbulent flux $\langle \zeta w \rangle^+$ as a red dash-dotted line. In the inset panel the premultiplied source term $y^+ S^+$ is reported. (b) Turbulent enstrophy budget terms: vortex stretching $\xi^+$ (black solid line), turbulent transport $T_{\zeta}^+$ in dash-dotted line, diffusive transport $D_{\zeta}^+$ in dotted line, production by mean vorticity gradient $P_{\zeta}^+$ in loosely dash-dotted lines and enstrophy destruction $- \chi^+$ in dashed line. The fully fluctuating contribution to the vortex stretching $ 2 \langle \omega_i \omega_j \partial u_i / \partial x_j \rangle^+$ is reported as a red solid line. The inset panel shows the budget terms in the outer region. All curves are reported in viscous units and refer to the friction Reynolds number $Re_{\tau} = 1501$. The vertical dotted line indicate the mean interface position.
\label{fig:enstr_budg}}
\end{figure}

Equation (\ref{eq:enstr_SP_TBL}) may be rewritten in conservative form by separating a net enstrophy source and a divergence of a flux
\begin{equation}
\begin{gathered}
        \frac{\partial \langle \zeta \rangle}{\partial t} + \frac{\partial \phi}{\partial z} = S 
\label{eq:enstr_SP_TBL_fluxes}
\end{gathered}
\end{equation}
where $\phi = \langle \zeta w\rangle - \nu \partial \langle \zeta \rangle / \partial z$ is the wall-normal component of the flux of enstrophy and $S = P_{\zeta} + \xi - \chi$ is the net source/sink of enstrophy. This formulation provides a clearer visualization of the regions of net enstrophy production and destruction and shows where enstrophy is transported on average. As shown in figure \ref{fig:enstr_budg}(a), enstrophy is generated at the wall and is injected in the domain through diffusion (i.e. through the boundary enstrophy flux \citep{Morton1984,Lighthill1963,Lyman1990,Terrington2023}). At the same time, the presence of the wall gives rise to strong vorticity gradients, making the near-wall region the primary location of enstrophy dissipation. As a result, the enstrophy flux exhibits its maximum at the wall and decreases rapidly as it passes through the viscous sublayer because of the draining effect of the local enstrophy dissipation. Nevertheless, the enstrophy flux remains positive throughout the whole boundary layer, indicating a net transfer of enstrophy from the wall to the outer regions. The flux exhibits a local minimum in the lower part of the buffer layer. Below this location, the flux is primarily sustained by viscous diffusion, whereas farther from the wall turbulent transport becomes the dominant contribution. Around the same wall-normal location, the net enstrophy source changes sign, marking the transition from the near-wall sink to a region of net enstrophy production. This positive source is strongest within the buffer layer and remains positive throughout the core of the boundary layer. To better highlight the relative contribution of the different wall-normal regions to the integrated enstrophy budget, the inset of figure \ref{fig:enstr_budg}(a) shows the source term premultiplied by the wall-distance coordinate.

By examining the individual terms of the enstrophy budget equation (\ref{eq:enstr_SP_TBL}), shown in figure \ref{fig:enstr_budg}(b), it is possible to obtain a more detailed picture of the mechanisms governing enstrophy dynamics and to compare the relative magnitude of production and transport terms. The largest contributions to the enstrophy budget are confined to the near-wall region. The dominant terms of the budget are: the diffusive transport of enstrophy $D_{\zeta}$ which is active up to the first few viscous units, its production through vortex stretching, $\xi$, and its viscous destruction, $\chi$. In contrast, the turbulent transport term, $T_{\zeta}$, remains small throughout the boundary layer and never becomes comparable to the other contributions. In the very-near wall region for $z^+ \lesssim 1$, the turbulent enstrophy entering the flow satisfies the simple balance between diffusive transport and viscous dissipation,
\begin{equation}
\nu \frac{\partial^2 \langle \zeta \rangle}{\partial z^2} = 2 \nu \langle \frac{\partial \omega_i}{\partial x_j} \frac{\partial \omega_i}{\partial x_j} \rangle \qquad \mbox{for} \,\,\,  z^+ \lesssim 1
\end{equation}
Away from the wall, diffusive transport becomes negligible, and the enstrophy dynamics is essentially governed by the balance between vortex stretching and viscous destruction, $\xi = \chi$, see again figure \ref{fig:enstr_budg}(b). Therefore, outside the viscous sublayer, the dominant mechanisms governing turbulent enstrophy are essentially local in space, with transport playing only a minor role. In particular, the budget takes a very simple form for  $z^+ \gtrsim 40$. At these wall-normal distances, the contributions to vortex stretching associated with the mean flow become negligible, so that enstrophy production is almost entirely due to the fluctuating component, $\xi \approx 2 \langle \omega_i \omega_j (\partial u_i/\partial x_j) \rangle$, shown in red in figure \ref{fig:enstr_budg}(b). Hence, the enstrophy budget reduces to the following local equilibrium condition,
\begin{equation}
\langle \omega_i \omega_j \frac{\partial u_i}{\partial x_j} \rangle = - \nu \langle \frac{\partial \omega_i}{\partial x_j} \frac{\partial \omega_i}{\partial x_j} \rangle \qquad \mbox{for} \,\,\,  z^+ \ge 40
\end{equation}

In summary, the turbulent enstrophy budget highlights two distinct regions, governed by different enstrophy dynamics. The first is a near-wall sink region, $z^+<10$, where enstrophy enters the flow diffusively from the wall and is subsequently dissipated. The second is the bulk turbulent region, $z^+>10$, where enstrophy is transported further away from the wall by turbulent mixing. In the buffer layer ($10<z^+<20$), the local production mechanisms exceed viscous dissipation. Farther away from the wall, the enstrophy dynamics is increasingly governed by spatially local processes, with production by vortex stretching reaching an approximate equilibrium with viscous dissipation.

\subsection{Confinement effects} \label{conf_eff}

In the current flow configuration, vorticity can only be generated by the shear forces introduced at the wall. Since the wall is a planar surface with normal aligned in the $z$-direction, it can only directly generate streamwise and spanwise vorticity component, corresponding to the enstrophy contributions $\zeta_x = \langle \omega_x \omega_x \rangle$ and $\zeta_y = \langle \omega_y \omega_y \rangle$. As mentioned in the previous subsection, the enstrophy generated  at the wall is then transported away from it, initially by diffusion, as the wall-normal velocity vanishes at the wall ($w \sim z^2$ in the near-wall region), and progressively by turbulent convective transport farther from the wall. During this process, the vortex stretching term $\xi$, which accounts not only for the stretching and compression of vortical elements but also for their tilting, amplifies enstrophy and redistributes it among the three vorticity components.

\begin{figure}[t!]
\centering
\includegraphics[width=0.5\linewidth]{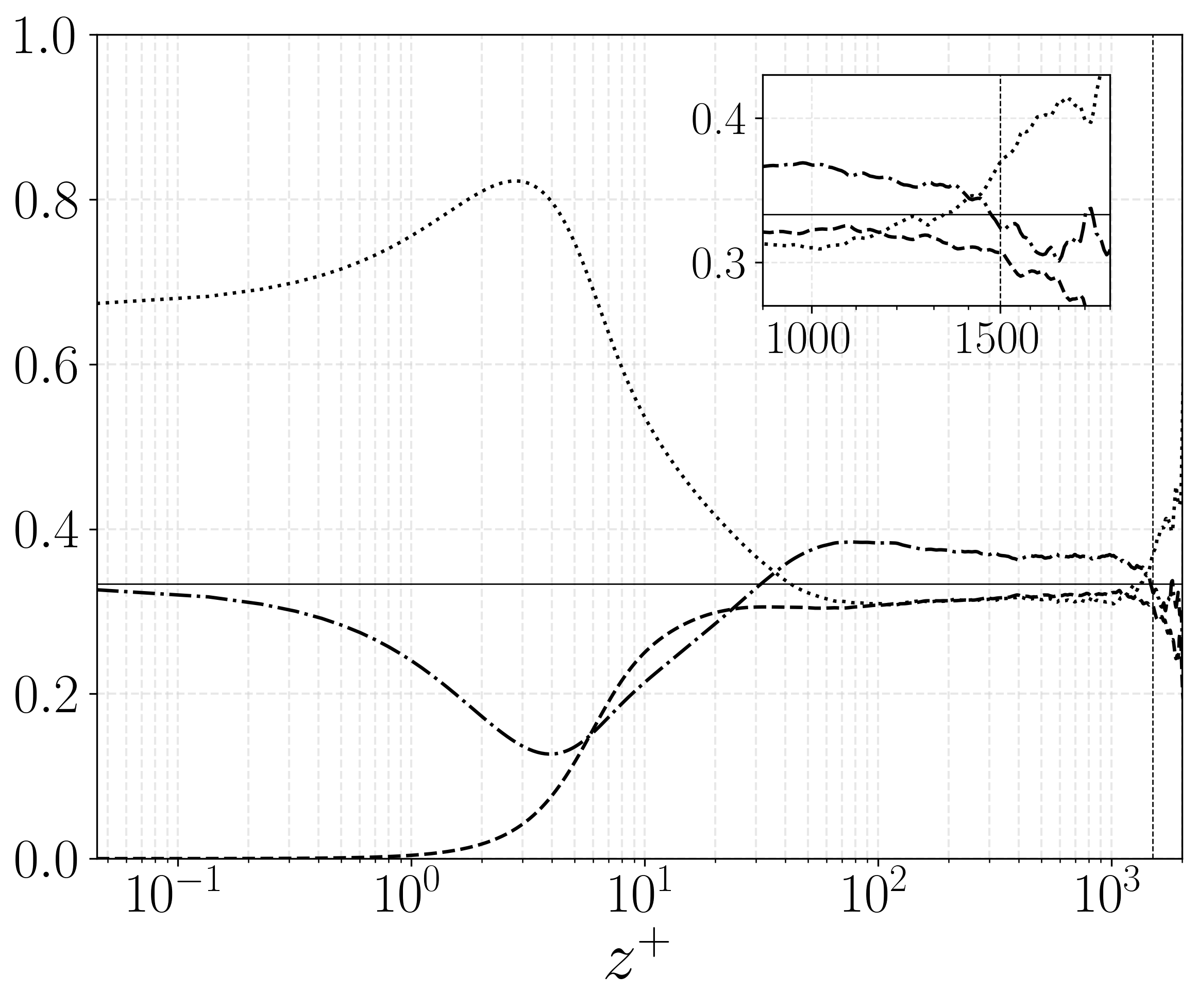}
        \caption{Distribution of the enstrophy content among its three contributions: streamwise $\zeta_x / \zeta$ (dash-dotted line), spanwise $\zeta_y / \zeta$ (dotted line) and wall-normal $\zeta_z / \zeta$ (dashed line). The reference value of $1/3$, corresponding to equal distribution is reported as a solid line, while the vertical dotted line indicates the mean interface position $z^+ = Re_{\tau}$. In the inset panel, an enlargement is reported to better highlight the interface region.}
\label{fig:enstr_contr}
\end{figure}

The redistribution process can be examined by considering the three enstrophy contributions associated with the streamwise, spanwise, and wall-normal vorticity components, namely $\zeta_x = \langle \omega_x \omega_x \rangle$, $\zeta_y = \langle \omega_y \omega_y \rangle$, and $\zeta_z = \langle \omega_z \omega_z \rangle$. Their wall-normal distribution is shown in figure \ref{fig:enstr_contr}, where each contribution is normalized by the total turbulent enstrophy. This normalization allows a clear visualization of the relative distribution of enstrophy among its contributions, highlighting the level of anisotropy (under isotropic conditions, each vorticity component contributes equally to the total enstrophy, $\zeta_x/\zeta = \zeta_y/\zeta = \zeta_z/\zeta = 1/3$). At the same time, this representation allows for a clear view of the enstrophy distribution among its three contributions across the entire boundary layer, despite the enstrophy intensity varying by several orders of magnitude with the wall distance.

As expected, the wall-normal enstrophy contribution is zero at the wall $\zeta_{z}|_w = 0$, such that the total turbulent enstrophy is entirely distributed between its spanwise and streamwise components. Specifically, the spanwise contribution accounts for two thirds of the total enstrophy, while the remaining one third is associated with the streamwise component ($\zeta_y / \zeta \approx 2/3$ and $\zeta_x / \zeta \approx 1/3$). The wall-normal enstrophy contribution remains negligible up to approximately $z^+ \approx 3$, indicating that the presence of the wall suppresses the development of a fully three-dimensional vorticity field throughout most of the viscous sublayer. As enstrophy is transported farther away from the wall and reaches the buffer layer, the vortex stretching and tilting term $\xi$ starts to redistribute enstrophy towards the wall-normal component thus leading to a fully three-dimensional vorticity field for $z^+ \ge 10$. Notice that $z^+ = 10$ has already been identified through the turbulent enstrophy budget as the crossover wall distance separating the diffusive enstrophy-sink layer from the turbulent enstrophy-source layer. For $z^+ \ge 10$, the streamwise contribution becomes slightly larger than the spanwise and wall-normal contributions, which remain of comparable magnitude. This behaviour is consistent with the fact that, in wall-turbulent flows, turbulent motions tend to show an elongated structure in the streamwise direction, which produces relatively weaker velocity gradients in the streamwise direction. Consequently, the streamwise enstrophy contribution $\zeta_x$, which is the only one not involving streamwise velocity gradients, becomes the largest of the three.

As we approach the TNTI region, for $z^+ = Re_{\tau} = 1501$, the enstrophy distribution starts to resemble that observed in the near-wall region, although less markedly. As highlighted in the inset of figure~\ref{fig:enstr_contr}, the spanwise contribution $\zeta_y$ becomes dominant once again, while the wall-normal contribution returns to be the weakest. This scenario is consistent with the idea proposed by \citet{cimarelli2015spectral, cimarelli2021spatially, cimarelli2024spatially} that, in the TNTI region, as in the near-wall region, the flow generates long, wide structures that dissipate energy through strong gradients in the vertical direction. This picture is further supported by the conditional TNTI analysis of \citet{zhou2020energycascade}, who showed that energy interscale transfer near the TNTI is associated with an increase in scale in the tangent to the interface direction, and a reduction in scale in the direction normal to the interface.

\begin{figure}[t!]
\centering
\includegraphics[width=0.49\linewidth]{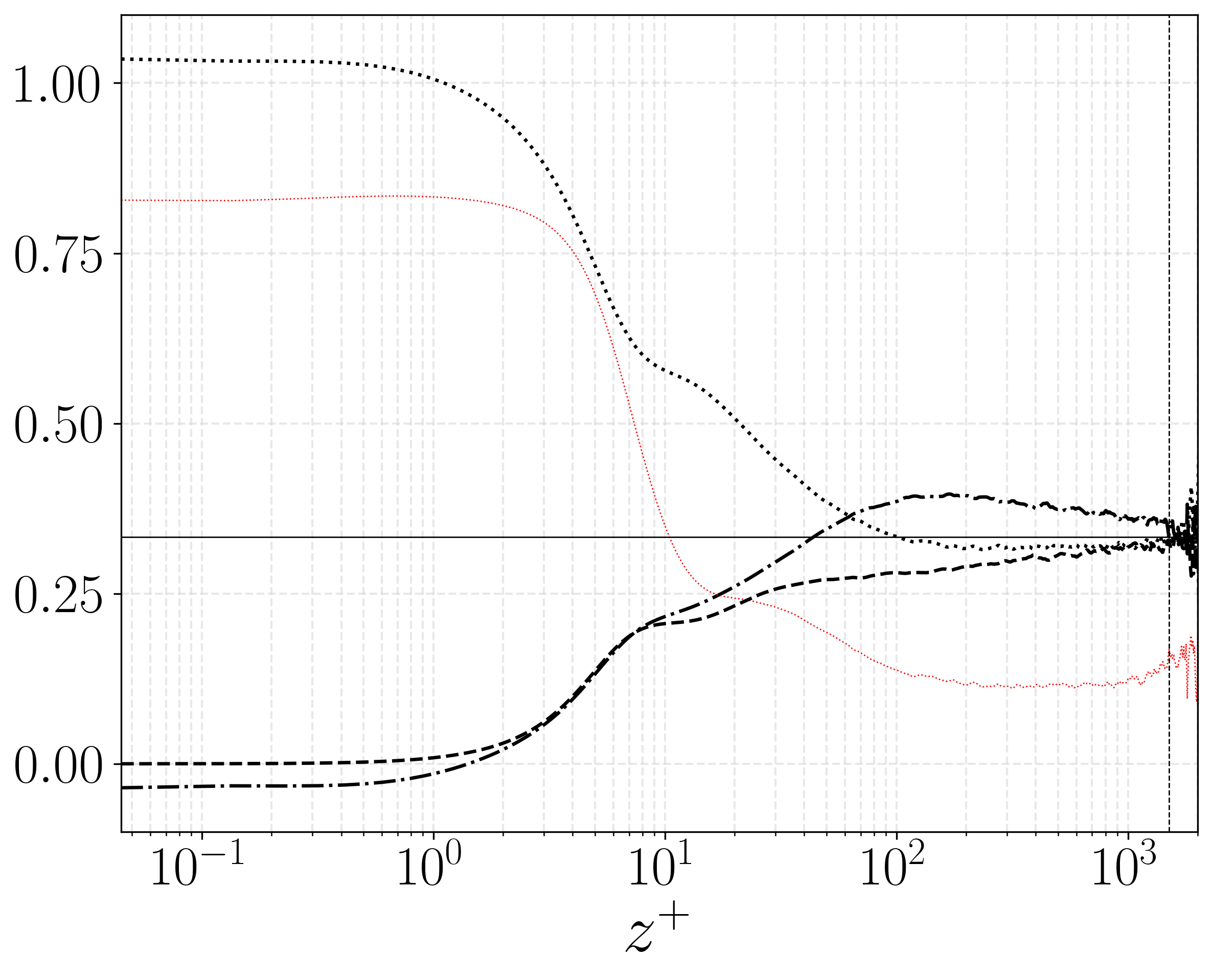}\mylab{-70mm}{50mm}{(\aaa)}
\includegraphics[width=0.49\linewidth]{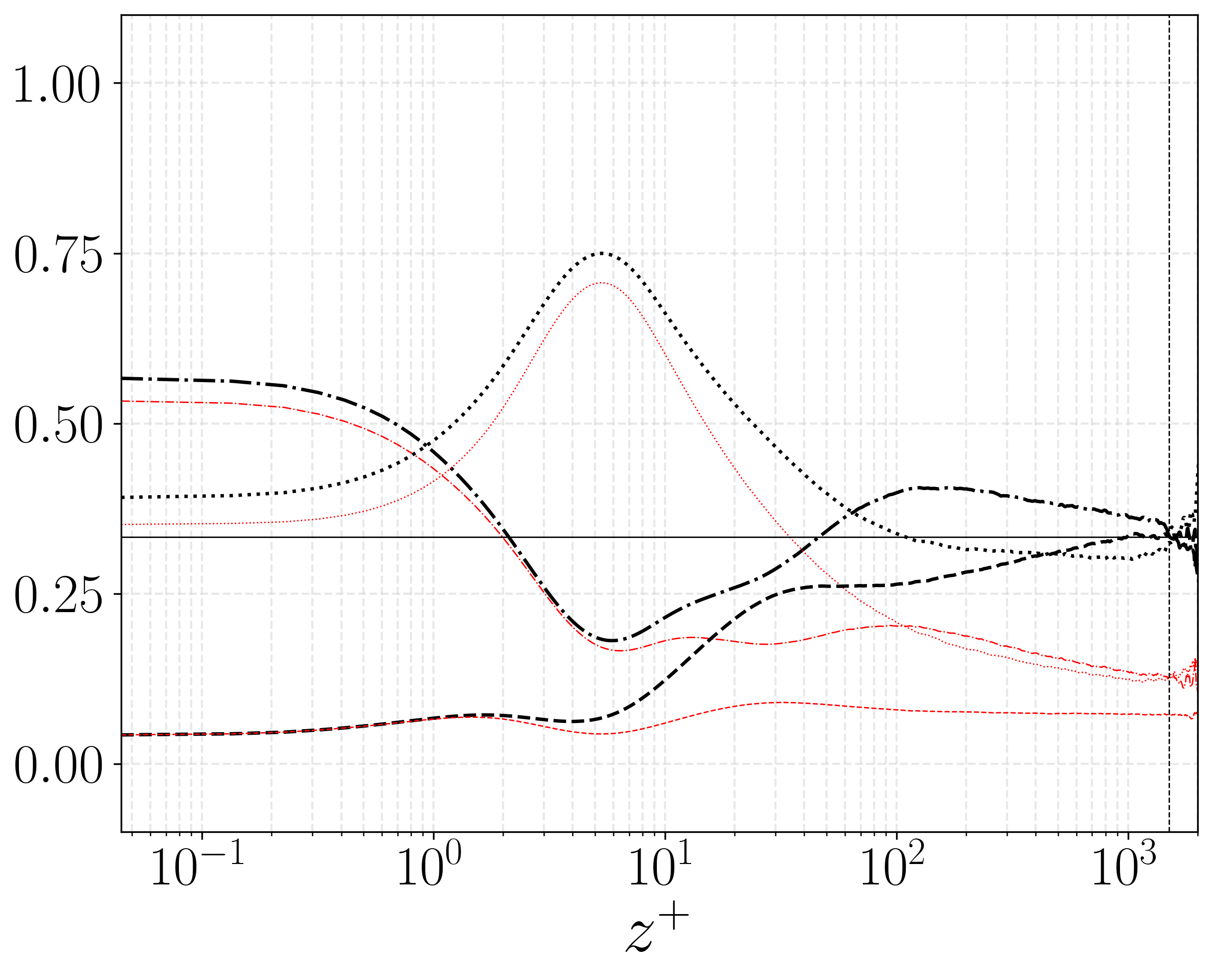}\mylab{2mm}{50mm}{(\bbb)}
        \caption{Relative contributions of (a) the vortex stretching term $\xi_x / \xi $ (dash-dotted line), $\xi_y / \xi$ (dotted line) and $\xi_z / \xi$ (dashed line) and (b) of the enstrophy destruction term $\chi_x / \chi$ (dash-dotted line), $\chi_y / \chi$ (dotted line) and $\chi_z / \chi$ (dashed line). In (a) the term $2 \Omega_y \langle \omega_y (\partial v/\partial y)\rangle + 2 \langle \omega_y \omega_y (\partial v/\partial y) \rangle$ is reported as a red dotted line. In (b) the terms $2 \nu \langle (\partial \omega_x / \partial z )^2 \rangle $ (dash-dotted), $2 \nu \langle (\partial \omega_y / \partial z )^2 \rangle $ (dotted) and $2 \nu \langle (\partial \omega_z / \partial z )^2 \rangle $ (dashed) are reported as red lines. The reference value of $1/3$, corresponding to an equal distribution, is reported as black solid lines in both (a) and (b), while the densely dotted vertical lines indicate the mean interface position $z^+ = Re_{\tau}$.}
\label{fig:VS_enstrDestr_distrib}
\end{figure}

While the decomposition discussed above provides insight into the local orientation of the vorticity field, it does not reveal the mechanisms responsible for the evolution of each contribution. To address this point, it is useful to derive separate transport equations for the three enstrophy components $\zeta_x$, $\zeta_y$ and $\zeta_z$. These equations identify the production, transport, and dissipation mechanisms governing each contribution individually and, together, recover the total enstrophy balance. They read, \begin{equation}
\begin{gathered}
\frac{\partial \langle \zeta_x \rangle}{\partial t} = \underbrace{- \frac{\partial \langle \omega_x \omega_x w \rangle}{\partial z}}_{T_{\zeta_x}} + \underbrace{\nu \frac{\partial^2 \langle \omega_x \omega_x \rangle}{\partial z^2}}_{D_{\zeta_x}}, \\
+ \underbrace{2 \langle \omega_x \omega_z \rangle \frac{\partial U}{\partial z} + 2 \Omega_y \langle \omega_x \frac{\partial u}{\partial y}\rangle + 2 \langle \omega_x \omega_j \frac{\partial u}{\partial x_j} \rangle}_{\xi_x} - \underbrace{2 \nu \langle \frac{\partial \omega_x}{\partial x_j} \frac{\partial \omega_x}{\partial x_j} \rangle}_{\chi_x},
\label{eq:enstr_SP_x}
\end{gathered}
\end{equation}
\begin{equation}
\begin{gathered}
\frac{\partial \langle \zeta_y \rangle}{\partial t} = \underbrace{ - \frac{\partial \langle \omega_y \omega_y w \rangle}{\partial z}}_{T_{\zeta_y}} + \underbrace{\nu \frac{\partial^2 \langle \omega_y \omega_y \rangle}{\partial z^2}}_{D_{\zeta_y}} + \underbrace{-2 \langle \omega_y w \rangle \frac{\partial \Omega_y}{\partial z}}_{P_{\zeta_y}}, \\
+ \underbrace{2 \Omega_y \langle \omega_y \frac{\partial v}{\partial y}\rangle + 2 \langle \omega_y \omega_j \frac{\partial v}{\partial x_j} \rangle}_{\xi_y} - \underbrace{2 \nu \langle \frac{\partial \omega_y}{\partial x_j} \frac{\partial \omega_y}{\partial x_j} \rangle}_{\chi_y},
\label{eq:enstr_SP_y}
\end{gathered}
\end{equation}
\begin{equation}
\begin{gathered}
\frac{\partial \langle \zeta_z \rangle}{\partial t} = \underbrace{- \frac{\partial \langle \omega_z \omega_z w \rangle}{\partial z}}_{T_{\zeta_z}} + \underbrace{\nu \frac{\partial^2 \langle \omega_z \omega_z \rangle}{\partial z^2}}_{D_{\zeta_z}}, \\
+ \underbrace{2 \Omega_y \langle \omega_z \frac{\partial w}{\partial y}\rangle + 2 \langle \omega_z \omega_j \frac{\partial w}{\partial x_j} \rangle}_{\xi_z} - \underbrace{2 \nu \langle \frac{\partial \omega_z}{\partial x_j} \frac{\partial \omega_z}{\partial x_j} \rangle}_{\chi_z} \, .
\label{eq:enstr_SP_y}
\end{gathered}
\end{equation}
As already shown in figure \ref{fig:enstr_budg}(b), the enstrophy dynamics is governed primarily by production through vortex stretching $\xi$ and destruction through vorticity gradients $\chi$ throughout most of the boundary layer. The only exception is the immediate vicinity of the wall, where diffusive transport $D_{\zeta}$ also plays a significant role. Accordingly, the following discussion focuses on these two dominant mechanisms and, in particular, on how they are partitioned among their streamwise, spanwise, and wall-normal contributions, namely $\xi_x$, $\xi_y$, and $\xi_z$ for vortex stretching, and $\chi_x$, $\chi_y$, and $\chi_z$ for enstrophy destruction. Their distributions are shown in figure \ref{fig:VS_enstrDestr_distrib}, where each contribution is normalized by the corresponding total term ($\xi$ or $\chi$) to facilitate comparison of their relative importance.

As discussed above, the immediate vicinity of the wall is characterized by a confined dynamics, in which the wall condition inhibits the development of a fully three-dimensional turbulent dynamics. As a result, the vortex stretching term $\xi$ also shows an uneven distribution among its contributions. Figure \ref{fig:VS_enstrDestr_distrib}(a) shows that, throughout the viscous sublayer, the only significant vortex stretching contribution is $\xi_y$, which amplifies the spanwise enstrophy contribution $\zeta_y$. This production is dominated by the ``pure'' stretching of the mean and fluctuating spanwise vorticity, namely the term $2 \Omega_y \langle \omega_y \partial v / \partial y \rangle + 2 \langle \omega_y \omega_y \partial v / \partial y \rangle$, reported in red in figure \ref{fig:VS_enstrDestr_distrib}(a). By contrast, although the streamwise enstrophy contribution $\zeta_x$ is the second largest in the viscous sublayer, it is not sustained by local vortex stretching. Instead, it is primarily dissipated through wall-normal vorticity gradients $-2 \nu \langle (\partial \omega_x / \partial z)^2 \rangle$, as indicated by the red dash-dotted curve in figure \ref{fig:VS_enstrDestr_distrib}(b), while the remaining part is transported away from the wall by viscous diffusion. Beyond the viscous sublayer, the vortex stretching contributions $\xi_x$, $\xi_y$, and $\xi_z$ all become significant, reflecting again the progressive recovery of a fully three-dimensional turbulent dynamics. From the overlap layer onwards, their relative magnitudes closely follow those of the corresponding enstrophy contributions discussed previously, with the streamwise contribution becoming the dominant one.

Having discussed the production of the three enstrophy contributions, we now examine their destruction. Unlike vortex stretching, the enstrophy destruction term exhibits a different partition among its contributions, particularly in the near-wall region. Figure \ref{fig:VS_enstrDestr_distrib}(b) reports the three destruction contributions, $\chi_x$, $\chi_y$, and $\chi_z$, normalized by the total enstrophy destruction rate $\chi$. As expected, all three enstrophy contributions are dissipated predominantly through wall-normal gradients in the vicinity of the wall (red curves in figure \ref{fig:VS_enstrDestr_distrib}(b)). Although the production of streamwise enstrophy, $\xi_x$, is negligible in the viscous sublayer compared with the spanwise contribution, $\xi_y$, its destruction is the largest of the three ($\chi_x > \chi_y$). This indicates that the streamwise vorticity generated at the wall is not sustained by local vortex stretching. Rather, it is  generated at the wall through spanwise velocity fluctuations $v$ and is subsequently transported away from the wall by viscous diffusion while being continuously dissipated by the strong wall-normal gradients. Only after reaching the buffer and outer layers, the three destruction contributions recover a distribution similar to that of the enstrophy contributions themselves.

In summary, the decomposition of enstrophy in its three contributions highlights a confinement of turbulent motions in the near-wall region ($z^+ < 10$) and at the turbulent/non-turbulent interface region ($z^+ \approx Re_\tau$). In both regions, the wall-normal vorticity component is weaker than the streamwise and spanwise components, although this difference is much more pronounced near the wall, where the wall-normal component is negligible and the vorticity field is therefore predominantly two-dimensional. In this region, spanwise vorticity is the most intense component and is the only one to exhibit amplification through vortex stretching. In contrast, streamwise vorticity is generated by wall-boundary effects and subsequently diffuses away from the wall, while being dissipated by the wall-normal gradients.


\section{Two-point turbulent enstrophy}\label{sec:TP}

The analysis presented in section \ref{sec:SP} provides a simple and synthetic view of how turbulent enstrophy is transported, produced, and dissipated within the boundary layer. However, it does not provide information on the structure and mechanisms that govern its transfer across scales of different dimensions. Instead, these aspects can be investigated through a two-point statistical tool. In the present section, we address this by using the formalism provided by the generalized Kolmogorov equation \citep{Hill2002exact}, applied to turbulent enstrophy. This formalism has already been used for the statistical symmetries of homogenous isotropic turbulence in \citet{Baj2022simultaneous,boga2026two} and in polymeric homogeneous and isotropic turbulence in \citet{Chiarini_2025}. The generalized Kolmogorov equation for enstrophy represents the exact evolution equation for the second-order structure function of fluctuating vorticity, defined as
\begin{equation}
\langle \delta \omega^2 \rangle = \langle \delta \omega_i \delta \omega_i \rangle,
\label{eq:enstr_struct_fun}
\end{equation}
where $\delta \omega_i = \omega_i (\bm{x'}, t) - \omega_i (\bm{x''}, t)$ denotes the vorticity increment between the two points $\bm{x'}$ and $\bm{x''}$. Within this framework, the space of scales is represented by the separation vector $\bm{r} = \bm{x'} - \bm{x''}$, while the physical space is represented by the midpoint $\bm{x_c} = (\bm{x'} + \bm{x''})/2$. Hence, the second-order vorticity structure function can be interpreted as a measure of the enstrophy intensity at different scales $\bm{r}$ and positions within the flow $\bm{x_c}$. For this reason, we will refer to it hereafter as ``scale enstrophy''.

The second-order vorticity structure function depends on a 7-dimensional space $\langle \delta \omega^2 \rangle = \langle \delta \omega^2 \rangle(\bm{x_c}, \bm{r}, t)$, comprising three physical-space coordinates ($\bm{x_c}$), three scale-space coordinates ($\bm{r}$), and time ($t$). In the flow settings of the temporal boundary layer considered here, the streamwise and spanwise spatial homogeneities reduce the problem to the 5-dimensional space $\langle \delta \omega^2 \rangle(r_x, r_y, r_z, z_c, t)$. To further reduce the dimensionality of the problem, we restrict the analysis of the space of scales to the sole two-dimensional space of homogenous scales by considering the hyperplane $r_z = 0$. The problem therefore reduces to a 4-dimensional space $\langle \delta \omega^2 \rangle(r_x, r_y, z_c, t)$. Applying the statistical symmetries of the temporal boundary layer and restricting the analysis to the subspace $(r_x,r_y,z_c,t)_{|r_z=0}$, the exact equation governing the scale enstrophy reads
\begin{equation}
\begin{gathered}
       \frac{\partial \langle \delta \omega^2 \rangle}{\partial t} = \underbrace{ - \frac{\partial \langle \delta \omega^2 \delta u_j \rangle}{\partial r_j}}_{T_r} + \underbrace{2 \nu \frac{\partial^2 \langle \delta \omega^2 \rangle}{\partial r_j \partial r_j}}_{D_r} \underbrace{- \frac{\partial \langle \delta \omega^2 \widetilde{w} \rangle}{\partial z_{c}} + \frac{\nu}{2} \frac{\partial^2 \langle \delta \omega^2 \rangle}{\partial z_c^2}}_{T_c}\\
       \underbrace{+ 2 \Omega_y \langle \delta \omega_i \,\delta \!\left( \frac{\partial u_i}{\partial y} \right) \rangle 
         + 2 \langle \delta \omega_x \delta \omega_z \rangle \left( \frac{\partial U}{\partial z} \right) + 2 \langle \delta \omega_i \widetilde{\omega}_j \, \delta \! \left( \frac{\partial u_i}{\partial x_j} \right) \rangle + 2 \langle \delta \omega_i \delta \omega_j \widetilde{\left( \frac{\partial u_i}{\partial x_j} \right)} \rangle}_{\xi_{\delta \omega^2}} \\
       \underbrace{- 2 \langle \delta \omega_y \delta w \rangle \left( \frac{\partial \Omega_y}{\partial z} \right)}_{\Pi_{\delta \omega^2}} - 2 \chi,
\label{eq:enstr_TP_TBL}
\end{gathered}
\end{equation}
where $\widetilde{\cdot}$ represents the two-point average, defined as $\widetilde{f} = ( f_i(\bm{x'},t) + f_i(\bm{x''},t) )/2$. The terms in equation \ref{eq:enstr_TP_TBL} represent, respectively: $T_r$ and $D_r$, the inertial and viscous transport of scale enstrophy across scales; $T_c$ the combined inertial and viscous transport of scale enstrophy in physical space (which, in the present flow configuration, reduces to the wall-normal direction); $\xi_{\delta \omega^2}$, the source of scale enstrophy related to vortex stretching; $\Pi_{\delta \omega^2}$, the production of scale enstrophy due to mean vorticity gradients; and $- 2 \chi$, the viscous destruction of scale enstrophy.

Similarly to the single-point enstrophy budget \ref{eq:enstr_SP_TBL}, the two-point enstrophy budget (\ref{eq:enstr_TP_TBL}) can be rewritten by grouping the source terms in a single net source and expressing the transport terms as the divergence of fluxes:
\begin{equation}
\begin{gathered}
       \frac{\partial \langle \delta \omega^2 \rangle}{\partial t} + \bm{\nabla_4} \cdot \bm{\phi_{\delta \omega^2}} = S_{\delta \omega^2} \,.
\label{eq:enstr_TP_TBL_fluxes}
\end{gathered}
\end{equation}
The 4-dimensional divergence operator comprises derivatives in both scale space and physical space $\bm{\nabla_4} = (\partial / \partial r_x, \partial / \partial r_y, \partial / \partial r_z, \partial / \partial z_c)$ and is applied to the field of fluxes $\bm{\phi_{\delta \omega^2}} = (\phi_{r_x}, \phi_{r_y}, \phi_{r_z}, \phi_{z_c})$ whose components are
\begin{equation}
\begin{gathered}
\phi_{r_i} = \langle \delta \omega^2 \delta u_i \rangle  - 2 \nu \frac{\partial \langle \delta \omega^2 \rangle}{\partial r_i}, \qquad  \phi_{z_c}  = \langle \delta \omega^2 \widetilde{w} \rangle - \frac{\nu}{2} \frac{\partial \langle \delta \omega^2 \rangle}{\partial z_c}.
\label{eq:TP_fluxes}
\end{gathered}
\end{equation}
The net source combines the contributions associated with vortex stretching, production due to mean vorticity gradients, and enstrophy destruction:
\begin{equation}
\begin{gathered}
S_{\delta \omega^2} = \xi_{\delta \omega^2} + \Pi_{\delta \omega^2} - 2 \chi.
\label{eq:enstr_source}
\end{gathered}
\end{equation}


The two-point enstrophy dynamics in the temporal boundary layer is examined as follows. We begin in section \ref{sec:TP_enstr_contr} by revisiting the confinement effects discussed in section \ref{conf_eff}, analysing how enstrophy is distributed among its three contributions at different wall-normal locations and characterizing their scale-space structure. We then investigate the two-point enstrophy budget in section \ref{sec:TP_enstr_3D}, providing an overall view of sources and fluxes of two-point enstrophy. Finally, a scale-by-scale analysis is performed at fixed wall distances in section \ref{sec:TP_enstr_1D}, yielding a simplified and more quantitative picture of the dominant mechanisms governing enstrophy dynamics throughout the boundary layer.

\section{Scale enstrophy and vorticity structure functions} \label{sec:TP_enstr_contr}

In the present subsection, we examine how enstrophy is distributed among its streamwise $\langle \delta \omega_x^2 \rangle$, spanwise $\langle \delta \omega_y^2 \rangle$, and wall-normal $\langle \delta \omega_z^2 \rangle$ contributions. For clarity, we restrict the discussion to four representative wall-normal locations: $z_c^+=0.5$, representative of the very near-wall region; $z_c^+=11$, within the buffer layer; $z_c^+=450$, in the outer region; and $z_c^+=Re_\tau=1501$, representative of the TNTI region.

\subsection{Very near-wall region}

\begin{figure}[t!]
\centering
\includegraphics[trim=3cm 1cm 60cm 15cm, clip=true, width=0.32\linewidth]{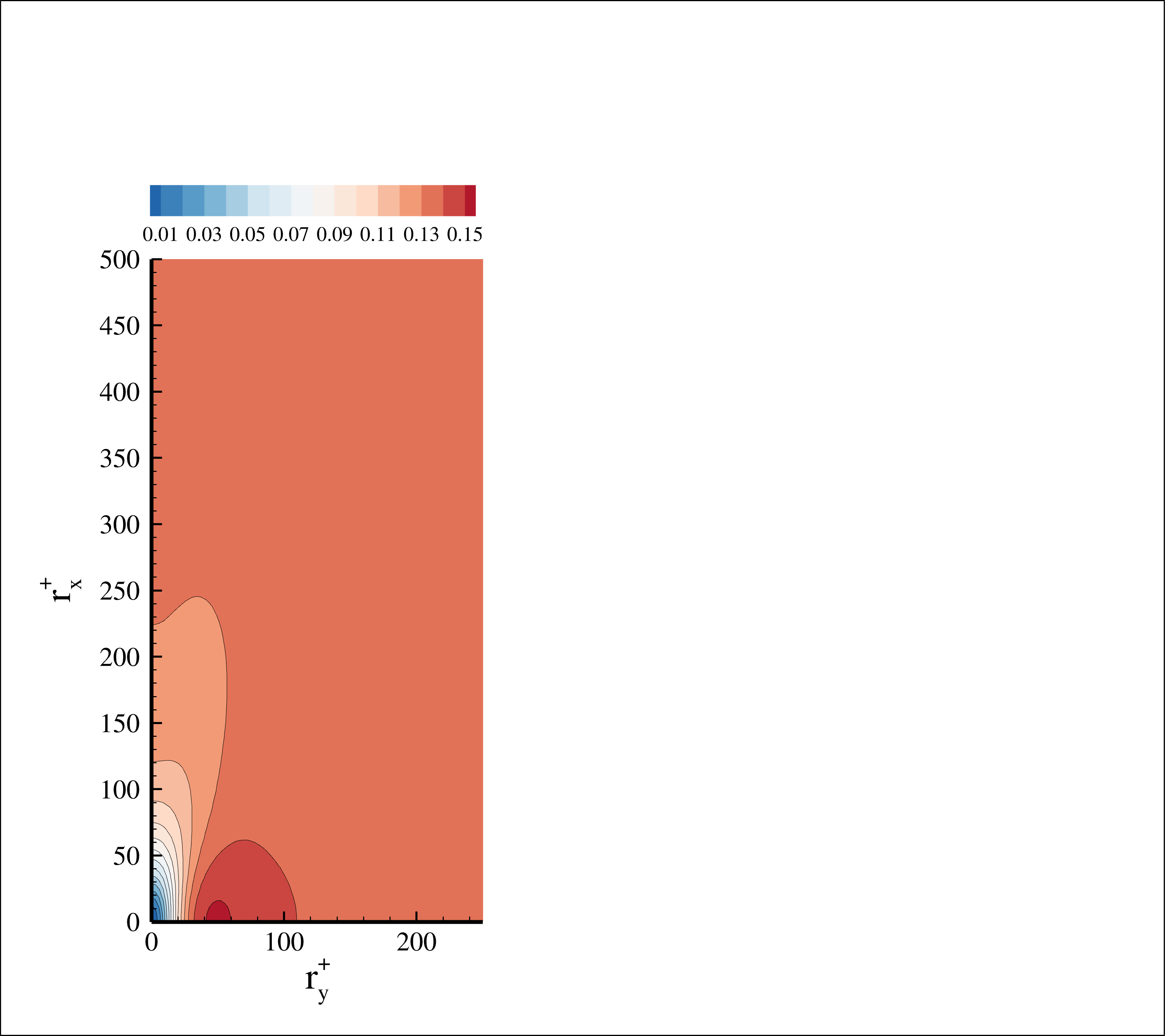}\mylab{-41mm}{74mm}{(\aaa)}
\includegraphics[trim=3cm 1cm 60cm 15cm, clip=true, width=0.32\linewidth]{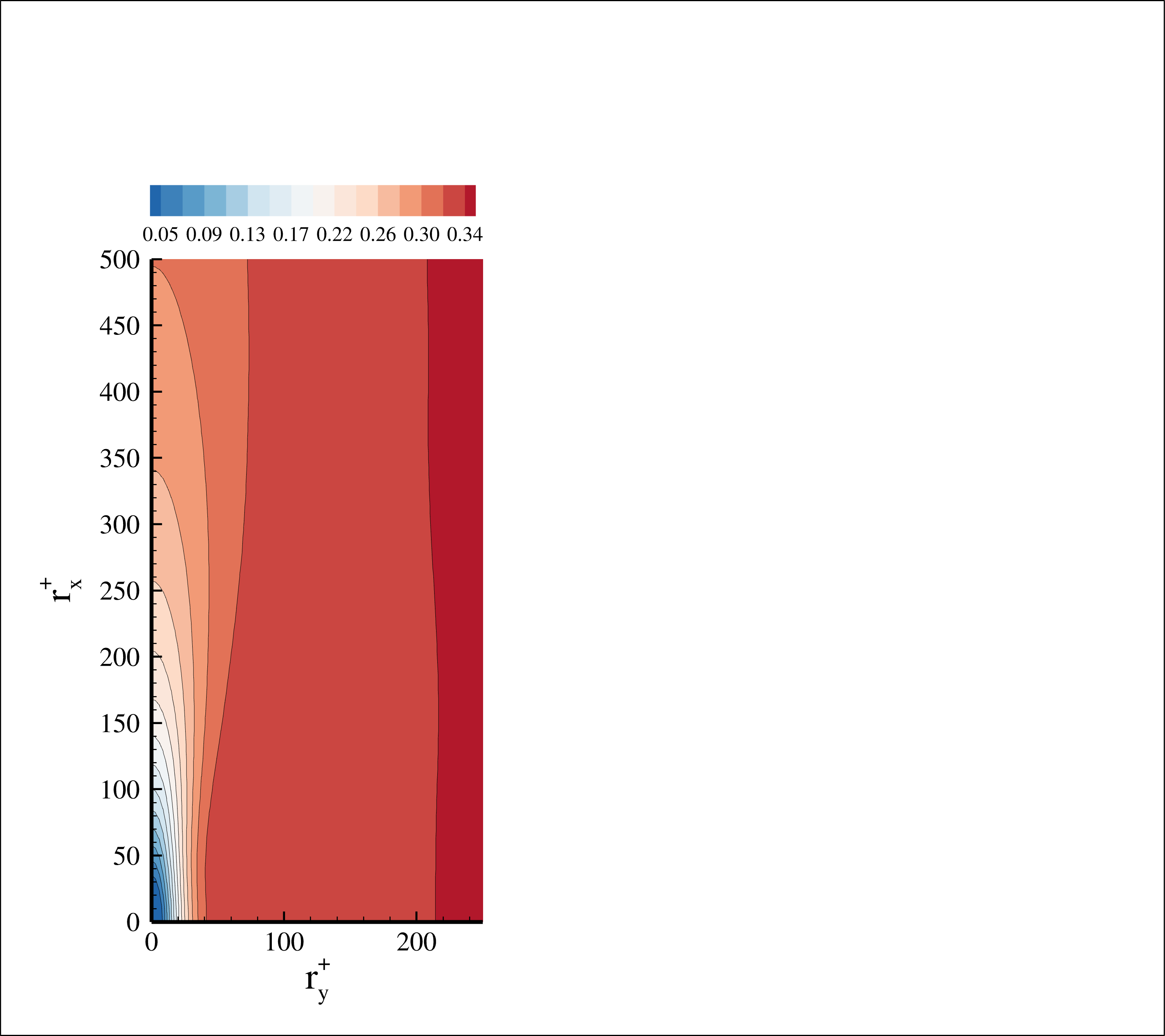}\mylab{-41mm}{74mm}{(\bbb)}
\includegraphics[trim=3cm 1cm 60cm 15cm, clip=true, width=0.32\linewidth]{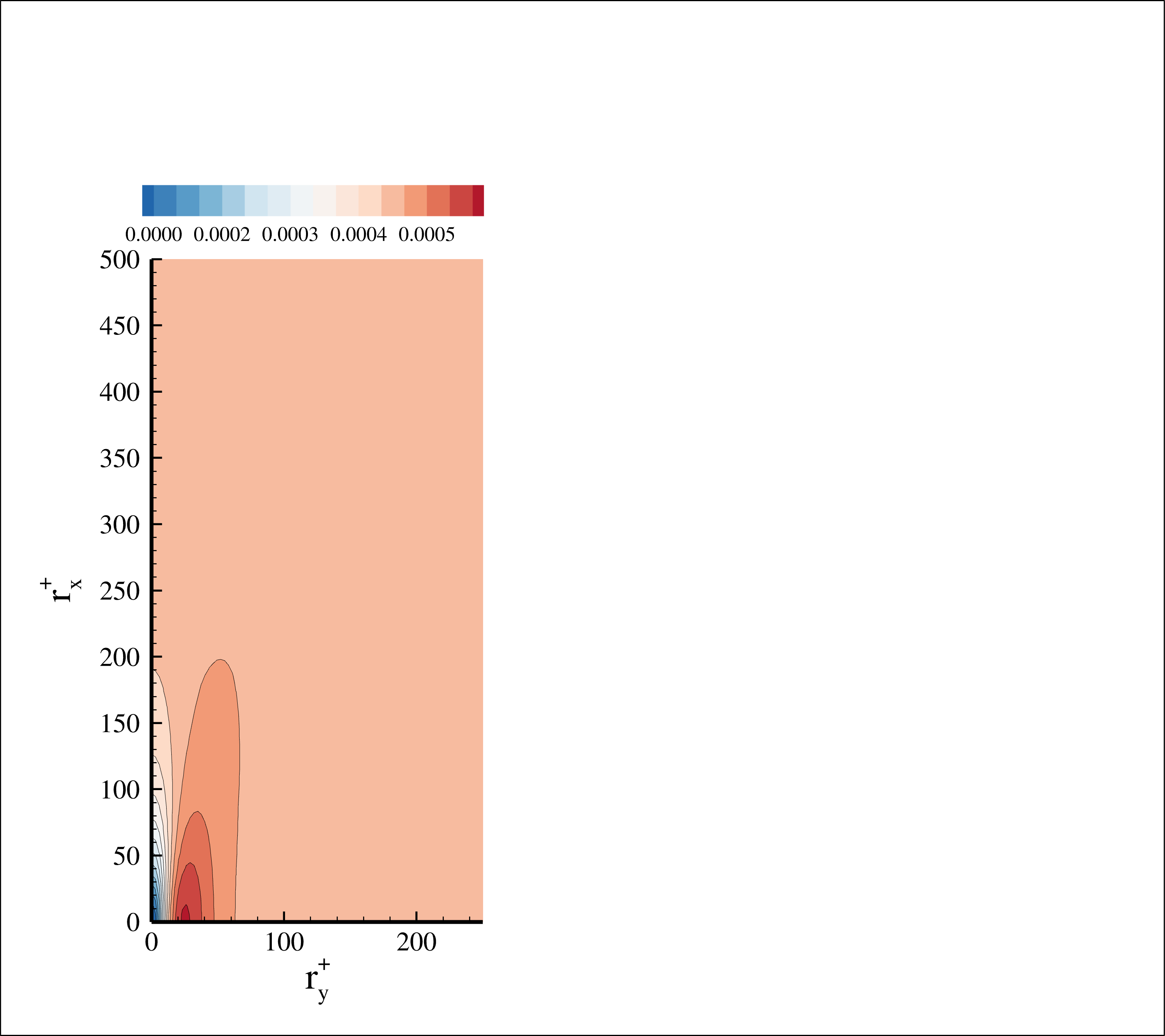}\mylab{-41mm}{74mm}{(\ccc)}
        \caption{Second-order structure functions of (a) streamwise $\langle \delta \omega_x^2 \rangle^+$, (b) spanwise $\langle \delta \omega_y^2 \rangle^+$ and (c) wall-normal $\langle \delta \omega_z^2 \rangle^+$ enstrophy contributions in viscous units at a wall-distance of $z_c^+ = 0.5$ in the $(r_x, r_y)|_{r_z = 0}$ subspace.
\label{fig:Sii_z05}}
\end{figure}

Motivated by the confinement effects discussed in section \ref{conf_eff}, we begin our analysis in the very near-wall region ($z_c^+=0.5$), where enstrophy is entirely constituted by the streamwise and spanwise vorticity components. The corresponding two-point structure functions provide a scale-space characterization of this near-wall enstrophy distribution. The results are shown in figure \ref{fig:Sii_z05}. Consistently with the single-point statistics shown in figure \ref{fig:enstr_contr}, the wall-normal contribution (figure \ref{fig:Sii_z05}(c)) is negligible with respect to the streamwise and spanwise contributions displayed in figure \ref{fig:Sii_z05}(a) and (b), respectively. The two non-negligible contributions exhibit the footprint of the coherent structures  populating the flow at larger wall-normal distances. In particular, the streamwise contribution, $\langle \delta \omega_x^2 \rangle$ (figure \ref{fig:Sii_z05}(a)), exhibits a local maximum at spanwise separations of approximately $r_y^+ \approx 50$, consistent with the characteristic spacing of the quasi-streamwise vortices residing in the buffer layer. We recall that streamwise vorticity at the wall can only be generated by spanwise velocity fluctuations, which induce a non-zero wall-normal gradient, $\partial v / \partial z$, through the no-slip condition. The observed peak can therefore be interpreted as the wall footprint of the spanwise velocity fluctuations induced by the counter-rotating quasi-streamwise vortices. These vortices generate alternating regions of positive and negative near-wall spanwise velocity $v$, which produce the corresponding streamwise vorticity fluctuations through the wall-normal velocity gradient. A similar argument applies to the spanwise enstrophy contribution $\langle \delta \omega_y^2 \rangle$. Spanwise vorticity at the wall is generated by streamwise velocity fluctuations which produce a non-zero $\partial u / \partial z$. Accordingly, spanwise vorticity (figure \ref{fig:Sii_z05}(b)) displays an elongated structure in the streamwise direction that reaches the saturation value only at very large streamwise and spanwise scales. This suggests that spanwise vorticity fluctuations at the wall primarily reflect the footprint of the high- and low-velocity streaks which populate the outer region. Although the wall-normal contribution, $\langle \delta \omega_z^2 \rangle$, is negligible in magnitude, it exhibits a peak at approximately the same spanwise scales $r_y^+$ as the one observed in the streamwise enstrophy contribution $\langle \delta \omega_x^2 \rangle$. As will be shown later, this feature is consistent with the vortex tilting mechanism acting on streamwise vorticity in the buffer layer.

\subsection{Buffer layer}

\begin{figure}[t!]
\centering
\includegraphics[trim=3cm 1cm 60cm 15cm, clip=true, width=0.32\linewidth]{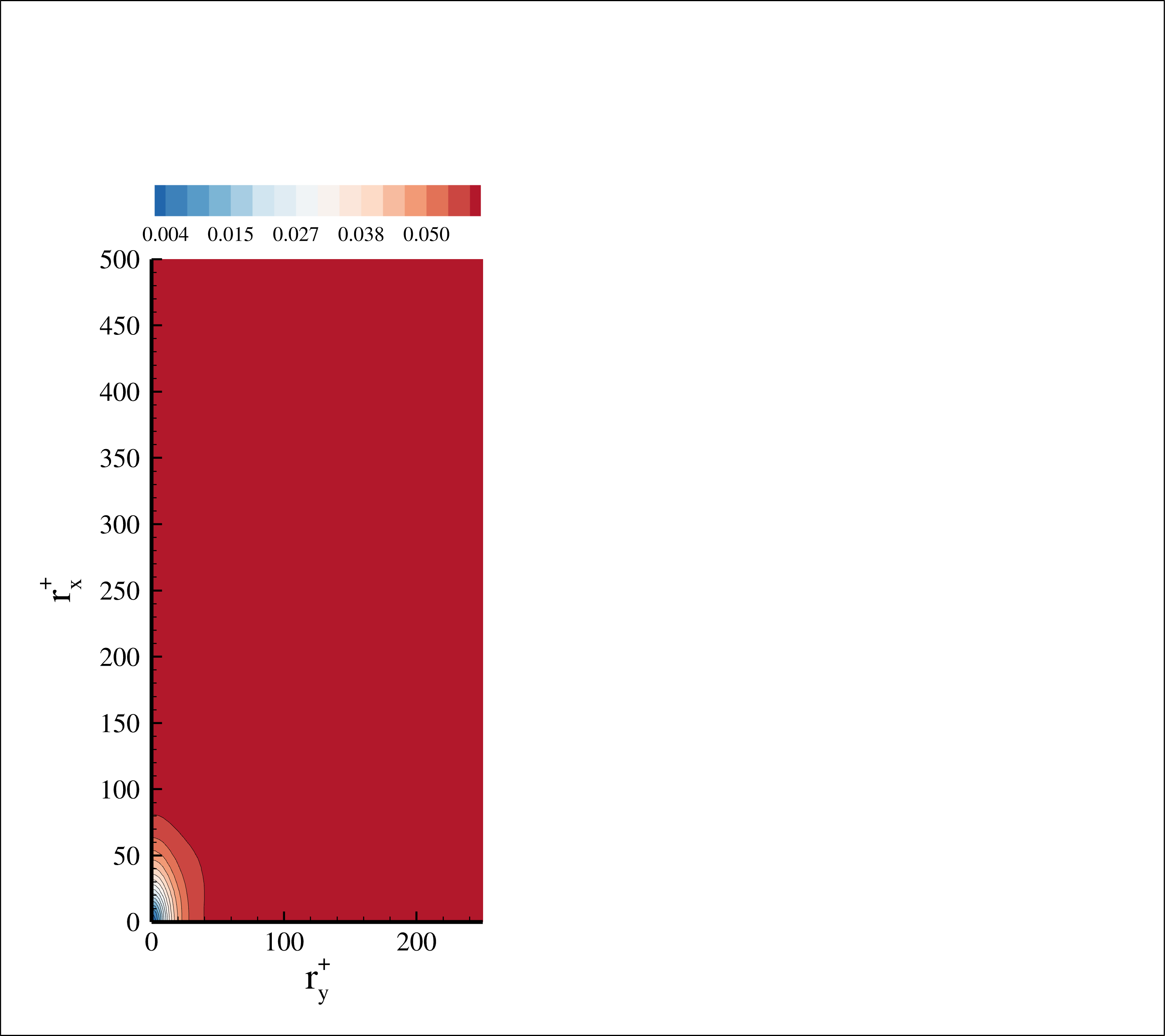}\mylab{-41mm}{74mm}{(\aaa)}
\includegraphics[trim=3cm 1cm 60cm 15cm, clip=true, width=0.32\linewidth]{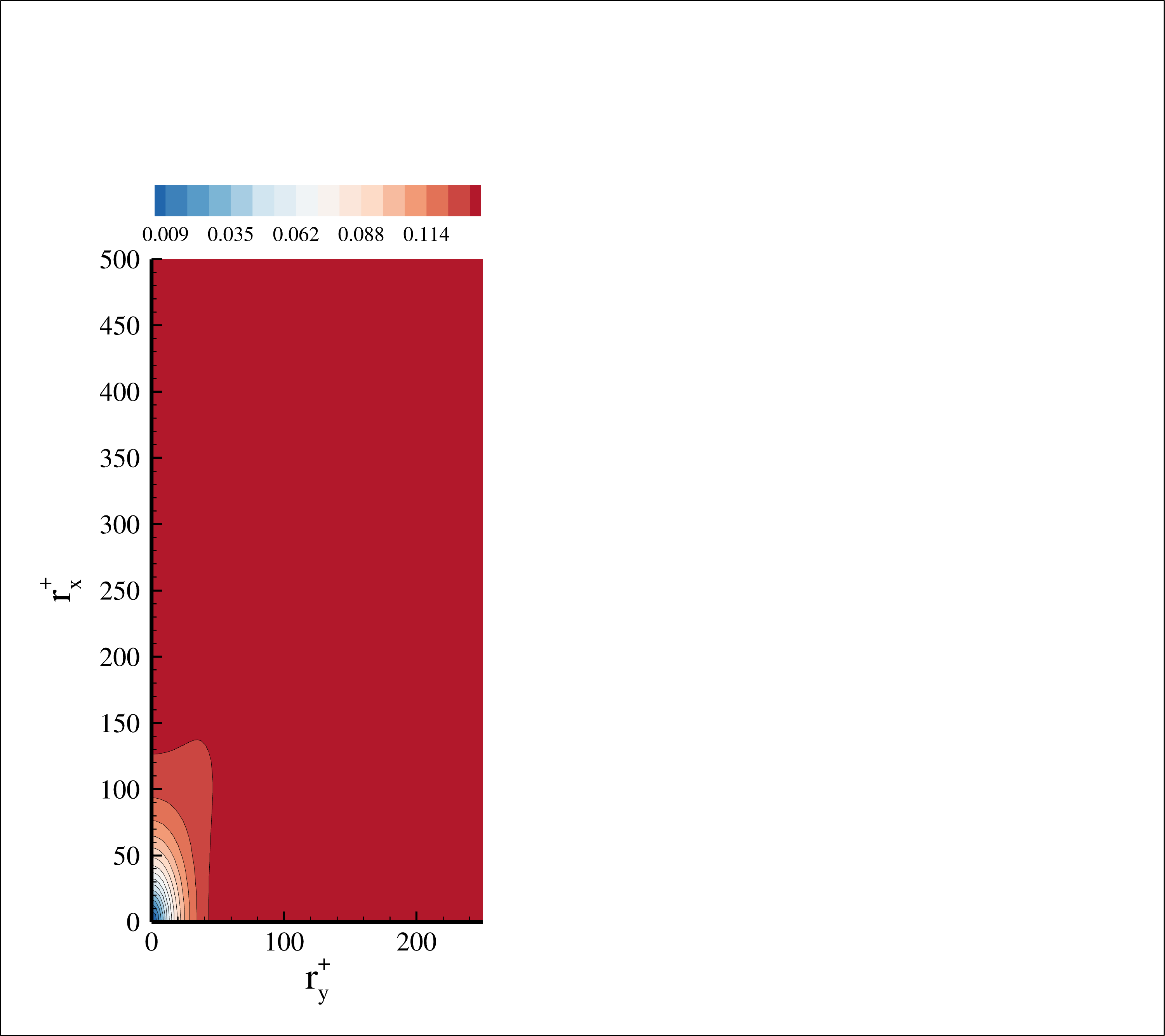}\mylab{-41mm}{74mm}{(\bbb)}
\includegraphics[trim=3cm 1cm 60cm 15cm, clip=true, width=0.32\linewidth]{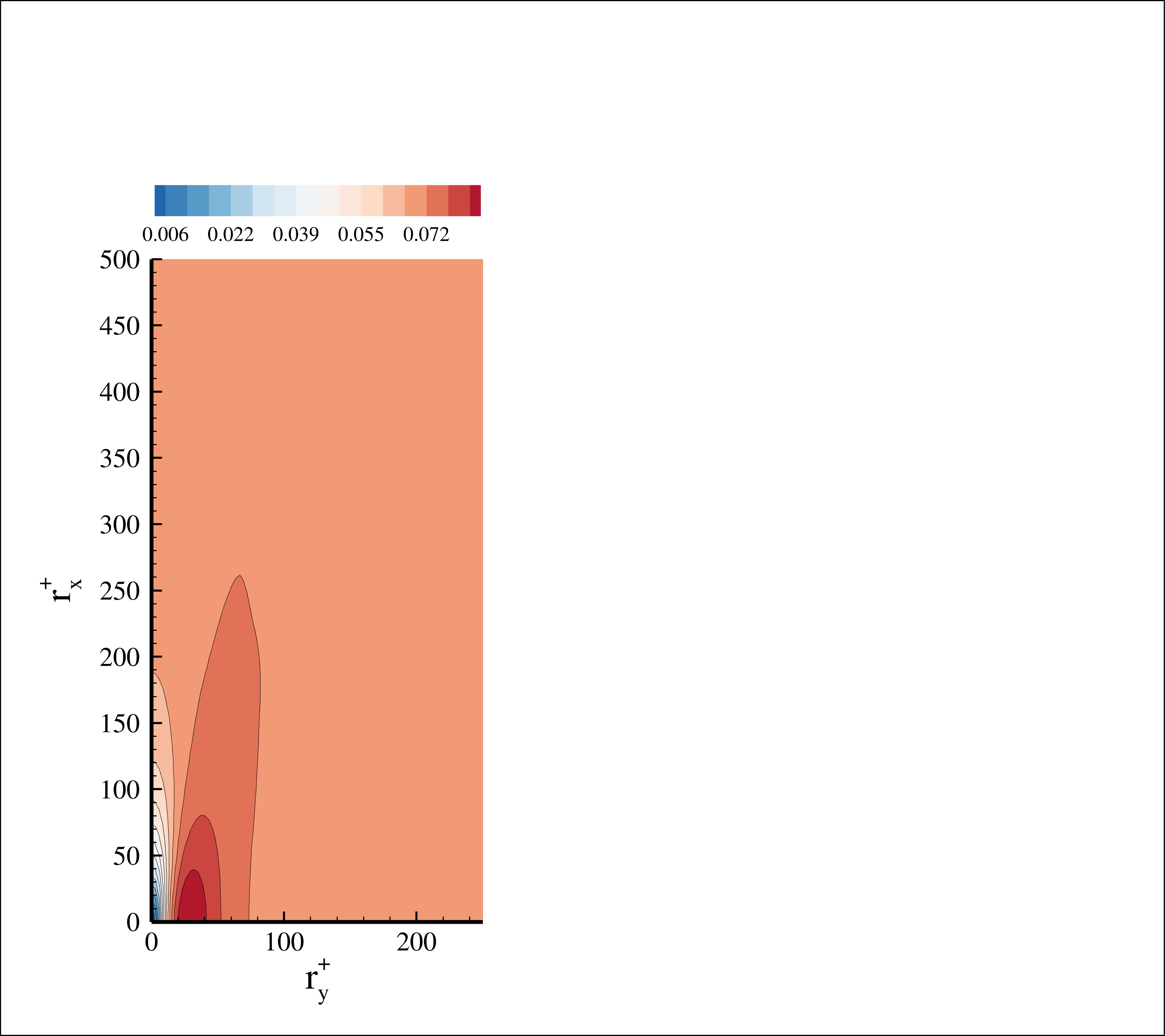}\mylab{-41mm}{74mm}{(\ccc)}
        \caption{Second-order structure functions of (a) streamwise $\langle \delta \omega_x^2 \rangle^+$, (b) spanwise $\langle \delta \omega_y^2 \rangle^+$ and (c) wall-normal $\langle \delta \omega_z^2 \rangle^+$ enstrophy contributions in viscous units at a wall-distance of $z_c^+ = 11$ in the $(r_x, r_y)|_{r_z = 0}$ subspace.
\label{fig:Sii_z11}}
\end{figure}

As enstrophy is transported away from the wall into the buffer layer, the picture shown in figure \ref{fig:Sii_z05} gradually evolves towards the scenario depicted in figure \ref{fig:Sii_z11}, corresponding to the wall distance $z_c^+ = 11$. As already shown by the single-point budget analysed in section \ref{sec:SP}, the enstrophy field starts showing an isotropic recovery from the buffer layer. In the multi-dimensional context of two-point statistics, isotropy manifests itself also among scales. In particular, the structure function of each vorticity component is expected to be elongated along and radially symmetric to its longitudinal separation direction. In the present case this corresponds to $\langle \delta \omega_x^2 \rangle$ being elongated along $r_x$, $\langle \delta \omega_y^2 \rangle$ along $r_y$ and $\langle \delta \omega_z^2 \rangle$ being radially symmetric in the $(r_x, r_y)|_{r_z = 0}$ plane and elongated in $r_z$ (not shown here). As shown in figure \ref{fig:Sii_z11}, the streamwise enstrophy contribution, $\langle \delta \omega_x^2 \rangle$, begins to recall the structure that would be expected under isotropic conditions, exhibiting a slight elongation along its longitudinal separation, i.e. the $r_x$ direction. On the other hand, the spanwise enstrophy contribution still exhibits an elongated structure along its transverse separation direction, $r_x$, indicating that it remains far from isotropic and continues to reflect the anisotropy inherited from the near-wall streaks. A fundamental difference that distinguishes the buffer layer from the very near-wall region is the emergence of a significant wall-normal vorticity component, $\langle \delta \omega_z^2 \rangle$. In fact, while the vorticity field can be considered to be 2-dimensional in the very near-wall region, being dominated by the streamwise and spanwise components, it recovers its three-dimensional character in the buffer layer as the wall-normal vorticity gains relevance. As shown in figure \ref{fig:Sii_z11}(c), the wall-normal enstrophy contribution $\langle \delta \omega_z^2 \rangle$ exhibits a local peak at spanwise scales that are comparable to those related to the quasi-streamwise vortices. This fact suggests that the three-dimensionalization of the vorticity field may be initiated by some mechanisms linked to the quasi-streamwise vortices.

To investigate the mechanism responsible for this behaviour, we consider the evolution equation of the wall-normal vorticity structure function, $\langle \delta \omega_z^2 \rangle$. Since the redistribution of enstrophy among the three vorticity components can only occur through the vortex tilting mechanism, the relevant contribution is the deviatoric part of the $\xi_{\delta \omega^2}$ term. For the symmetries of the temporal boundary layer and by restricting the analysis to the subspace $r_z=0$, the evolution equation of $\langle \delta \omega_z^2 \rangle$ reads:
\begin{equation}
\begin{gathered}
       \frac{\partial \langle \delta \omega_z^2 \rangle}{\partial t} = \underbrace{- \frac{\partial \langle \delta \omega_z^2 \delta u_j \rangle}{\partial r_j}}_{\langle T_{r,\delta \omega_z^2} \rangle} + \underbrace{2 \nu \frac{\partial^2 \langle \delta \omega_z^2 \rangle}{\partial r_j \partial r_j}}_{\langle D_{r,\delta \omega_z^2} \rangle} \underbrace{- \frac{\partial \langle \delta \omega_z^2 \widetilde{w} \rangle}{\partial z_{c}} + \frac{\nu}{2} \frac{\partial^2 \langle \delta \omega_z^2 \rangle}{\partial z_c^2}}_{\langle T_{c,\delta \omega_z^2} \rangle}\\
       \underbrace{+ 2 \Omega_y \langle \delta \omega_z \, \delta \! \left( \frac{\partial w}{\partial y} \right) \rangle + 2 \langle \delta \omega_z \widetilde{\omega}_j \, \delta \!\left( \frac{\partial w}{\partial x_j} \right) \rangle + 2 \langle \delta \omega_z \delta \omega_j \widetilde{\left( \frac{\partial w}{\partial x_j} \right)} \rangle}_{\langle \xi_{\delta \omega_z^2} \rangle} - \underbrace{4\langle \frac{\partial \omega_z}{\partial x_j} \frac{\partial \omega_z}{\partial x_j} \rangle}_{\chi_{\omega_z^2}}.
\label{eq:enstr_z_TP_TBL}
\end{gathered}
\end{equation}
The vortex stretching term $\xi_{\delta \omega_z^2}$, which acts overall as a source for the wall-normal vorticity structure function $\langle \delta \omega_z^2 \rangle$, can be decomposed into the following contributions:
\begin{itemize}
    \item $2 \Omega_y \langle \, \delta \! \omega_z \delta \left( \partial w / \partial y \right) \rangle$ representing the generation of $\langle \delta \omega_z^2 \rangle$ through the tilting of the mean spanwise vorticity $\Omega_y$ into the wall-normal direction caused by non-uniform wall-normal velocity fluctuations $w$ in the spanwise direction $y$;
    \item $2 \langle \delta \omega_z \widetilde{\omega}_j \, \delta \! \left( \partial w / \partial x_j \right) \rangle + 2 \langle \delta \omega_z  \delta  \omega_j \widetilde{\left( \partial w / \partial x_j \right)} \rangle$, with $j=1,2$, representing the generation of $\langle \delta \omega_z^2 \rangle$ through the tilting of streamwise and spanwise vorticity fluctuations $\omega_x$ and $\omega_y$ in the wall-normal direction caused by non-uniform wall-normal velocity fluctuations $w$ in the $j$ direction;
    \item $2 \langle \delta \omega_z \widetilde{\omega}_z \, \delta \! \left( \partial w / \partial z \right) \rangle + 2 \langle \delta \omega_z \delta \omega_z \widetilde{\left( \partial w / \partial z \right)} \rangle$, representing the amplification of the existing wall-normal vorticity through stretching along the wall-normal direction.
\end{itemize}

\begin{figure}[t!]
\centering
\includegraphics[width=0.44\linewidth]{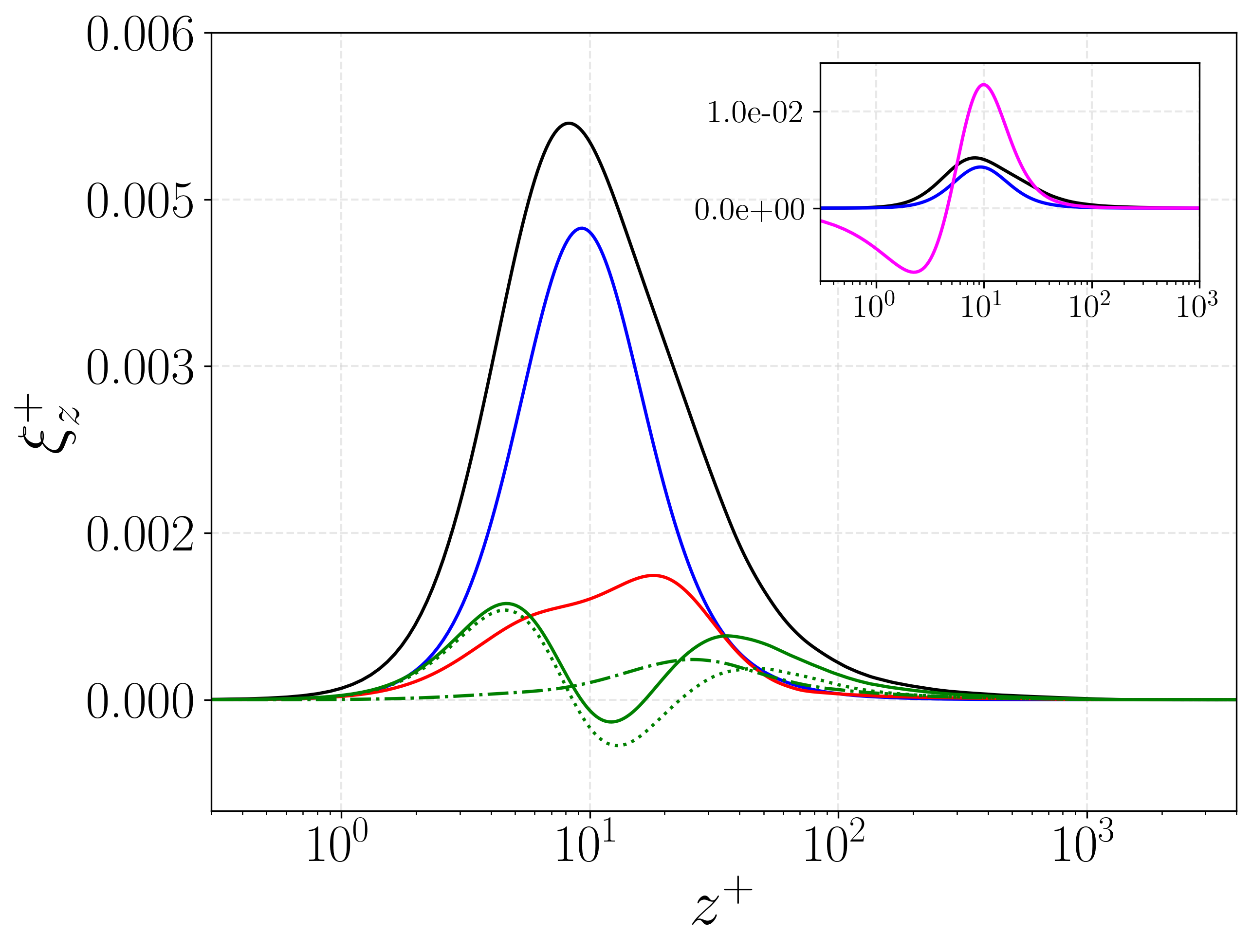}\mylab{-64mm}{45mm}{(\aaa)}
\includegraphics[trim=4cm 0.5cm 2.5cm 1cm, clip=true, width=0.55\linewidth]{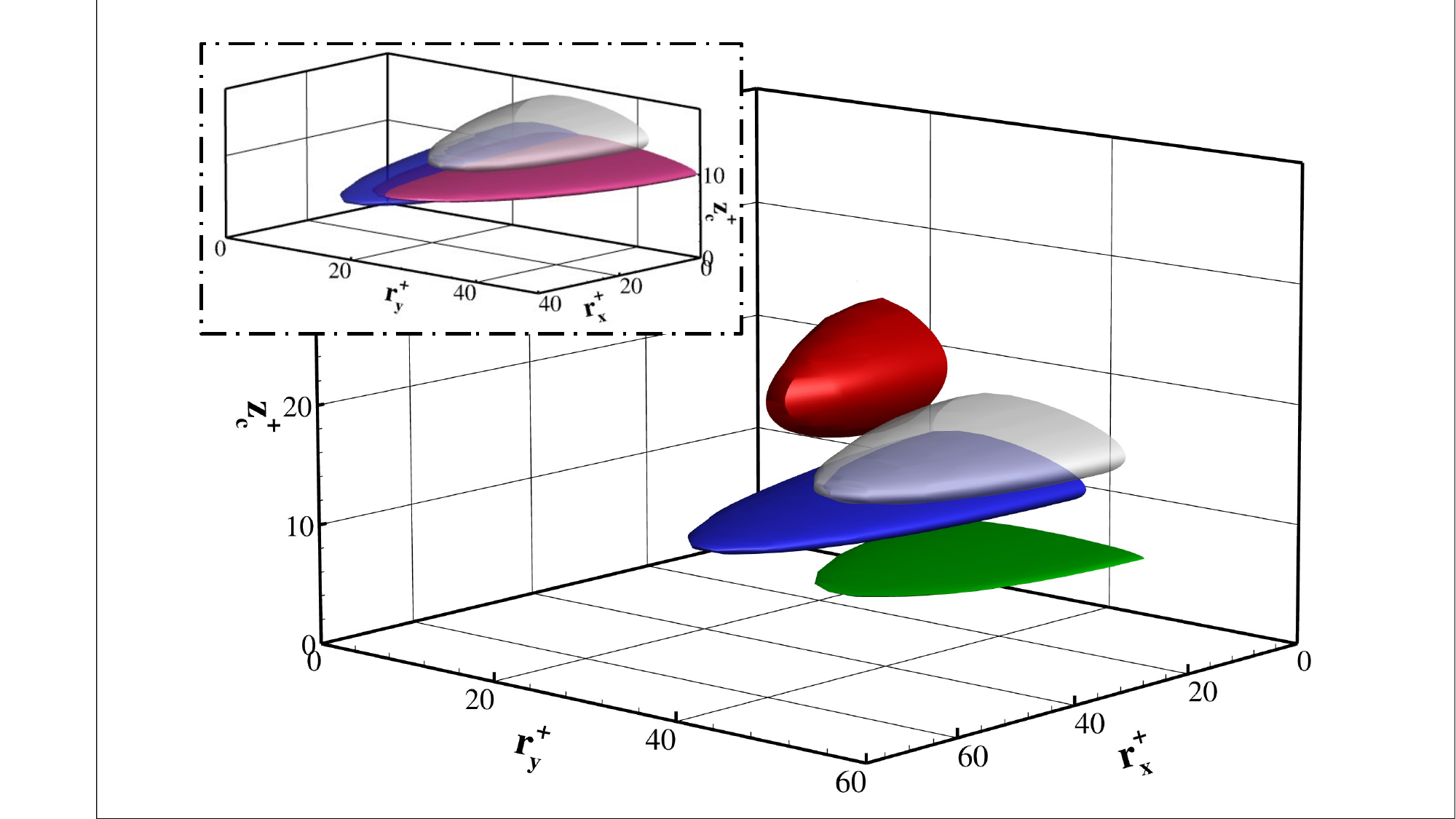}\mylab{-5mm}{45mm}{(\bbb)}
        \caption{(a) Wall-normal vortex stretching contribution $\xi_{z}$ (solid black line), wall-normal tilting of mean spanwise vorticity $2 \Omega_y \langle \omega_z (\partial w / \partial y) \rangle$ (solid blue line), pure stretching of wall-normal vorticity $2 \langle \omega_z \omega_z (\partial w / \partial z) \rangle$ (solid red line), tilting of streamwise and spanwise vorticity fluctuations $2 \langle \omega_z \omega_{x_i} (\partial w / \partial x_i) \rangle$, with $i=1,2$, (solid green line). The terms $2 \langle \omega_z \omega_x (\partial w / \partial x) \rangle$ (dash-dotted green line) and $2 \langle \omega_z \omega_y (\partial w / \partial y) \rangle$ (dotted green line) are also reported separately. The inset compares the streamwise tilting of wall-normal vorticity by mean velocity gradient $2 \langle \omega_x \omega_z \rangle \partial U / \partial z$ (solid fuchsia line) with the terms shown in the main panel. (b) Wall-normal enstrophy contribution isosurface $\langle \delta \omega_z^2 \rangle = 0.085$ (transparent white), wall-normal tilting of mean spanwise vorticity isosurface $2 \Omega_y \langle \delta \omega_z \delta \left( \partial w / \partial y \right) \rangle = 0.01$ (in blue), pure stretching of wall-normal vorticity isosurface $2 \langle \delta \omega_z \widetilde{\omega}_z \delta \left( \partial w / \partial z \right) \rangle + 2 \langle \delta \omega_z \delta \omega_z \widetilde{\left( \partial w / \partial z \right)} \rangle = 0.003$ (in red), tilting of vorticity fluctuations isosurface $2 \langle \delta \omega_z \widetilde{\omega}_j \delta \left( \partial w / \partial x_j \right) \rangle + 2 \langle \delta \omega_z \delta \omega_j \widetilde{\left( \partial w / \partial x_j \right)} \rangle = 0.002$, with $j=1,2$, (in green). The inset additionally shows the isosurface of the streamwise tilting of wall-normal vorticity by the mean velocity gradient,  $2 \langle \delta \omega_x \delta \omega_z \rangle \partial U / \partial z = 0.0285$ (in fuchsia), alongside the terms shown in the main panel.
\label{fig:VSz_comp}}
\end{figure}

The relative importance of these three terms is shown in figure \ref{fig:VSz_comp}. Figure \ref{fig:VSz_comp}(a) reports their single-point counterparts, while figure \ref{fig:VSz_comp}(b) shows their distribution over the streamwise and spanwise separation scales, $r_x$ and $r_y$. The overall picture indicates that the redistribution of vorticity toward the wall-normal component, $\omega_z$, is predominantly active within the buffer layer. The leading mechanism is the tilting of the mean spanwise vorticity $\Omega_y$ into the wall-normal direction by wall-normal velocity fluctuations that are non-uniform in the spanwise direction $\partial w /\partial y$. This effect is captured by the term $2 \Omega_y \langle \omega_z \partial w / \partial y \rangle$ in the single-point formulation (figure \ref{fig:VSz_comp}(a)) and by its two-point counterpart $2 \Omega_y \langle \delta \omega_z \delta (\partial w / \partial y ) \rangle$ in figure \ref{fig:VSz_comp}(b). As shown in blue in figure \ref{fig:VSz_comp}(a) and (b), this is the dominant contribution within the buffer layer. Moreover, it is the only contribution whose maximum occurs at the same scales as the peak of the wall-normal enstrophy contribution, $\langle \delta \omega_z^2 \rangle$ (transparent white isosurface in figure \ref{fig:VSz_comp}(b)). The close overlap between the two isosurfaces (blue and transparent white in figure \ref{fig:VSz_comp}(b)) suggests that wall-normal vorticity is primarily injected at such scales through the tilting of the mean spanwise vorticity $\Omega_y$ into the wall-normal direction induced by wall-normal velocity fluctuations that are non-uniform in the spanwise direction $\partial w /\partial y$.

The remaining contributions are comparatively weaker and reach their maximum intensity at different wall-distances or scales. The pure stretching of wall-normal vorticity (represented by the term $2 \langle \omega_z \omega_z ( \partial w / \partial z ) \rangle$ in the single-point view and shown in red in figure \ref{fig:VSz_comp}) is strictly positive and becomes active only in regions where $\omega_z$ already attains a relevant intensity, since it requires a pre-existing wall-normal vorticity to undergo stretching. Although it amplifies the existing $\omega_z$, it never becomes the dominant source within the buffer layer. The final contribution corresponds to the tilting of the fluctuating streamwise and spanwise vorticity components into the wall-normal direction. This term is not dominant in the buffer layer and reaches its maximum intensity closer to the wall, at $z^+ \approx 5$. Notably, while the tilting of streamwise vorticity ($2 \langle \omega_z \omega_x ( \partial w / \partial x ) \rangle$) consistently contributes positively to the generation of wall-normal vorticity, the tilting of spanwise vorticity fluctuations ($2 \langle \omega_z \omega_y ( \partial w / \partial y ) \rangle$) exhibits a sign change with increasing wall distance, see figure \ref{fig:VSz_comp}(a). In particular, it acts as a positive source of wall-normal enstrophy in the viscous sublayer and in the outer region, while predominantly acting as a sink within the buffer layer. The scales at which these vorticity redistribution mechanisms operate, which ultimately render the vorticity field fully three-dimensional, are consistent with the characteristic size of the structures populating the buffer layer \citep{Kline1967,JimenezMoin1991,Hamilton1995,Jimenez2013,kumar2023vorticity}.

Furthermore, the observed pattern of vorticity reorientation is consistent with hairpin-vortex dynamics \citep{Hamilton1995,Zhou1999,Adrian2007}. The insets in figure \ref{fig:VSz_comp} show the tilting of wall-normal vorticity into the streamwise direction by the mean velocity gradient, represented by $2 \langle \omega_x \omega_z \rangle \partial U / \partial z$ in the single-point formulation and by $2 \langle \delta \omega_x \delta \omega_z \rangle \partial U / \partial z$ in the two-point formulation (fuchsia in figures \ref{fig:VSz_comp}(a) and \ref{fig:VSz_comp}(b), respectively). This term acts as a source of streamwise vorticity and exhibits a pronounced peak at $z^+ \approx 10$, at wall distances and scales comparable to those at which the wall-normal tilting of mean spanwise vorticity, $2 \Omega_y \langle \delta \omega_z \delta(\partial w / \partial y) \rangle$, and wall-normal enstrophy contribution, $\langle \delta \omega_z^2 \rangle$, show their maxima. This correspondence is consistent with the hairpin-vortex picture, in which the newly generated wall-normal vorticity generated through the tilting of spanwise vorticity is further tilted into the streamwise direction by the mean shear $\partial U / \partial z$. In the present flow configuration, since $\Omega_y = \partial U / \partial z$ and $\delta \omega_x = \delta(\partial w / \partial y) - \delta(\partial v / \partial z)$, the two terms ($2 \Omega_y \langle \delta \omega_z \delta (\partial w / \partial y) \rangle$ and $2 \langle \delta \omega_x \delta \omega_z \rangle \partial U / \partial z$) can be rewritten as 
\begin{equation}
    2 \Omega_y \langle \delta \omega_z \delta (\partial w / \partial y) \rangle \qquad \mathrm{and} \qquad 2 \Omega_y \langle \delta \omega_z \delta(\partial w / \partial y - \partial v / \partial z) \rangle
\end{equation}
respectively. This highlights that the two terms are also mathematically partially overlapping, as they share a common contribution. This suggests that they might be the manifestation of a single phenomenon.


\subsection{Overlap layer}

\begin{figure}[t!]
\centering
\includegraphics[trim=3cm 1cm 60cm 15cm, clip=true, width=0.32\linewidth]{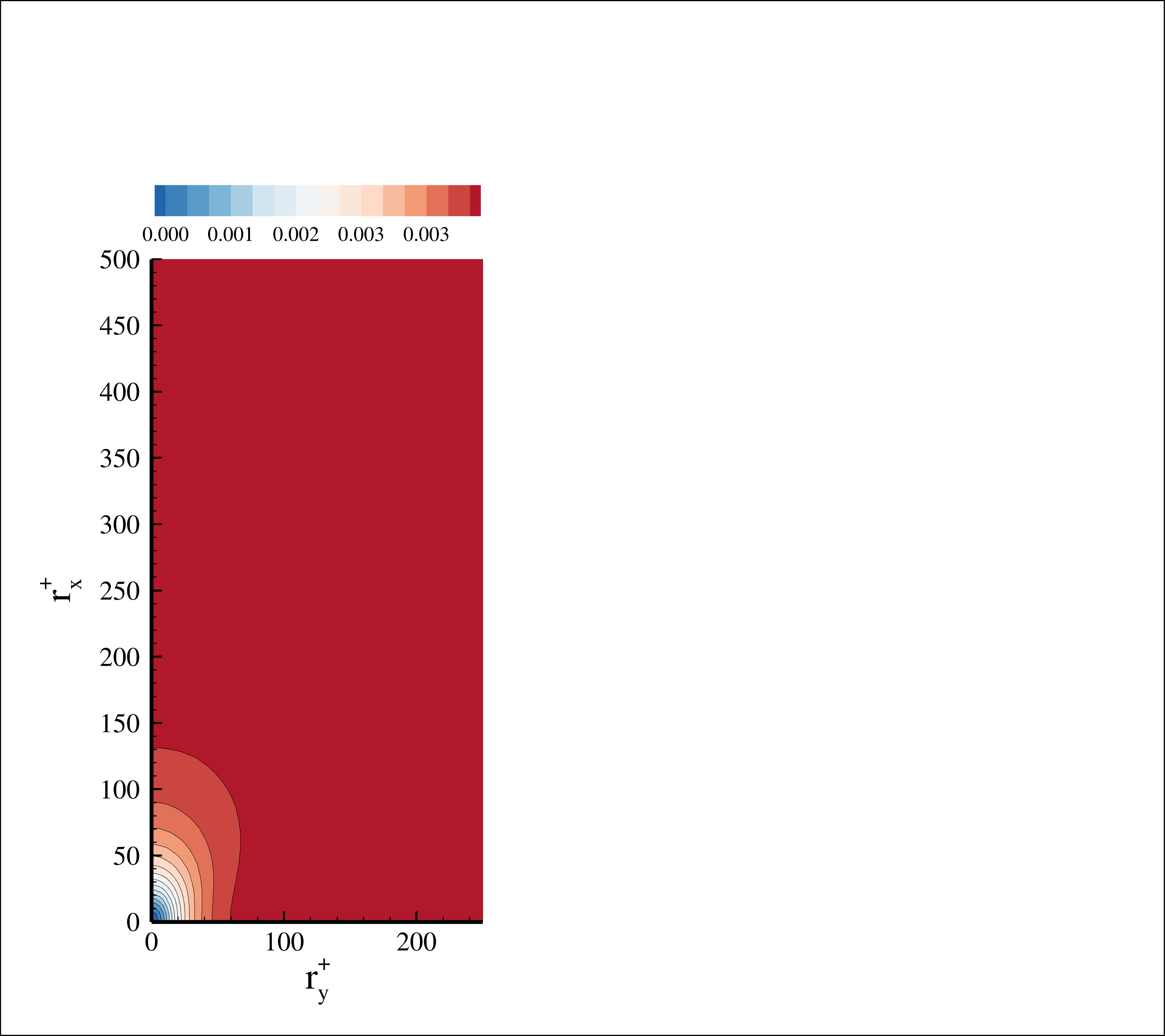}\mylab{-41mm}{74mm}{(\aaa)}
\includegraphics[trim=3cm 1cm 60cm 15cm, clip=true, width=0.32\linewidth]{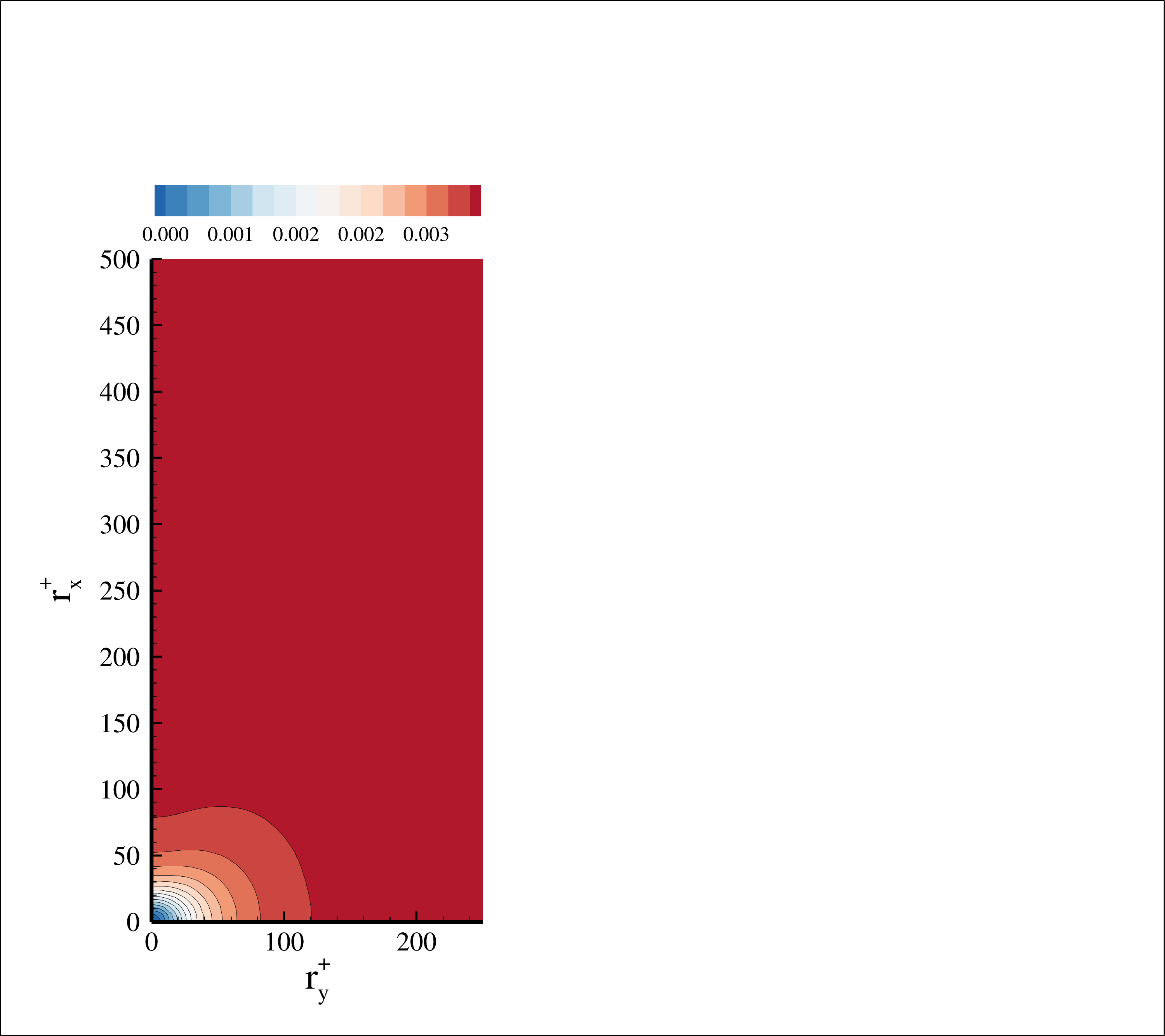}\mylab{-41mm}{74mm}{(\bbb)}
\includegraphics[trim=3cm 1cm 60cm 15cm, clip=true, width=0.32\linewidth]{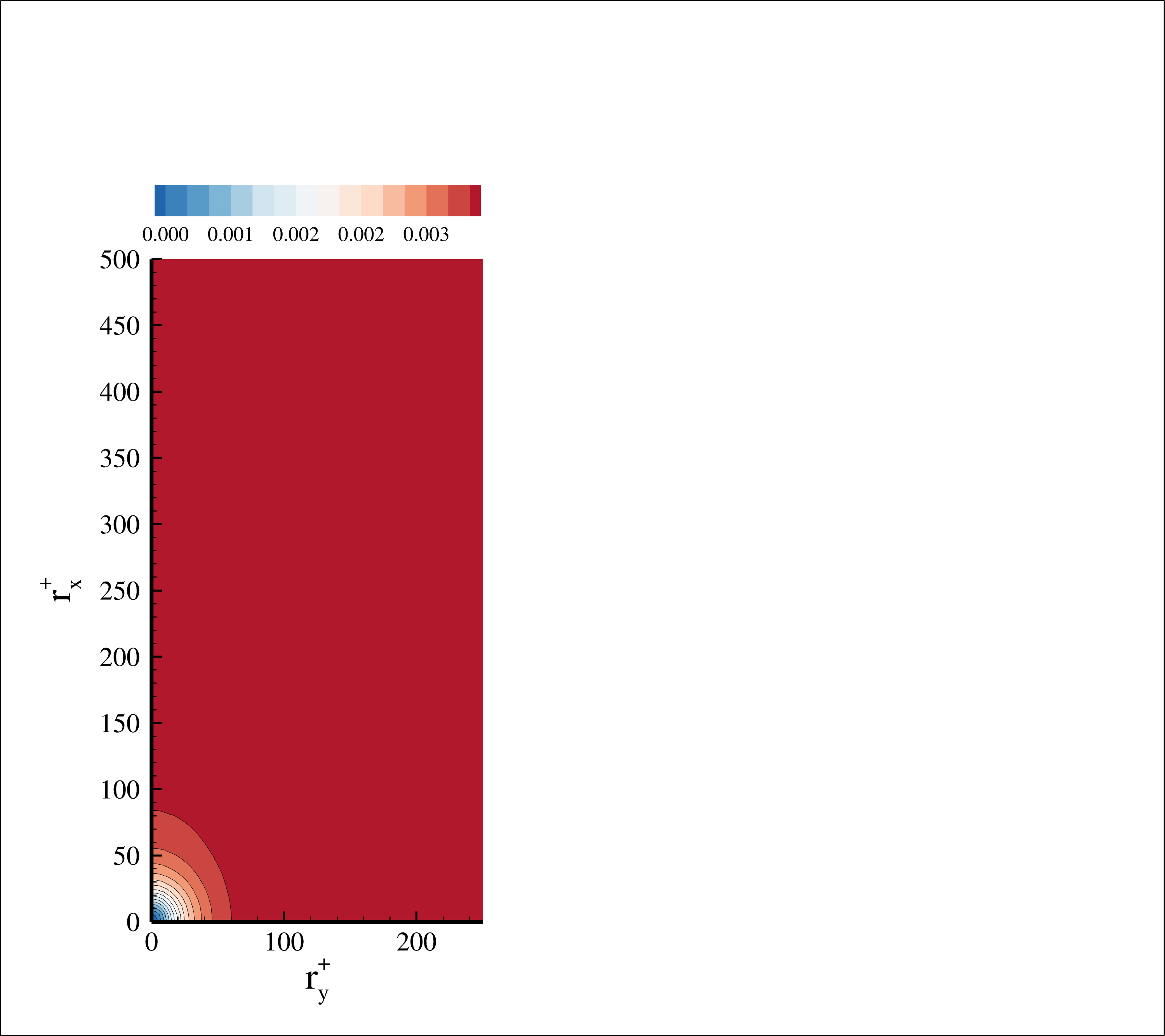}\mylab{-41mm}{74mm}{(\ccc)}
        \caption{Second-order structure functions of (a) streamwise $\langle \delta \omega_x^2 \rangle^+$, (b) spanwise $\langle \delta \omega_y^2 \rangle^+$ and (c) wall-normal $\langle \delta \omega_z^2 \rangle^+$ enstrophy contributions in viscous units at a wall-distance of $z_c^+ = 450$ in the $(r_x, r_y)|_{r_z = 0}$ subspace.
\label{fig:Sii_z450}}
\end{figure}

Moving away from the wall, beyond the buffer layer, the vorticity field, which has already recovered its three-dimensional character, progressively approaches a state of near-isotropy. The distribution of the two-point enstrophy contributions at $z_c^+ = 450$ is shown in figure \ref{fig:Sii_z450}.

At this wall distance, representative of the overlap layer, the flow exhibits many of the features expected for isotropic turbulence. Indeed, enstrophy is distributed almost uniformly among the three vorticity components, as indicated by the comparable intensities of $\langle \delta \omega_x^2 \rangle$, $\langle \delta \omega_y^2 \rangle$ and $\langle \delta \omega_z^2 \rangle$, in figure \ref{fig:Sii_z450}(a), (b) and (c), respectively. Also their structure is consistent with that expected in isotropic conditions. In particular, the structure function of each of the three vorticity components is elongated along its longitudinal separation direction: along $r_x$ for $\langle \delta \omega_x^2 \rangle$, along $r_y$ for $\langle \delta \omega_y^2 \rangle$ and approximately radially symmetric in the $(r_x,r_y)$-plane for $\langle \delta \omega_z^2 \rangle$. This picture is approximately maintained throughout most of the outer region of the boundary layer, with the main difference being a progressive increase of the characteristic scales as the distance from the wall increases. However, a different picture emerges when approaching the region affected by the turbulent/non-turbulent interface. 

\subsection{Interface region}

\begin{figure}[t!]
\centering
\includegraphics[trim=3cm 1cm 60cm 15cm, clip=true, width=0.32\linewidth]{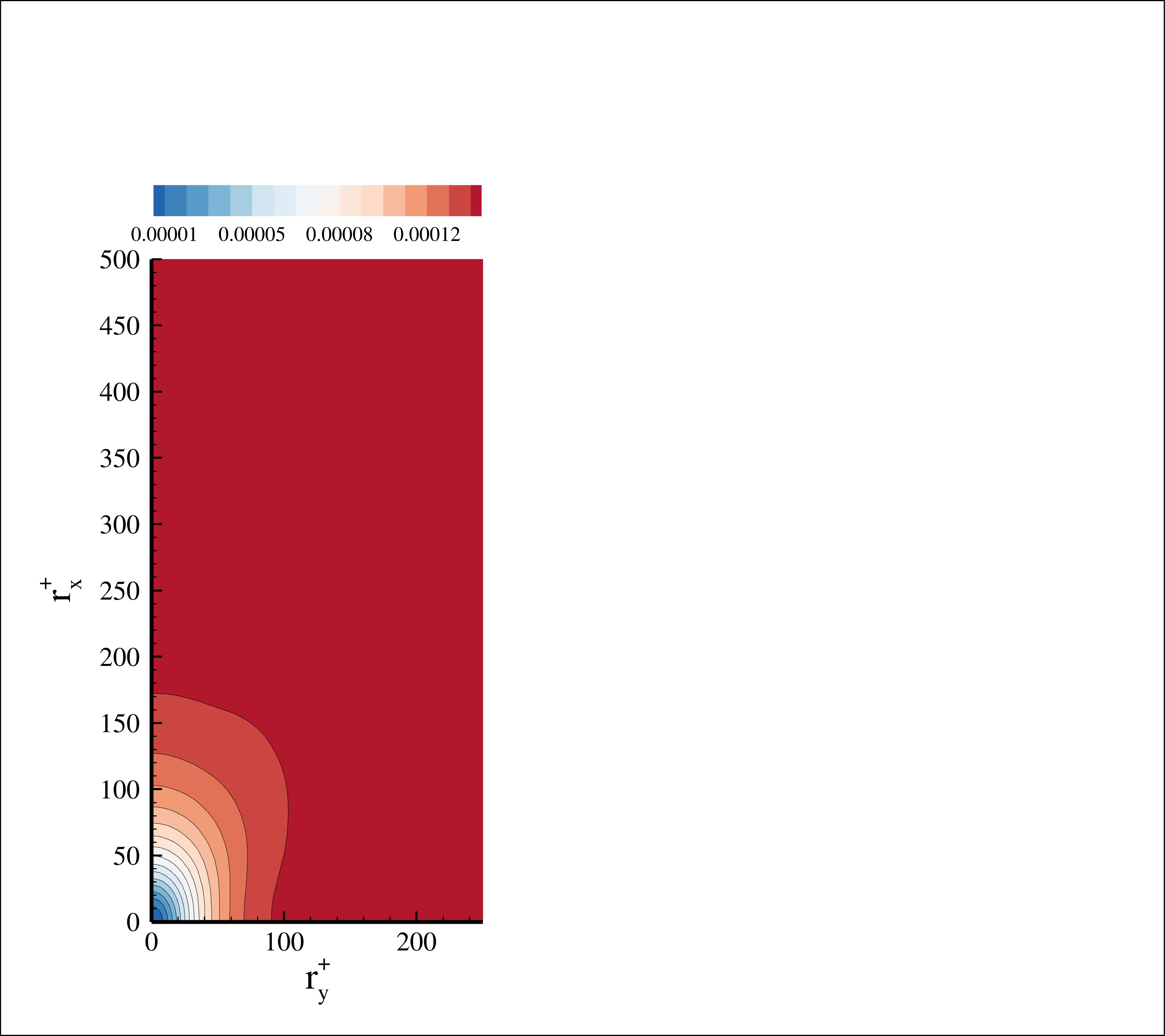}\mylab{-41mm}{74mm}{(\aaa)}
\includegraphics[trim=3cm 1cm 60cm 15cm, clip=true, width=0.32\linewidth]{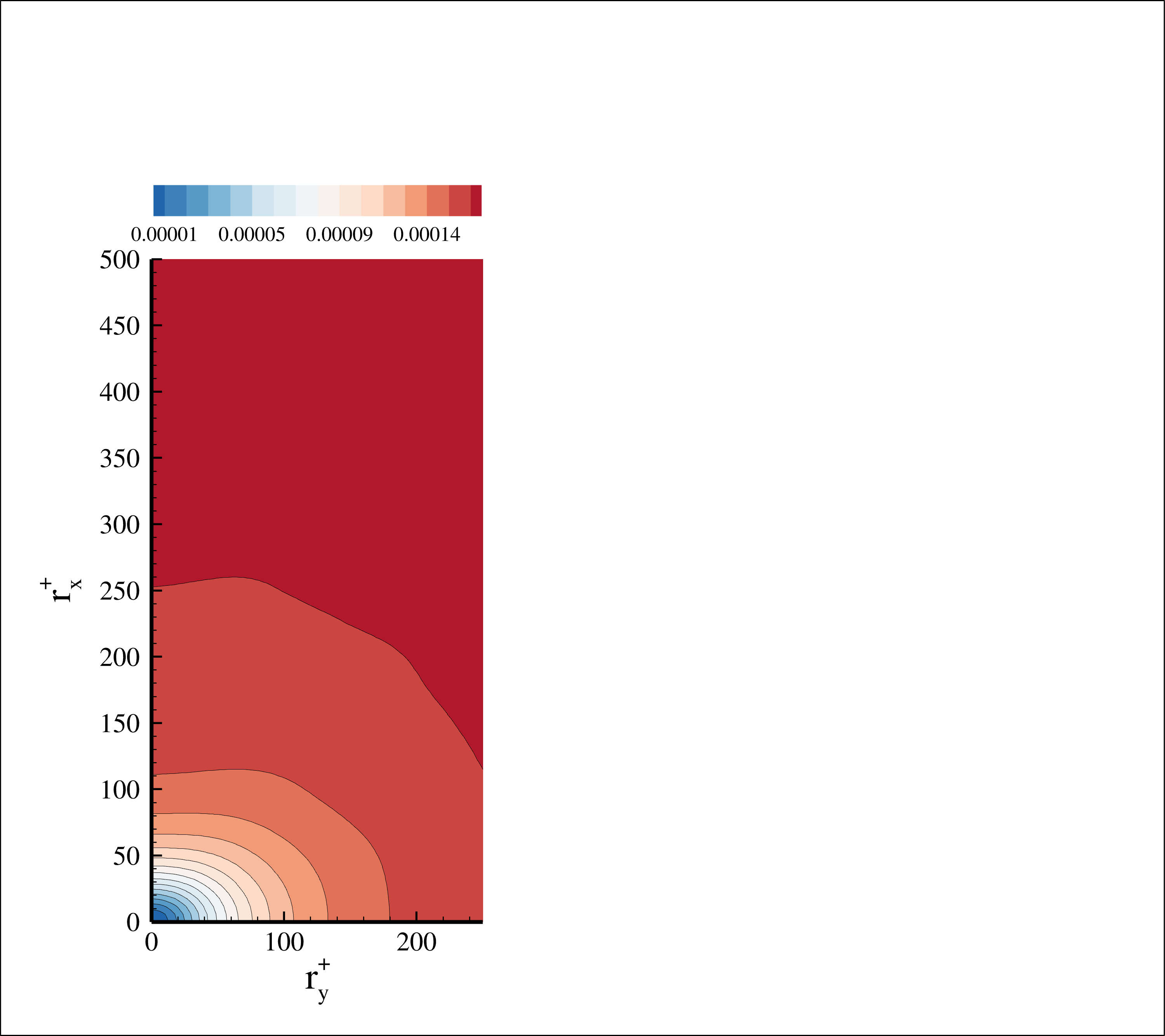}\mylab{-41mm}{74mm}{(\bbb)}
\includegraphics[trim=3cm 1cm 60cm 15cm, clip=true, width=0.32\linewidth]{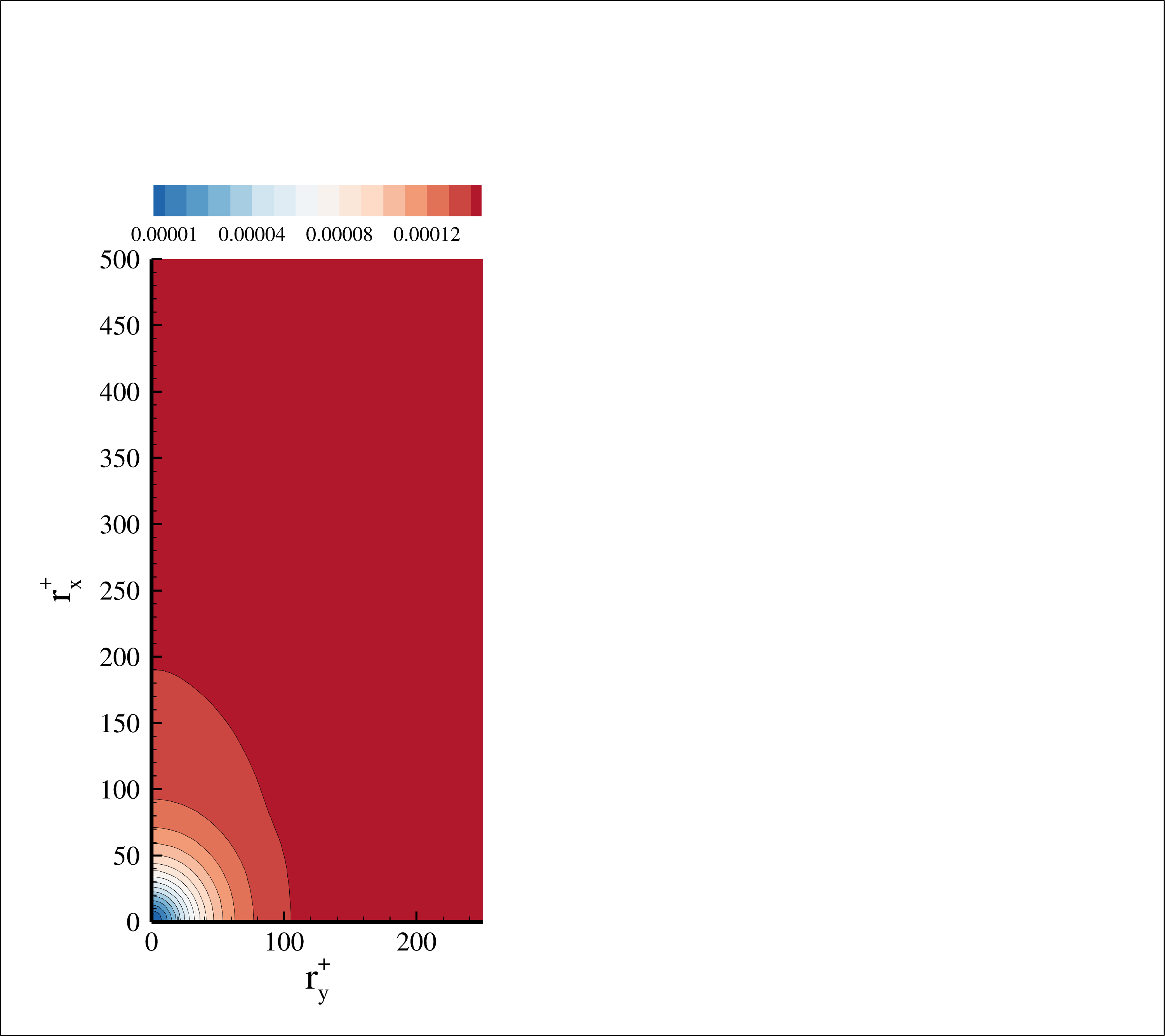}\mylab{-41mm}{74mm}{(\ccc)}
        \caption{Second-order structure functions of (a) streamwise $\langle \delta \omega_x^2 \rangle^+$, (b) spanwise $\langle \delta \omega_y^2 \rangle^+$ and (c) wall-normal $\langle \delta \omega_z^2 \rangle^+$ enstrophy contributions in viscous units at a wall-distance of $z_c^+ = Re_{\tau}$ in the $(r_x, r_y)|_{r_z = 0}$ subspace.}
\label{fig:Sii_z1500}
\end{figure}

As already shown in the single-point view in section \ref{conf_eff}, the flow in this region is influenced by confinement effects associated with the interface, although these are less pronounced than those induced by the solid wall. The corresponding scale enstrophy distribution at the mean interface position $z_c^+ = Re_{\tau} = 1501$ is shown in figure \ref{fig:Sii_z1500}.

The isotropy recovered in the outer region becomes unbalanced again in the vicinity of the TNTI region. As shown in figure \ref{fig:Sii_z1500}, this is reflected by the fact that $\langle \delta \omega_y^2 \rangle$ is distributed over a wider range of scales than the other contributions $\langle \delta \omega_x^2 \rangle$ and $\langle \delta \omega_z^2 \rangle$, and exhibits a slightly higher intensity. This observation is consistent with the presence of structures that are elongated in the interface-parallel horizontal directions and characterized by sharp gradients in the interface-normal vertical direction in agreement with the physical scenario proposed by \citet{cimarelli2015spectral, cimarelli2021spatially, cimarelli2024spatially} and \citet{zhou2020energycascade}. In analogy with the near-wall region, dissipation in this zone is expected to be primarily accomplished through these intense interface-normal gradients. Although the statistical framework employed in the present work does not distinguish between directions normal and tangential to the TNTI, since it relies on classical averages, the imprint of these mechanisms remains visible in the wall-normal plane considered here. This signature is likely weaker than it would appear in an analysis conducted using interface-conditioned statistics (see \citet{zhou2020energycascade}).

\subsection{Summary}

The results shown in the present section can be summarized as follows. In the immediate vicinity of the wall, the vorticity field is essentially two-dimensional and is primarily associated with some canonical flow structures located further away from the wall, rather than with motions sustained locally. In other words, enstrophy near the wall is a wall footprint of the interaction between the motion induced by larger coherent structures away from the wall and the wall boundary condition. In particular, high- and low-speed streaks generate large-scale spanwise vorticity fluctuations, whereas quasi-streamwise vortices give rise to intermediate-scale ($r_y^+ \approx 50$) streamwise vorticity fluctuations. As enstrophy is transported away from the wall by viscous diffusion, it progressively recovers its three-dimensional character within the buffer layer, mainly through the tilting of the mean spanwise vorticity, $\Omega_y$, into the wall-normal direction. This is accompanied by a tilting of the newly generated wall-normal vorticity into the streamwise direction induced by the mean velocity gradient $\partial U / \partial z$. The turbulent transport accomplishes the remaining transport away from the wall and enstrophy recovers near-isotropic statistics in the outer region. Finally, in the interface region, confinement effects analogous to the one imposed by the wall are observed, breaking the isotropy and giving rise to structures that are elongated in the wall-parallel directions and confined in the interface-normal vertical direction. The relation between enstrophy and turbulent dissipation allows us to extend this phenomenological view. In particular, the buffer and overlap layers are characterized by a turbulent dissipation accomplished by local small-scale and isotropic structures. On the other hand, turbulent dissipation in the very near-wall and interface regions is not associated with local flow structures, but rather with the interaction between motions induced by coherent structures in the bulk of the flow and the no-slip condition at the wall, or with the irrotationality of the flow outside the boundary layer at the TNTI. As a result, dissipation is accomplished by vertical shearing phenomena occurring in a thin vertical layer over a wide range of horizontal scales.

\section{Sources and fluxes of scale enstrophy in the compound space of scales and positions} \label{sec:TP_enstr_3D}

Section \ref{sec:TP_enstr_contr} examined the enstrophy structure function and its scale-space distribution at selected wall-normal locations representative of the different boundary layer regions. We now turn to the corresponding two-point enstrophy budget to characterize the mechanisms governing the generation, transfer and dissipation of enstrophy in the compound space of scales and wall distances.

\begin{figure}[t!]
\centering
\includegraphics[trim=20cm 0.5cm 20cm 1cm, clip=true, width=0.42\linewidth]{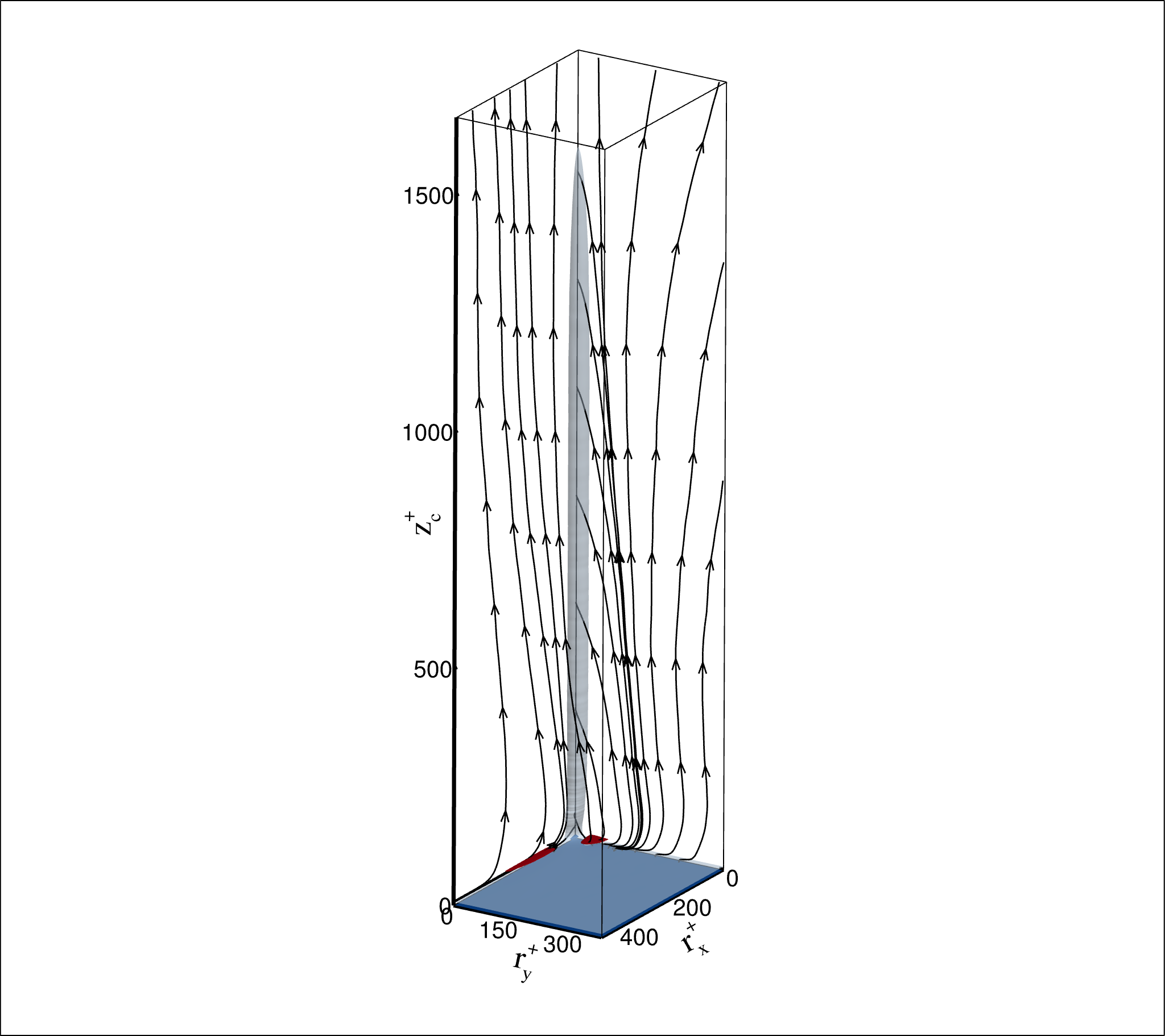}\mylab{-50mm}{100mm}{(\aaa)}
\includegraphics[trim=3cm 0.5cm 10cm 1cm, clip=true, width=0.56 \linewidth]{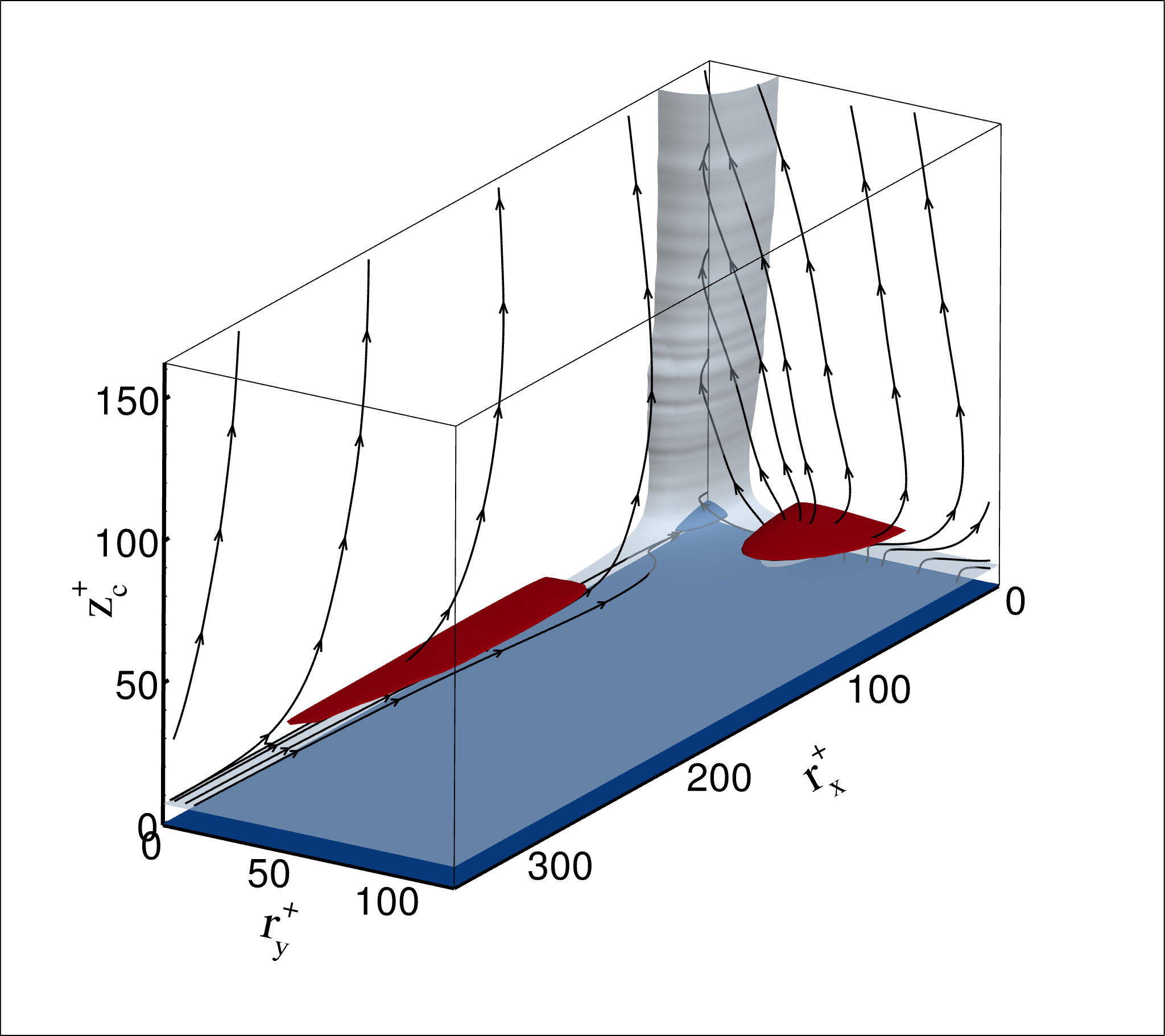}\mylab{-70mm}{60mm}{(\bbb)}
        \caption{Sources and fluxes of scale enstrophy. (a) Isosurfaces of positive net source of enstrophy $S_{\delta \omega^2}^+ = 0.008$ (red) and negative net source $S_{\delta \omega^2}^+ = -0.05$ (solid blue) and $S_{\delta \omega^2}^+ = -5 \times 10^{-6}$ (transparent grey). The flux $(\phi_{r_x}, \phi_{r_y}, \phi_{z_c})|_{r_z=0}$ is reported as black lines and arrows. (b) Enlarged view of the same quantities.
\label{fig:GKE_3D}}
\end{figure}

The results are presented in figure \ref{fig:GKE_3D}, where the sources and sinks are represented with isosurfaces, while the field of fluxes $(\phi_{r_x}, \phi_{r_y}, \phi_{z_c})$ is reported with streamlines. Red isosurfaces denote regions of positive net source $S_{\delta \omega^2} > 0$, corresponding to enstrophy production, while the blue/grey isosurfaces indicate regions where $S_{\delta \omega^2} < 0$, representing enstrophy sinks with different intensities. Two positive source peaks can be identified in the near-wall region. As shown more clearly in figure \ref{fig:GKE_3D}(b), both peaks are located within the buffer layer, which remains the region of highest net enstrophy production even excluding these local maxima. These two source regions are elongated in the $r_x$-direction and are centred on the planes $r_x = 0$ and $r_y = 0$, respectively. The most intense peak is the one centred on the $r_x = 0$ plane, where the source term reaches a maximum value of $S_{\delta \omega^2}^+ \approx 0.0088$ at $( r_x^+, r_y^+, z_c^+ ) \approx ( 0, 43, 12 )$. The second source region exhibits a comparable intensity, with a local maximum of $S_{\delta \omega^2}^+ = 0.0083$ at $( r_x^+, r_y^+, z_c^+ ) = ( 145, 0, 12 )$, at the same wall-normal distance. The streamwise and spanwise scales of the two peaks of scale enstrophy source, $r_x^+ = 145$ and $r_y^+ = 43$, match those of quasi-streamwise vortices populating the buffer layer. Accordingly, we may conjecture that both source peaks of enstrophy are related to the action of a single flow coherent pattern resembling that of quasi-streamwise vortices.

Although the buffer layer constitutes the primary region of net enstrophy production, it is important to highlight that, as in the single-point budget, all enstrophy originates from the wall boundary rather than the buffer layer itself. At the same time, the wall is also the region of highest rate of enstrophy destruction, as indicated by the solid blue isosurfaces in figure \ref{fig:GKE_3D}. Hence, the field of fluxes shown in figure \ref{fig:GKE_3D}, emerges from the highest scale enstrophy sink at the wall and the role of the vortex stretching and enstrophy production mechanisms of the buffer layer is to progressively deflect and amplify this ascending field of fluxes. In particular, the induced deviation of the peak of source located at $r_x= 0$ is such that the field of fluxes diverges in the space of scales thus forming a split enstrophy transfer towards both larger and smaller scales. Eventually, the field of fluxes vanishes in the secondary sink region of scale enstrophy that is the classical small-scale dissipative range distributed among all wall distances, as represented by the translucent grey isosurface in figure \ref{fig:GKE_3D}. This enstrophy sink extends throughout the entire thickness of the boundary layer and appears to be primarily sustained by fluxes diverging from the source region centred on the $r_x = 0$ plane. Alternatively, the field of fluxes ends up in the outer region feeding a wide range of long and wide scales in the interface region.

\begin{figure}[t!]
\centering
\includegraphics[trim=20cm 0.5cm 12cm 1cm, clip=true, width=0.45\linewidth]{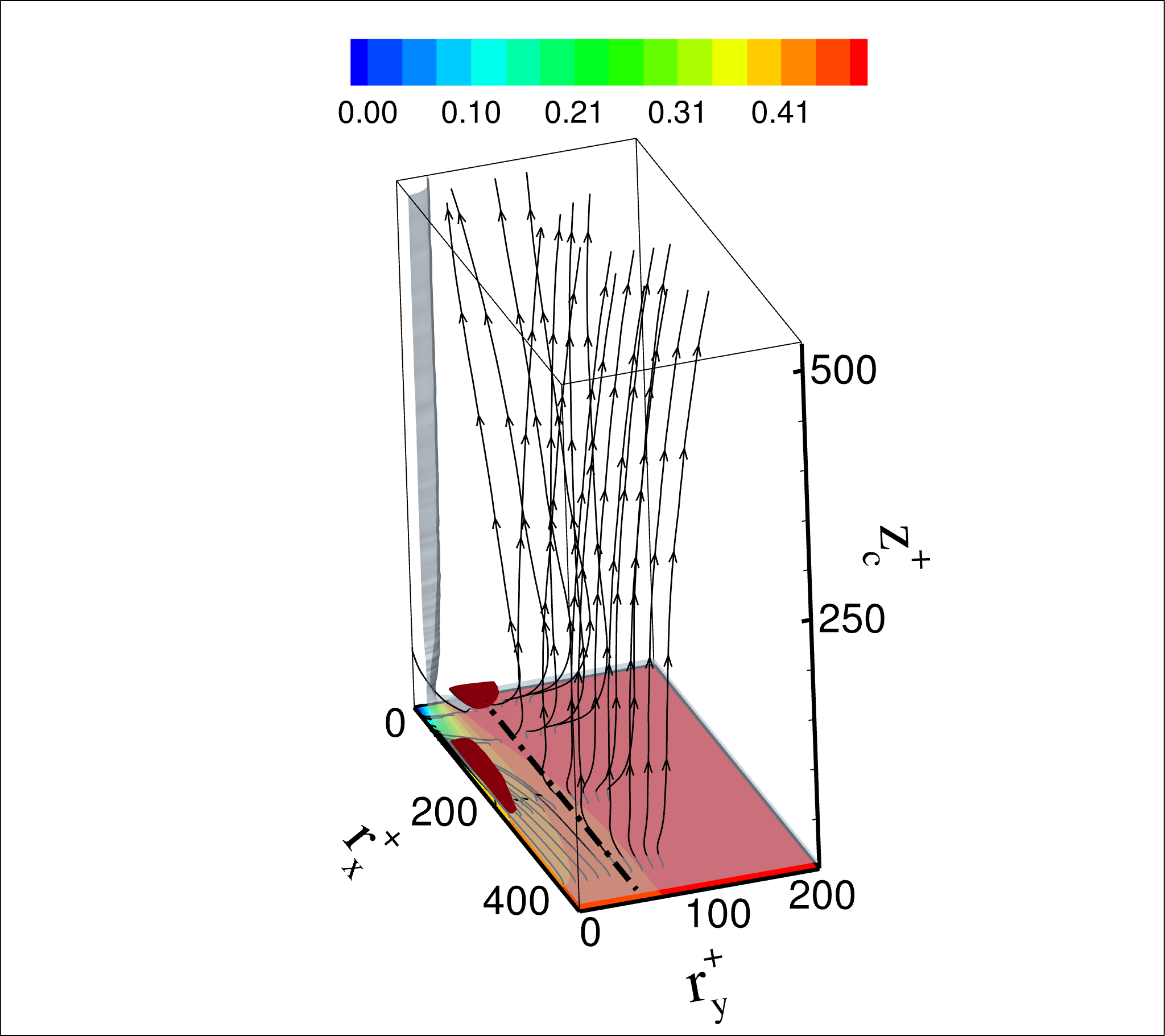}
        \caption{Presence of a separation scale in the enstrophy fluxes. Isosurfaces of positive net source of enstrophy $S_{\delta \omega^2}^+ = 0.008$ (red) and negative net source $S_{\delta \omega^2}^+ = -5 \times 10^{-6}$ (transparent grey). The flux $(\phi_{r_x}, \phi_{r_y}, \phi_{z_c})|_{r_z=0}$ is reported as black lines and arrows. Isocontours of $\langle \delta \omega^2 \rangle^+$ are reported at the wall. The dash-dotted line is placed at $(r_y^+,z_c^+) = (55,12)$ for comparison.
\label{fig:GKE_3D_div_scale}}
\end{figure}

Figure \ref{fig:GKE_3D_div_scale} provides a more detailed view of the enstrophy fluxes originating at the wall. This visualization reveals the presence of a characteristic spanwise scale, $r_y^+ \approx 55$, which separates two distinct enstrophy transport behaviours. In particular, enstrophy generated at spanwise scales smaller than this separation scale, $r_y^+ \lesssim 55$, remains confined below the buffer layer. In this region, scale enstrophy undergoes a direct interscale transfer towards smaller scales and ultimately dissipates near the $z_c$-axis, without experiencing a significant spatial transport away from the near-wall region. Accordingly, this family of fluxes statistically describes local dynamics of enstrophy that directly pertain to flow mechanisms within the viscous sublayer. On the other hand, scale enstrophy generated at the wall at spanwise scales larger than this separation scale, $r_y^+ \gtrsim 55$, is able to pass through the buffer layer and reach the outer region of the boundary layer. Unlike the first small-scale family of fluxes, this second family of fluxes is not confined to the near-wall region. Accordingly, this family of fluxes statistically describes non local interactions between large-scale structures populating the buffer and outer regions of the flow and the wall boundary condition.

The isocontour at the wall shown in figure \ref{fig:GKE_3D_div_scale} indicates the intensity of scale enstrophy $\langle \delta \omega^2 \rangle$. This representation highlights that the scale enstrophy associated to the first family of fluxes, which remains confined below the buffer layer, is characterized by relatively low intensity. This feature further supports the idea that this branch of fluxes represents the statistical evidence of local dynamics near the wall, involving low-intensity motions whose dimensions are small enough to be accommodated in the viscous sublayer. In contrast, the scale enstrophy associated to the second family of fluxes is more intense, see again the isocontour at the wall shown in figure \ref{fig:GKE_3D_div_scale}. As already shown in section \ref{sec:TP_enstr_contr}, this intense enstrophy generated at the wall is the near-wall footprint of the turbulent structures populating the buffer and overlap layers. This feature further supports the idea that the second family of fluxes is statistically representative of the presence of non-local interactions between the large scale and intense motions spanning the boundary layer thickness with the wall boundary condition. The ability of this portion of enstrophy to cross the buffer layer may be related to its sufficiently large spanwise scale, which allows it to correlate effectively with the large-scale motions that are most effective in driving turbulent transport. Hence, the turbulent flux $\langle \delta \omega^2 \widetilde{w} \rangle$ is more likely to be effective when the quantity transported, $\delta \omega^2$, has a characteristic scale as the motions transporting it, $\widetilde{w}$. In this context, the structures that populate the buffer layer and the outer region not only dominate the enstrophy generation through their footprint at the wall, which constitutes the main enstrophy feature within the viscous sublayer, but also seem to actively interact with the near-wall region itself, by transporting enstrophy away from the wall.

To put this into perspective, the observed behaviour highlights the role of ``non-local'' interactions between the hierarchy of flow structures populating the turbulent core and the constraints imposed by the wall and the outer, irrotational, region. This interpretation is consistent with the absence of a direct energy cascade in the viscous sublayer, where the dynamics is dominated by ``non-local structures'' whose evolution is not compatible with a conventional transfer of energy towards smaller scales. Energy must therefore be dissipated across the range of horizontal scales over which it is distributed.

\subsection{The $(r_x, z_c)$ and $(r_y, z_c)$ subspaces} \label{TP_enstr_2D}

The analysis of the enstrophy budget in the $(r_x,r_y,z_c)|_{r_z=0}$ space highlighted distinct behaviours in the streamwise and spanwise scales. To better isolate the dynamics in these two directions and facilitate their physical interpretation, the $(r_x,z_c)|_{r_y=r_z=0}$ and $(r_y,z_c)|_{r_x=r_z=0}$ subspaces are analysed separately here. Particular attention is devoted to the spanwise scale dynamics, as the previous analysis revealed a characteristic spanwise scale, located at $r_y^+ \approx 55$ and approximately independent of $r_x$, that separates two qualitatively different families of enstrophy transport.

\begin{figure}[t!]
\centering
\includegraphics[trim=6cm 2.5cm 4cm 1cm, clip=true, width=0.98\linewidth]{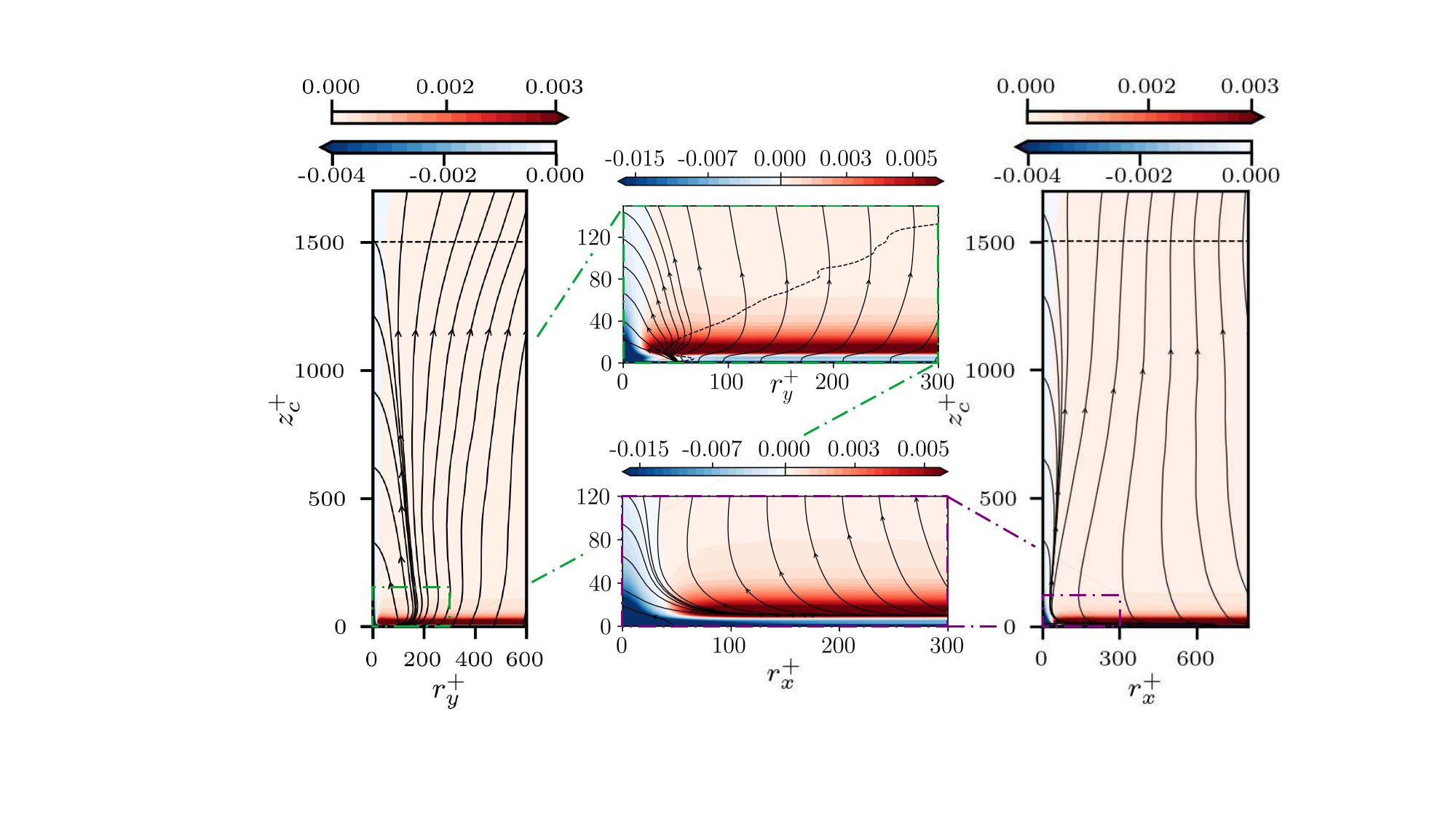}\mylab{-125mm}{84mm}{(\aaa)}\mylab{-35mm}{84mm}{(\bbb)}\mylab{-85mm}{74mm}{(\ccc)}\mylab{-85mm}{36mm}{(\ddd)}
        \caption{(a) Source $S_{\delta \omega^2}^+$ and flux $(\phi_{r_y},\phi_{z_c})_{r_x=r_z=0}^+$ of two-point enstrophy in the subspaces $(r_y^+, z_c^+)|_{r_x=r_z=0}$. (b) Source $S_{\delta \omega^2}^+$ and flux $(\phi_{r_x},\phi_{z_c})_{r_y=r_z=0}^+$ of two-point enstrophy in the subspaces $(r_x^+, z_c^+)|_{r_y=r_z=0}$. Negative values of $S_{\delta \omega^2}^+ < 0$ are reported on scale of blues, while positive values $S_{\delta \omega^2}^+ > 0$ on scale of reds. The fluxes are reported as black lines and arrows. In (c) and (d) are reported enlarged views of the near-wall region of (a) and (b), respectively. In (c), the crossover scale $\ell_{\delta \omega^2}$ is reported as a black dashed line.
\label{fig:GKE_2D}}
\end{figure}

The sources and fluxes of scale enstrophy in the $(r_y,z_c)|_{r_z=r_x=0}$ and $(r_x,z_c)|_{r_z=r_y=0}$ subspaces are shown in figure \ref{fig:GKE_2D}. Following the same representation adopted above, the enstrophy fluxes $\boldsymbol{\phi_{\delta \omega^2}}$ are represented by black arrows, while the net source term $S_{\delta \omega^2}$ is shown by isocontours. Blues contours denote a net sink of two-point enstrophy, while red contours indicate a positive net source. In this simplified view, arrows pointing towards the right indicate a reverse interscale enstrophy transfer, feeding larger scales, while arrows pointing towards the left indicate a direct interscale enstrophy transfer, towards smaller scales.

From a macroscopic perspective, the main features observed in the $(r_x,r_y,z_c)|_{r_z=0}$ subspace (in section \ref{sec:TP_enstr_3D}) remain clearly visible in both $(r_y,z_c)|_{r_x=r_z=0}$ and $(r_x,z_c)|_{r_y=r_z=0}$ subspaces. In both subspaces, the net source term $S_{\delta \omega^2}$ exhibits strong negative values in the near-wall region, constituting the main sink of enstrophy, and shows a milder sink region at small scales throughout the boundary layer thickness. In contrast, high enstrophy production through vortex stretching is observed within the buffer layer and persists, with a lower intensity, throughout the boundary layer core, except at the smallest scales where enstrophy is dissipated.

The enstrophy fluxes originating from the wall exhibit markedly different behaviours in the two subspaces. In the $(r_x,z_c)|_{r_y=r_z=0}$ plane (figure \ref{fig:GKE_2D}(b)), the interscale transfer in the near-wall region is characterized by a strong direct transfer towards smaller scales, see also the enlarged view reported in figure \ref{fig:GKE_2D}(d). Further away from the wall, the fluxes can be divided into two distinct branches: one feeding the smaller scales while ascending through the boundary layer, whereas the other transporting enstrophy towards the outer region. The former branch, while ascending, exhibits a mild reverse transfer before bending towards the $z_c$-distributed dissipative range of small scales. The latter branch, while ascending, exhibits first a mild direct transfer towards small scales before eventually becoming vertical, thus highlighting a very mild interscale transport with respect to the spatial one.

The $(r_y,z_c)|_{r_x=r_z=0}$ plane displayed in figure \ref{fig:GKE_2D}(a), clearly shows the two family of fluxes discussed in section \ref{sec:TP_enstr_3D}. The family of fluxes originating at the wall for scales $r_y^+ \lesssim 55$, undergoes a direct transfer towards smaller scales while moving away from the wall, finally reaching the smallest scales where scale enstrophy is dissipated. Conversely, the family of fluxes originating at the wall for scales $r_y^+ \gtrsim 55$, initially undergoes a reverse transfer towards larger spanwise scales while moving away from the wall. This reverse interscale transfer is the result of the intense enstrophy production region located in the buffer layer, which acts as a repulsion for the fluxes deflecting them towards larger scales, as highlighted by the enlarged view reported in figure \ref{fig:GKE_2D}(c). This picture is consistent with the ``near-wall cycle'' described by \citet{Jimenez1999autonomous}, in which quasi-streamwise vortices induce a spanwise velocity fluctuation $v$ near the wall, generating a secondary streamwise vorticity component, $\omega_x$, which is subsequently transported away from the wall towards the buffer layer. This transport is expected to be most effective for enstrophy associated with scales comparable to those of the buffer-layer structures.

Once the enstrophy fluxes overcome the buffer layer production region, they continue to move away from the wall and further split into two distinct branches. The first undergoes a direct transfer towards smaller spanwise scales, finally reaching the $z_c$-distributed small scale dissipative region. The second branch continues to ascend through the boundary layer until it approaches the interface region. In the vicinity of the interface, a slight deflection of the flux trajectories towards larger spanwise scales can be observed, indicating a weak reverse enstrophy transfer feeding larger spanwise scales. As will be shown in the next section \ref{sec:TP_enstr_1D}, this reverse interscale transfer is accompanied by a direct transfer in the wall-normal scales $r_z$. This behavior further supports the presence of confined dynamics in the turbulent/non-turbulent interface region similar to the ones observed for the wall region as already mentioned in section \ref{conf_eff}. Indeed, a direct transfer towards smaller wall-normal scales, together with a reverse transfer towards larger horizontal scales (predominantly in the spanwise direction $r_y$), is consistent with a scenario in which large-scale structures in the turbulent core induce shear in the interface and near-wall regions. These shearing motions develop across thin wall-normal layers and span a hierarchy of horizontal scales, consistent with the observations from the structure function analysis in section \ref{conf_eff}. This scenario was originally proposed by \citet{cimarelli2015spectral, cimarelli2021spatially, cimarelli2024spatially} for both the near-wall and interface regions and finds strong support from the scale enstrophy behavior here reported.

\subsection{Relation between energy and enstrophy interscale transfers}

The split enstrophy interscale transfer analysed in the previous section and shown in figure \ref{fig:GKE_2D}, shares some similarities with the split energy cascade found in several previous works \citep{cimarelli2013paths, cimarelli2015sources, cimarelli2016cascades, cimarelli2024spatially}. In order to highlight the relation between forward and reverse transfers of enstrophy and energy, we compare the range of spanwise scales involved in the two processes. In particular, we introduce a cross-over scale separating regions dominated by direct interscale transfer from those characterized by inverse interscale transfer. For scale enstrophy, this cross-over scale is denoted as $\ell_{\delta \omega^2}$ and is defined such that $\phi_{r_y}(\ell_{\delta \omega^2},z_c) = 0$. Analogously, for scale energy, this cross-over scale is denoted as $\ell_{\delta q^2}$ and is defined as the spanwise separation scale at which the scale energy flux changes sign, $\phi_{\delta q^2, r_y}(\ell_{\delta q^2},z_c) = 0$, where $\phi_{\delta q^2, r_y} = \langle \delta q^2 \delta v \rangle -2\nu \partial \langle \delta q^2 \rangle / \partial r_y$ is the scale energy flux and $\delta q^2 = \delta u_i \delta u_i$ is the second-order structure function of velocity (see \citet{cimarelli2024spatially} for further details).

\begin{figure}[t!]
\centering
\includegraphics[width=0.65\linewidth]{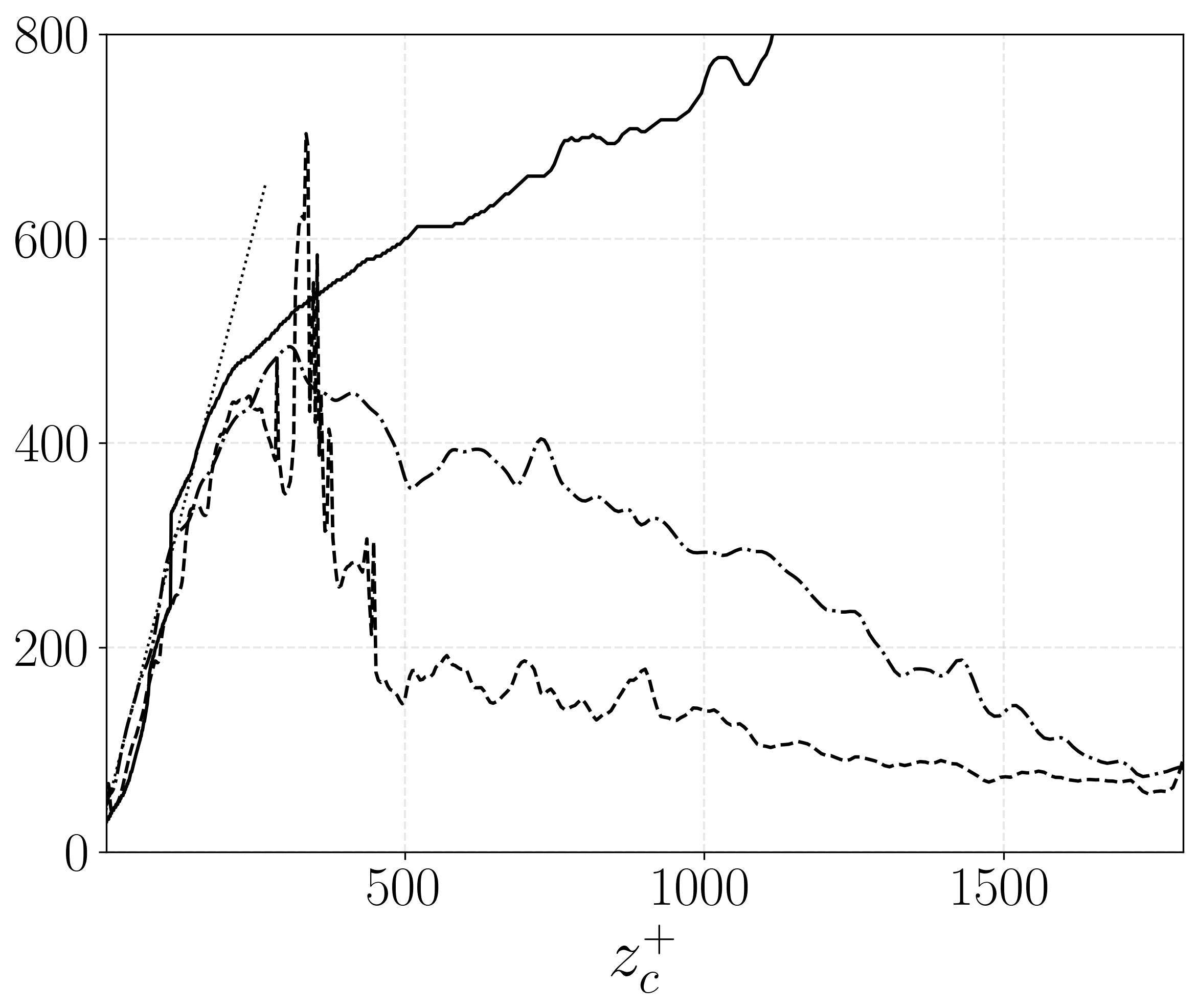}
        \caption{Crossover scales of scale enstrophy $\ell_{\delta \omega^2}^+ (z_c^+)$ (dashed line) and scale energy $\ell_{\delta q^2}^+ (z_c^+)$ (dash-dotted line) as a function of the wall distance $z_c^+$. The energy injection scale $\ell_{\Pi}^+ (z_c^+)$ is reported as a solid line. The dotted line reports the self-similar scaling $\ell_{\delta q^2}^+ \approx \ell_{\delta \omega^2}^+ \approx 2.3 z_c^+ +40$.
\label{fig:crossover}}
\end{figure}

The comparison between the enstrophy and energy crossover scales, $\ell_{\delta \omega^2}$ and $\ell_{\delta q^2}$, is shown in figure \ref{fig:crossover}. Both crossover scales exhibit a very similar linear scaling up to $z_c^+ \approx 200$, suggesting that the split energy transfer observed in scale-space dynamics, which \citet{cimarelli2013paths, cimarelli2016cascades, cimarelli2024spatially} associated with the presence of attached eddies, is also reflected in the enstrophy dynamics. This behaviour may be interpreted as follows. The forward (direct) energy transfer, often associated with the break-up of detached eddies \citep{cimarelli2013paths}, generates progressively finer scales in both the velocity and enstrophy fields. Conversely, the inverse energy transfer, linked to the growth and organization of attached-eddy-like motions as the wall-normal distance increases \citep{cimarelli2016cascades}, promotes the formation of more relaxed velocity gradients, which in turn shape the enstrophy distribution. In this sense, enstrophy appears to mimic the behaviour and scale distribution of turbulent energy.

Further support for this connection, figure \ref{fig:crossover} also compares these crossover scales with the energy-injection scale $\ell_\Pi$, which has been associated with the characteristic size of attached eddies and is defined as the scale at which the energy production term, $\Pi = -2\langle \delta u \delta w \rangle (\partial U / \partial z)$ (for $r_z=0$), reaches its maximum value: $\Pi(\ell_\Pi,z_c) = \text{max}_{r_y} \Pi(r_y,z_c)$. The close agreement between these scales, as shown in figure \ref{fig:crossover}, further supports the link between the enstrophy dynamics and the characteristic scales of attached-eddy motions.

Hence, the confinement effect given by the wall also extends into the core of the boundary layer, affecting both energy and enstrophy dynamics, as reflected in the attached-eddy-like scaling observed in both fields. In fact, above the buffer layer, the inverse energy transfer can be interpreted as a consequence of the confinement given by the wall: energy-containing motions generated by the mean shear can either transfer energy to larger scales, thus extending farther away from the wall, or break down into smaller scales performing a direct energy transfer. The portion of energy transferred towards larger motions, which necessarily move away from the wall (since their wall distance cannot be smaller than their dimension), continues to interact with the wall and with the generation and transport of turbulent enstrophy. On the other hand, the portion of energy transferred towards smaller scales is no longer directly influenced by the presence of the wall and stops interacting with it, performing a direct transfer up to small dissipative scales.

\section{Scale-by-scale budgets of scale enstrophy} \label{sec:TP_enstr_1D}

The analysis presented so far has examined the structure of two-point enstrophy, its source and dissipation regions, and the topology of the multidimensional field of fluxes in the $(r_x,r_y,z_c)|_{r_z=0}$, $(r_x,z_c)|_{r_y=r_z=0}$, and $(r_y,z_c)|_{r_x=r_z=0}$ subspaces. While these multidimensional representations provide a global picture of enstrophy dynamics, the physical role of the individual budget terms is more clearly assessed by restricting the analysis to the spanwise separation scale $r_y$ at $r_x=r_z=0$ and examining each budget term separately at selected wall-normal locations $z_c$. To this end, the following subsections examine the scale-by-scale budget, providing a more quantitative picture of the relative importance of the different terms in the very near-wall region (represented by $z_c^+ = 0.5$), in the buffer layer (represented by $z_c^+ = 11$), in the outer region (represented by $z_c^+ = 450$), and in the TNTI region (represented by $z_c^+ = Re_{\tau} = 1501$).

\subsection{Very near-wall region}

\begin{figure}[t!]
\centering
    \includegraphics[width=0.487 \linewidth]{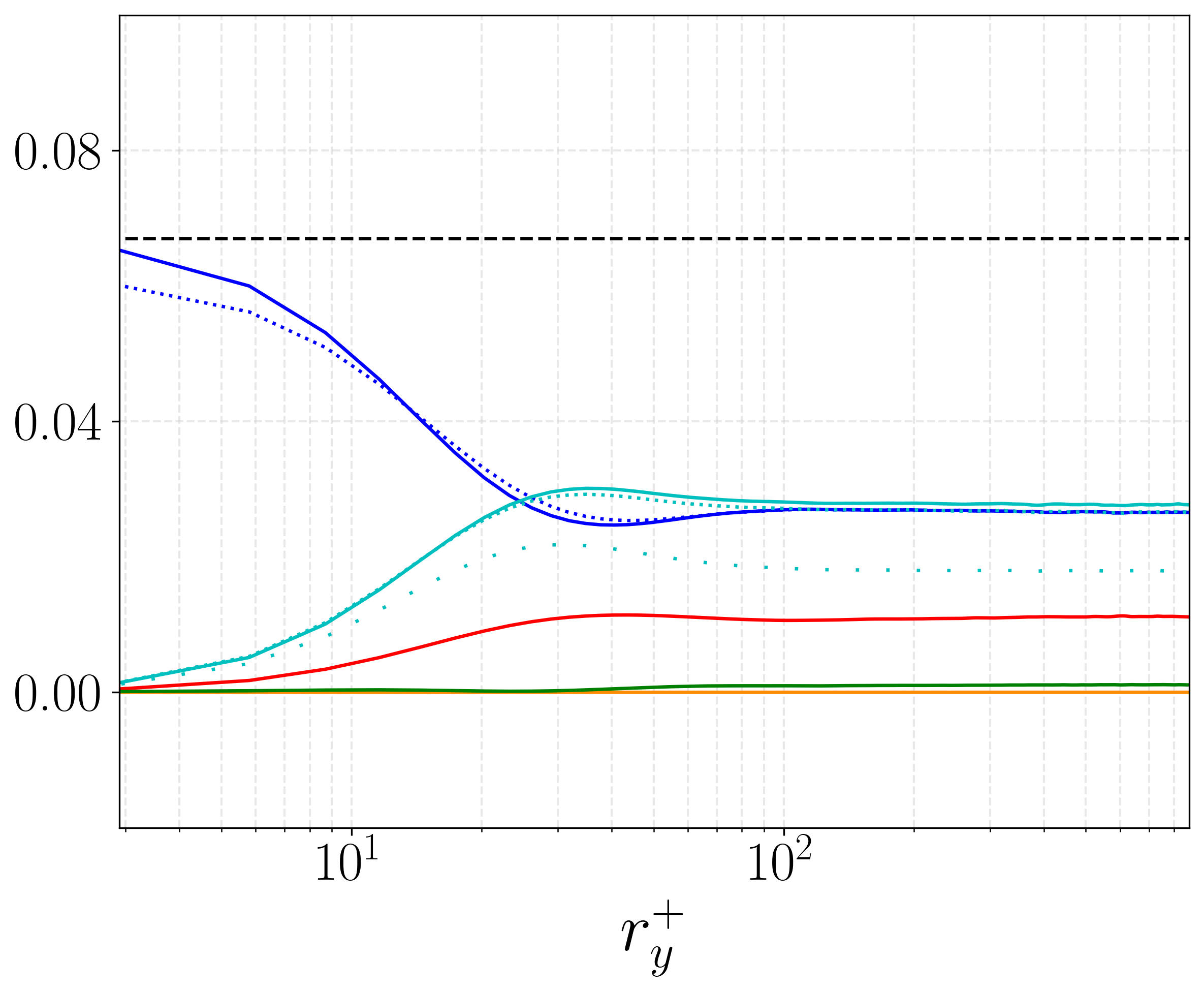}\mylab{-58mm}{50mm}{(\aaa)}
\includegraphics[width=0.493 \linewidth]{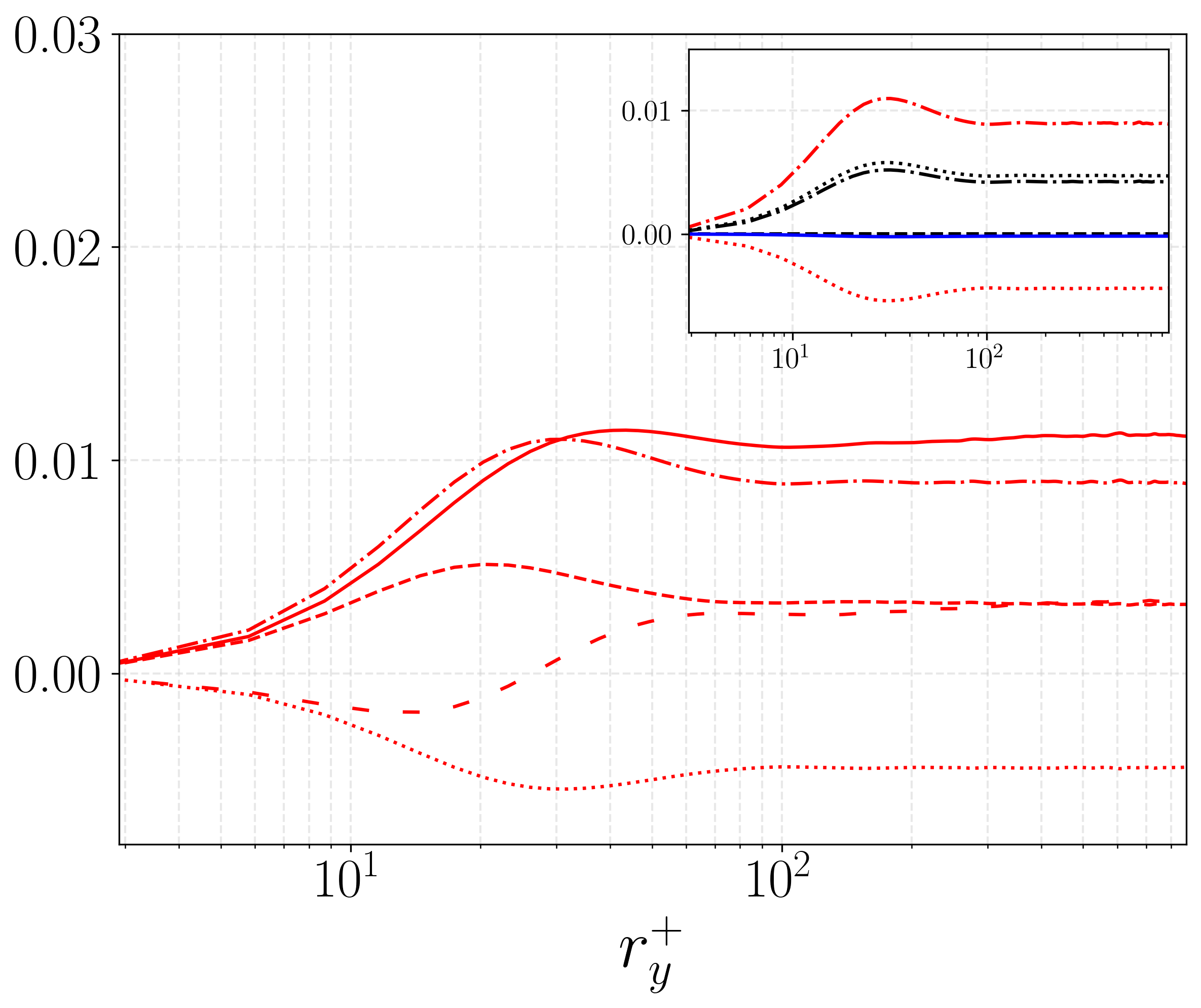}\mylab{-58mm}{50mm}{(\bbb)}
        \caption{Scale-by-scale enstrophy budget terms in the spanwise scales $r_y^+$ evaluated at $z_c^+ = 0.5$. (a) Complete budget, with $T_{r}^+$ (solid green line), $D_{r}^+$ (solid blue line), $T_{c}^+$ (solid cyan line), $\xi_{\delta \omega^2}^+$ (solid red line), $\Pi_{\delta \omega^2}^+$ (solid orange line), $2 \chi^+$ (dashed black line). Some relevant contributions are highlighted: $(2 \nu (\partial^2 \langle \delta \omega^2 \rangle / \partial r_z^2))^+$ (dashed blue line), $(\nu / 2 (\partial^2 \langle \delta \omega^2 \rangle / \partial z_c^2))^+$ (densely-dashed cyan line) and $(\nu / 2 (\partial^2 \langle \delta \omega_x^2 \rangle / \partial z_c^2))^+$ (loosely-dotted cyan line). (b) Vortex stretching contributions: total $\xi_{\delta \omega^2}^+$ (red solid line), $2 \Omega_y^+ \langle \delta \omega_i \delta \left( \partial u_i / \partial y \right) \rangle^+$ (red dash-dotted line), $2 \langle \delta \omega_x \delta \omega_z \rangle^+ \partial U^+ / \partial z^+$ (red dotted line), $2 \langle \delta \omega_i \widetilde{\omega}_j \delta \left( \partial u_i / \partial x_j \right) \rangle^+$ (red dashed line), $2 \langle \delta \omega_i \delta \omega_j \widetilde{\left( \partial u_i / \partial x_j \right)} \rangle^+$ (red loosely-dashed line). In the inset panel the partial cancellation between $2 \Omega_y^+ \langle \delta \omega_i \delta \left( \partial u_i / \partial y \right) \rangle^+$ (red dash-dotted line) and $2 \langle \delta \omega_x \delta \omega_z \rangle^+ \partial U^+ / \partial z^+$ (red dotted line) is shown: the sum $2 \Omega_y^+ \langle \delta \omega_x \delta \left( \partial u / \partial y \right) \rangle^+ + 2 \langle \delta \omega_x \delta \omega_z \rangle^+ \partial U^+ / \partial z^+$ is reported in solid blue line; $2 \Omega_y^+ \langle \delta \omega_x \delta \left( \partial u / \partial y \right) \rangle^+$ in black dash-dotted line; $2 \Omega_y^+ \langle \delta \omega_y \delta \left( \partial v / \partial y \right) \rangle^+$ in black dotted line and $2 \Omega_y^+ \langle \delta \omega_z \delta \left( \partial w / \partial y \right) \rangle^+$ in black dashed line.}
\label{fig:GKE_1D_z05}
\end{figure}

We begin by examining the scale-by-scale enstrophy budget at $z_c^+=0.5$, reported in figure \ref{fig:GKE_1D_z05}(a), in the immediate vicinity of the wall, where vorticity is generated. The budget is dominated by diffusion processes. In particular, the transport among scales is dominated by viscous diffusion $D_r$ being the inertial transport always negligible $T_r \approx 0$, and the spatial transport is almost exclusively determined by the diffusive term, $T_c \approx \nu / 2 (\partial^2 \langle \delta \omega^2 \rangle / \partial z_c^2)$. On the other hand, the only non-zero source terms are those related to vortex stretching $\xi_{\delta \omega^2}$ and to the viscous dissipation $2\chi$, being the production associated with the mean vorticity gradient $\Pi_{\delta \omega^2}$ negligible. The detailed analysis of each single process follows.

The diffusive transport in the physical-space, $\nu / 2 (\partial^2 \langle \delta \omega^2 \rangle / \partial z_c^2)$, exhibits a mild peak at moderate spanwise scales, $r_y^+ \approx 30\text{--}40$. This suggests that enstrophy is injected at these scales mainly by wall-normal diffusion, driven by the curvature of two-point enstrophy field in the $z_c$ direction. The occurrence of this peak may be related to the local maximum of streamwise vorticity observed in figure \ref{fig:Sii_z05}(a), which is likely associated with the wall footprint of the quasi-streamwise vortices. Consistently, the streamwise vorticity contribution to the wall-normal diffusion term, $\nu / 2 (\partial^2 \langle \delta \omega_x^2 \rangle / \partial z_c^2)$, is found to be the dominant contribution, as shown by the loosely-dotted cyan line in figure \ref{fig:Sii_z05}(a). Interestingly, despite the larger magnitude of the spanwise vorticity component, the streamwise vorticity contribution is more effectively diffused in the domain through wall-normal diffusion.

The transport among scales shows that small spanwise scales are primarily fed by diffusive interscale transport $D_{r}$. In particular, the dominant contribution at this wall-distance is provided by the wall-normal contribution $2\nu \partial^2 \langle \delta \omega^2 \rangle / \partial r_z^2$ (reported as the dotted blue line in figure \ref{fig:GKE_1D_z05}(a)). This indicates that, at small scales, the main interscale transfer of enstrophy in the very near-wall region is driven by the strongly inhomogeneous distribution of the enstrophy with respect to the wall-normal separation scales $r_z$.

Even at such a small wall-normal distance ($z_c^+ = 0.5$), the vortex stretching term $\xi_{\delta \omega^2}$ is non-negligible (solid red line in figure \ref{fig:GKE_1D_z05}(a)). As shown in equation (\ref{eq:enstr_TP_TBL}), this term can be decomposed into two purely fluctuating contributions, $2 \langle \delta \omega_i \widetilde{\omega}_j \delta ( \partial u_i / \partial x_j ) \rangle$ and $2 \langle \delta \omega_i \delta \omega_j \widetilde{( \partial u_i / \partial x_j )} \rangle$, and two contributions involving the mean flow, $2 \Omega_y \langle \delta \omega_i \delta ( \partial u_i / \partial y ) \rangle$ and $2 \langle \delta \omega_x \delta \omega_z \rangle \partial U / \partial z$. The behaviour of these contributions is reported in figure \ref{fig:GKE_1D_z05}(b). Interestingly, while the two fully fluctuating terms approximately recover the behaviour observed in homogeneous isotropic conditions \citep{boga2026two}, the two mean-flow-related terms provide opposite contributions, leading to a partial cancellation in their net effect. This partial cancellation can be traced back to a specific component of the term $2 \Omega_y \langle \delta \omega_i \delta ( \partial u_i / \partial y ) \rangle$, whose decomposition is reported in the inset panel of figure \ref{fig:GKE_1D_z05}(b). At this wall-normal location, the contribution $2 \Omega_y \langle \delta \omega_z \delta ( \partial w / \partial y ) \rangle$, associated with the generation of wall-normal vorticity $\langle \delta \omega_z^2 \rangle$ through tilting of the mean spanwise vorticity remains negligible. The dominant contributions are therefore those associated with the streamwise tilting and the pure stretching of the mean spanwise vorticity $\Omega_y$, $2 \Omega_y \langle \delta \omega_x \delta ( \partial u / \partial y ) \rangle$ and $2 \Omega_y \langle \delta \omega_y \delta ( \partial v / \partial y ) \rangle$, respectively. These two terms have comparable amplitudes (black lines in the inset of figure \ref{fig:GKE_1D_z05}(b)) and both exhibit a maximum around $r_y^+\approx 30$. This characteristic scale is likely associated with the wall signature of quasi-streamwise vortices, whose counter-rotating motion generates alternating spanwise variations of the velocity field. In particular, the rotation directly induces fluctuations in the spanwise velocity $v$, while the associated wall-normal velocity transports high-speed fluid from the buffer layer towards the wall, generating corresponding variations in the streamwise velocity $u$. These velocity fluctuations locally stretch and tilt the mean spanwise vorticity $\Omega_y$ enhancing two-point enstrophy production at the corresponding spanwise scales.

The term responsible for the cancellation with $2 \langle \delta \omega_x \delta \omega_z \rangle \partial U / \partial z$ is $2 \Omega_y \langle \delta \omega_x \delta ( \partial u / \partial y ) \rangle$, whose sum with the former contribution is reported as the solid blue line in the inset of figure \ref{fig:GKE_1D_z05}(b)). Both terms contribute to the streamwise vorticity $\langle \delta \omega_x^2 \rangle$ budget (not shown), and represent the tilting of the mean vorticity $\Omega_y$ through fluctuations of $\partial u / \partial y$ and the tilting of fluctuating $\omega_z$ through the mean shear $\partial U / \partial z$, respectively. This cancellation can be explained by noting that, in the present flow configuration, $\Omega_y = \partial U / \partial z$. Moreover, using the definition of spanwise vorticity fluctuation, $\delta \omega_z = \delta (\partial v / \partial x - \partial u / \partial y)$, and assuming an ``$x$-elongated'' velocity field, i.e. $\partial v / \partial x \ll \partial u / \partial y$, one obtains $\delta \omega_z \approx - \delta (\partial u / \partial y)$, hence $2 \Omega_y \langle \delta \omega_x \delta ( \partial u / \partial y ) \rangle \approx -2 \langle \delta \omega_x \delta \omega_z \rangle \partial U / \partial z$. From a physical point of view, the cancellation between these two terms may result from the dominance of sheet-like vortical structures in the near-wall region (e.g. elongated shear-layer–like formations), where dissipation and enstrophy are strongly spatially correlated (\citet{Bechlars_Sandberg_2017}), causing the two contributions to become nearly identical.

Overall, the above reasoning allow us to write a reduced-order scale-by-scale budget for scale enstrophy in the very near-wall region,
\begin{equation}
\begin{gathered}
       -2 \nu \frac{\partial^2 \langle \delta \omega^2 \rangle}{\partial r_z^2} - \frac{\nu}{2} \frac{\partial^2 \langle \delta \omega^2 \rangle}{\partial z_c^2} \approx \\  2 \Omega_y \langle \delta \omega_y \delta \left( \frac{\partial v}{\partial y} \right) \rangle 
       + 2 \langle \delta \omega_i \widetilde{\omega}_j \delta \left( \frac{\partial u_i}{\partial x_j} \right) \rangle + 2 \langle \delta \omega_i \delta \omega_j \widetilde{\left( \frac{\partial u_i}{\partial x_j} \right)} \rangle - 2 \chi.
\label{eq:enstr_TP_TBL_z05}
\end{gathered}
\end{equation}

\subsection{Buffer layer}

\begin{figure}[t!]
\centering
\includegraphics[width=0.49 \linewidth]{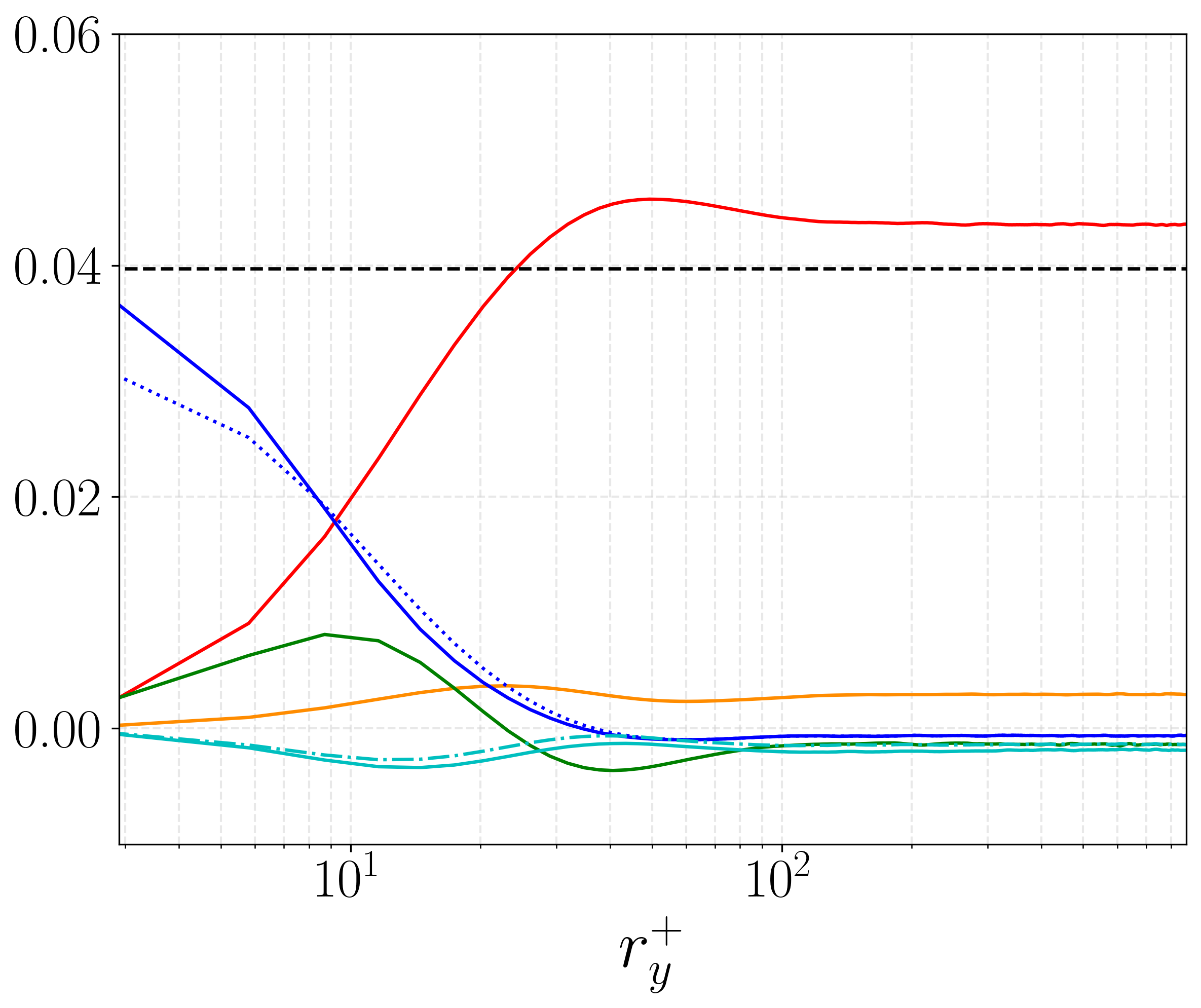}\mylab{-58mm}{50mm}{(\aaa)}
\includegraphics[width=0.49 \linewidth]{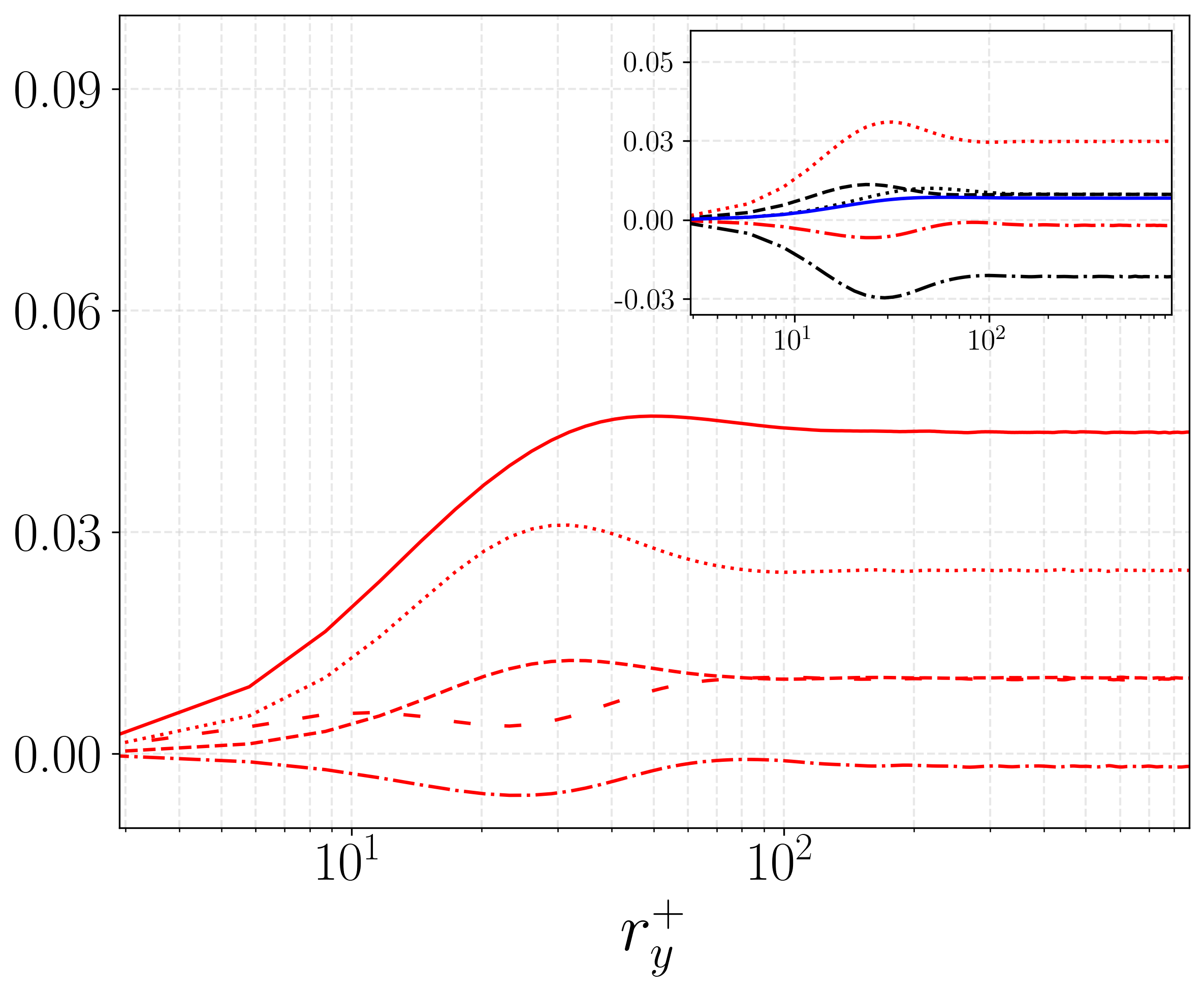}\mylab{-58mm}{50mm}{(\bbb)}
        \caption{Scale-by-scale enstrophy budget terms in the spanwise scales $r_y^+$ evaluated at $z_c^+ = 11$. (a) Complete budget, with $T_{r}^+$ (solid green line), $D_{r}^+$ (solid blue line), $T_{c}^+$ (solid cyan line), $\xi_{\delta \omega^2}^+$ (solid red line), $\Pi_{\delta \omega^2}^+$ (solid orange line), $2 \chi^+$ (dashed black line). Some relevant contributions are highlighted: $(2 \nu (\partial^2 \langle \delta \omega^2 \rangle / \partial r_z^2))^+$ (dashed blue line), and $-(\partial \langle \delta \omega^2 \widetilde{w} \rangle / \partial z_c)^+$ (dash-dotted cyan line). (b) Vortex stretching contributions: total $\xi_{\delta \omega^2}^+$ (red solid line), $2 \Omega_y^+ \langle \delta \omega_i \delta \left( \partial u_i / \partial y \right) \rangle^+$ (red dash-dotted line), $2 \langle \delta \omega_x \delta \omega_z \rangle^+ \partial U^+ / \partial z^+$ (red dotted line), $2 \langle \delta \omega_i \widetilde{\omega}_j \delta \left( \partial u_i / \partial x_j \right) \rangle^+$ (red dashed line), $2 \langle \delta \omega_i \delta \omega_j \widetilde{\left( \partial u_i / \partial x_j \right)} \rangle^+$ (red loosely-dashed line). In the inset panel the partial cancellation between $2 \Omega_y^+ \langle \delta \omega_i \delta \left( \partial u_i / \partial y \right) \rangle^+$ (red dash-dotted line) and $2 \langle \delta \omega_x \delta \omega_z \rangle^+ \partial U^+ / \partial z^+$ (red dotted line) is shown: the sum $2 \Omega_y^+ \langle \delta \omega_x \delta \left( \partial u / \partial y \right) \rangle^+ + 2 \langle \delta \omega_x \delta \omega_z \rangle^+ \partial U^+ / \partial z^+$ is reported in solid blue line; $2 \Omega_y^+ \langle \delta \omega_x \delta \left( \partial u / \partial y \right) \rangle^+$ in black dash-dotted line; $2 \Omega_y^+ \langle \delta \omega_y \delta \left( \partial v / \partial y \right) \rangle^+$ in black dotted line and $2 \Omega_y^+ \langle \delta \omega_z \delta \left( \partial w / \partial y \right) \rangle^+$ in black dashed line.}
\label{fig:GKE_1D_z11}
\end{figure}

Having examined the immediate vicinity of the wall, where vorticity is generated, we now move our attention to the buffer layer. This region exhibits the maximum net source of enstrophy $S_{\delta \omega^2}$ and therefore plays an important role in the enstrophy dynamics. As a representative wall-normal location, we consider $z_c^+=11$, and examine the corresponding scale-by-scale budget shown in figure \ref{fig:GKE_1D_z11}(a).

Overall, the budget in the buffer layer appears to recover, at least qualitatively, a structure comparable to that observed in homogeneous isotropic conditions \citep{Chiarini_2025,boga2026two}. In particular, the interscale diffusive transport $D_r$ dominates the budget at small scales, whereas the vortex stretching term $\xi_{\delta\omega^2}$ becomes dominant across essentially all larger scales. From a quantitative perspective, unlike the homogeneous isotropic case, the vortex stretching term $\xi_{\delta \omega^2}$ exceeds enstrophy destruction $2 \chi$ and exhibits a mild peak at scales $r_y^+ \approx 50$, which is consistent with the typical structures populating the buffer layer.

Despite the macroscopic similarity, the decomposition of the individual budget terms strongly differs from the homogeneous isotropic case. In particular, the interscale diffusive transport $D_r$, is still primarily accomplished through gradients across wall-normal scales, $2 \nu (\partial^2 \langle \delta \omega^2 \rangle / \partial r_z^2)$, as indicated by the blue dotted line in figure \ref{fig:GKE_1D_z11}(a). Moreover, the vortex stretching term $\xi_{\delta \omega^2}$ still retains a significant contribution associated with the mean flow. In fact, in addition to the fully fluctuating contributions, the term $2 \langle \delta \omega_x \delta \omega_z \rangle (\partial U / \partial z)$ (which feeds the streamwise vorticity $\langle \delta \omega_x^2 \rangle$) remains highly relevant. As shown in the inset of figure \ref{fig:GKE_1D_z11}(b), $2 \langle \delta \omega_x \delta \omega_z \rangle (\partial U / \partial z)$ still displays a partial cancellation with $2 \Omega_y \langle \delta \omega_x \delta (\partial u / \partial y) \rangle$, under the same approximations discussed in the previous subsection. However, since we are now examining a location farther away from the wall, these approximations are less accurate and the cancellation between these two term is considerably less precise than in the viscous sublayer. In contrast to the very near-wall region, the term $2 \Omega_y \langle \delta \omega_z \delta (\partial w / \partial y) \rangle$, associated with the tilting of $\Omega_y$ in the wall-normal direction $z$, becomes significant. As discussed in section
\ref{sec:TP_enstr_contr}, this term can be interpreted as a key mechanism for the redistribution of enstrophy from the mean spanwise vorticity into the wall-normal vorticity component $\omega_z$, whose contribution is negligible in the viscous sublayer. In contrast, the contribution associated with the ``pure stretching'' of the mean spanwise vorticity remains comparable to that observed in the immediate vicinity of the wall.

These observations suggest the following physical interpretation for the enstrophy dynamics in the buffer layer. The mean spanwise vorticity, $\Omega_y$, is first tilted in the wall-normal direction by non-uniform wall-normal velocity fluctuations $w$, thus generating wall-normal vorticity. This newly generated $\omega_z$ is subsequently tilted and stretched by the mean velocity gradient, $\partial U/\partial z$, through the term $2 \langle \delta \omega_x \delta \omega_z \rangle (\partial U / \partial z)$, which constitutes the strongest contribution to enstrophy production in this region and reaches its maximum around $r_y^+ \approx 30$. The overall dynamics described and the characteristic scales observed are consistent with the hairpin vortex regeneration mechanism described in the literature \citep{Hamilton1995,Jimenez1999autonomous,Zhou1999,Adrian2007}.

Finally, unlike in the very near-wall region, both the enstrophy production associated with the mean vorticity gradient $\Pi_{\delta \omega^2}$ and the inertial interscale transport $T_r$ become non-negligible in the buffer layer. In particular, $T_r$ exhibits a local minimum at $r_y^+ \approx 40$, approximately coinciding with the spanwise scale at which the vortex stretching term $\xi_{\delta \omega^2}$ reaches its maximum. This correspondence suggests that the excess enstrophy generated locally by vortex stretching is redistributed across scales through the inertial interscale transfer. Nevertheless, neither term emerges as a leading contributor to the budget, which remains primarily governed by the interscale diffusive transport $D_r$ and the vortex stretching term $\xi_{\delta \omega^2}$.

\subsection{Outer layer}

\begin{figure}[t!]
\centering
\includegraphics[width=0.49 \linewidth]{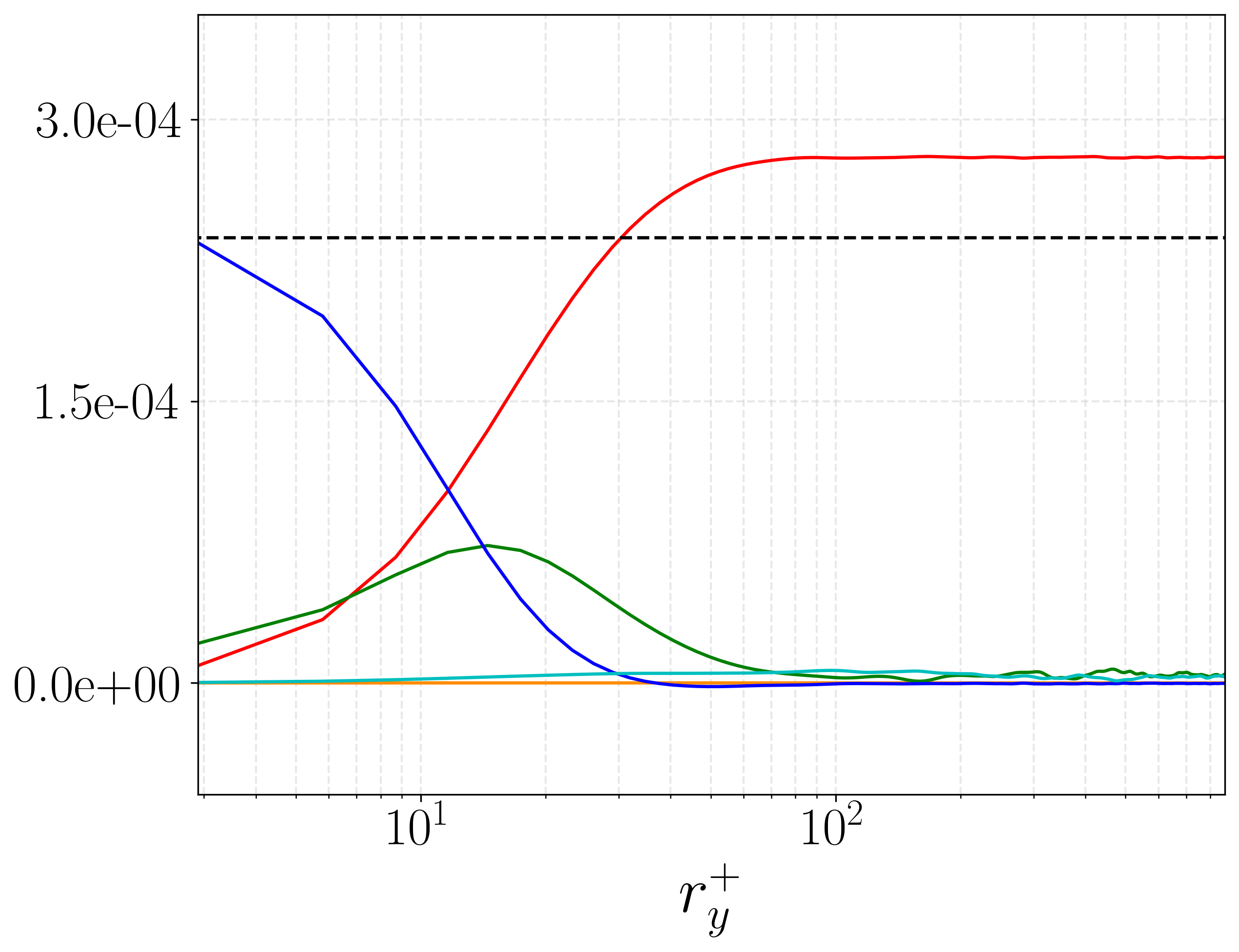}\mylab{-55mm}{46mm}{(\aaa)}
\includegraphics[width=0.49 \linewidth]{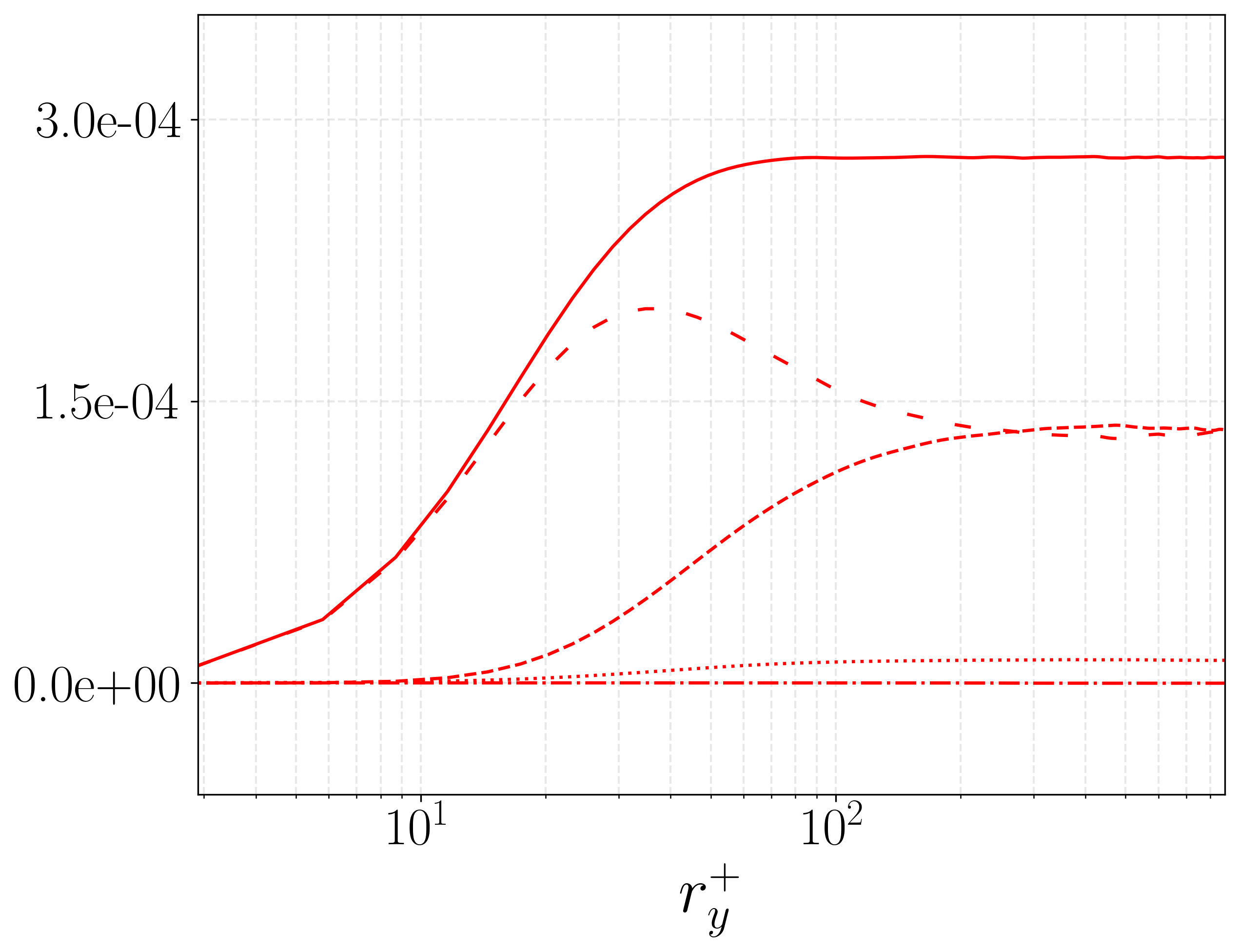}\mylab{-55mm}{46mm}{(\bbb)}
        \caption{Scale-by-scale enstrophy budget terms in the spanwise scales $r_y^+$ evaluated at $z_c^+ = 450$. (a) Complete budget, with $T_{r}^+$ (solid green line), $D_{r}^+$ (solid blue line), $T_{c}^+$ (solid cyan line), $\xi_{\delta \omega^2}^+$ (solid red line), $\Pi_{\delta \omega^2}^+$ (solid orange line), $2 \chi^+$ (dashed black line). (b) Vortex stretching contributions: total $\xi_{\delta \omega^2}^+$ (red solid line), $2 \Omega_y^+ \langle \delta \omega_i \delta \left( \partial u_i / \partial y \right) \rangle^+$ (red dash-dotted line), $2 \langle \delta \omega_x \delta \omega_z \rangle^+ \partial U^+ / \partial z^+$ (red dotted line), $2 \langle \delta \omega_i \widetilde{\omega}_j \delta \left( \partial u_i / \partial x_j \right) \rangle^+$ (red dashed line), $2 \langle \delta \omega_i \delta \omega_j \widetilde{\left( \partial u_i / \partial x_j \right)} \rangle^+$ (red loosely-dashed line).}
\label{fig:GKE_1D_z450}
\end{figure}

As shown in section \ref{sec:TP_enstr_3D}, only the enstrophy generated at the wall at spanwise scales $r_y^+ \gtrsim 55$ manages to correlate with the structures of the buffer layer which effectively transport enstrophy towards the outer region. Moreover, as enstrophy is transported through the buffer layer, it is redistributed among its three vorticity contributions, with the wall-normal vorticity, $\omega_z$, becoming increasingly significant (figure \ref{fig:enstr_contr}). Consequently, the enstrophy budget in the outer region substantially differs from that observed in the near-wall region, where the confinement effects caused by the presence of the solid wall are pronounced. The scale-by-scale budget of enstrophy at a wall distance $z_c^+ = 450$, representing the outer region, is shown in figure \ref{fig:GKE_1D_z450}(a).

To a first approximation, the enstrophy dynamics in the outer region of the boundary layer is very similar to the one observed in homogeneous and isotropic conditions \citep{boga2026two}. Indeed, that spatial transport related to vertical inhomogeneity is negligible and the budget is dominated by interscale diffusive transport at small scales, whereas vortex stretching becomes the leading contribution at larger scales. The only other non-negligible term reported is the inertial interscale transport, which acts over intermediate scales without ever becoming dominant, thereby preventing the establishment of a pure inertial range of enstrophy cascade (see \citet{boga2026two}).

For the present flow settings, the vortex stretching term $\xi_{\delta \omega^2}$ saturates at a value that exceeds that of enstrophy dissipation $2 \chi$. This imbalance arises because the boundary layer studied in the present work is temporally evolving. Specifically, the excess enstrophy production observed implies a positive local enstrophy growth rate, i.e. $\partial \langle \delta \omega^2 \rangle / \partial t > 0$, at the wall-normal location considered.

The decomposition of the vortex stretching term among its contributions reported in figure \ref{fig:GKE_1D_z450}(b), also closely resembles that observed in homogeneous isotropic turbulence \citep{boga2026two}. In particular, the contributions related to the mean flow, $2 \langle \delta \omega_x \delta \omega_z \rangle (\partial U / \partial z)$ and $2 \Omega_y \langle \delta \omega_i \delta ( \partial u_i / \partial y ) \rangle$, are negligible, whereas the purely fluctuating contributions account for essentially the entire vortex stretching term $\xi_{\delta \omega^2}$. The behaviour of the purely fluctuating components of the vortex stretching term, $2 \langle \delta \omega_i \delta \omega_j \widetilde{(\partial u_i/ \partial x_j)} \rangle$ and $2 \langle \delta \omega_i \widetilde{\omega}_j \delta (\partial u_i/ \partial x_j) \rangle$, is also consistent with that observed in homogeneous isotropic turbulence. Specifically, the former contribution exhibits a pronounced peak at intermediate scales, whereas the latter increases gradually towards its asymptotic value. At the largest scales, the two contributions converge, as required by their asymptotic values.

In conclusion, the scale-by-scale enstrophy budget of the outer region can be approximated by that of unsteady homogeneous isotropic conditions:
\begin{equation}
\begin{gathered}
       \frac{\partial \langle \delta \omega^2 \rangle}{\partial t} + \frac{\partial \langle \delta \omega^2 \delta u_j \rangle}{\partial r_j} - 2 \nu \frac{\partial^2 \langle \delta \omega^2 \rangle}{\partial r_j \partial r_j} \approx \\
       2 \langle \delta \omega_i \widetilde{\omega}_j \delta \left( \frac{\partial u_i}{\partial x_j} \right) \rangle + 2 \langle \delta \omega_i \delta \omega_j \widetilde{\left( \frac{\partial u_i}{\partial x_j} \right)} \rangle - 2 \chi.
\label{eq:enstr_TP_TBL_z450}
\end{gathered}
\end{equation}

\subsection{TNTI region}

We conclude the scale-by-scale analysis by examining the enstrophy budget in the region where the mean interface is located, hence $z_c^+ = Re_{\tau} =1501$. As discussed in the previous sections, the present analysis is based on classical statistics. Consequently, the budget terms are averaged over a wall-parallel plane that, at this wall-normal location, intersects both irrotational and rotational regions. The ``irrotational'' portions of the plane give a negligible contribution to the enstrophy budget, whereas the turbulent bulges intercepted by the plane contain essentially all of the enstrophy and, therefore, account for nearly the entire budget.

\begin{figure}[t!]
\centering
\includegraphics[width=0.49 \linewidth]{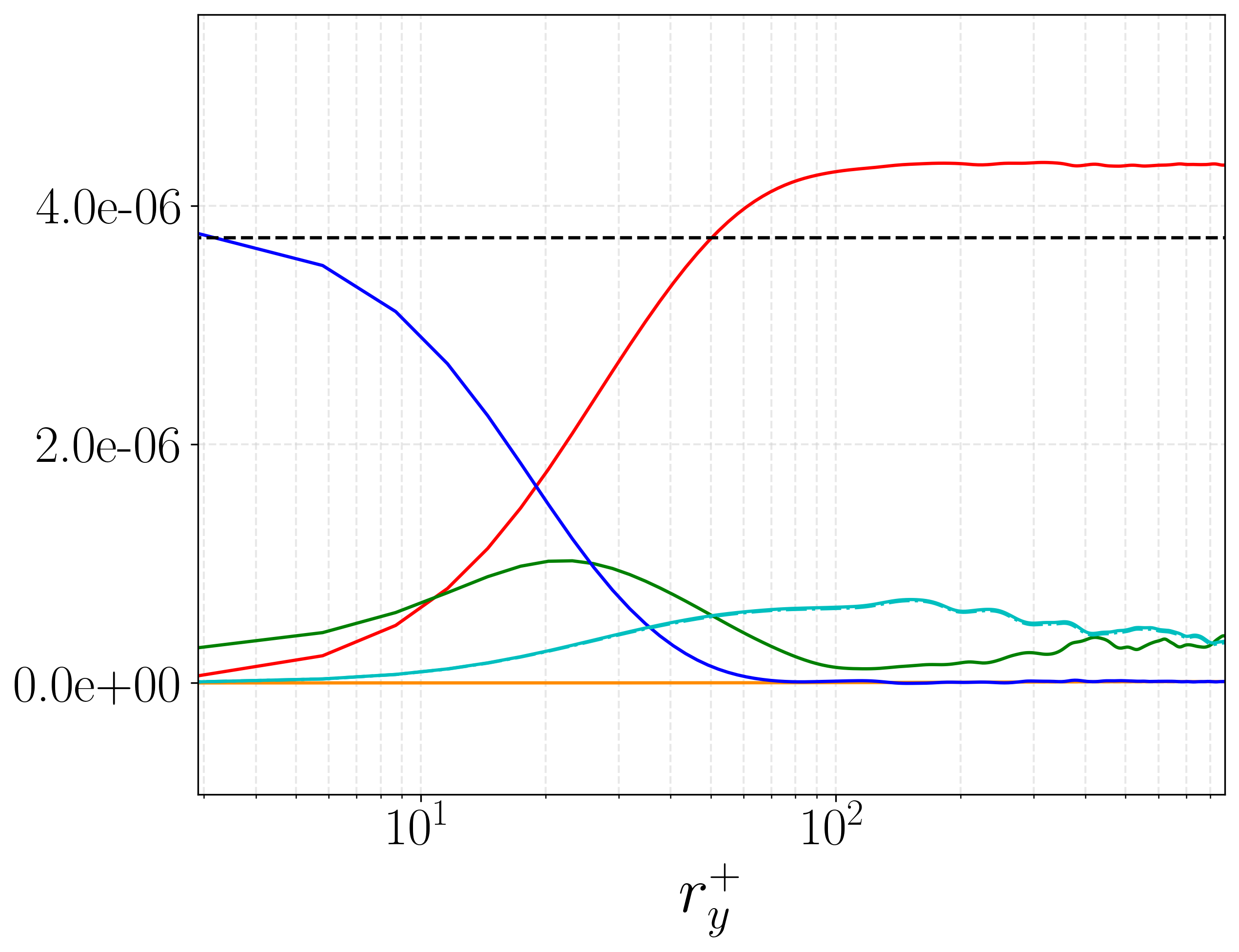}\mylab{-55mm}{46mm}{(\aaa)}
\includegraphics[width=0.49 \linewidth]{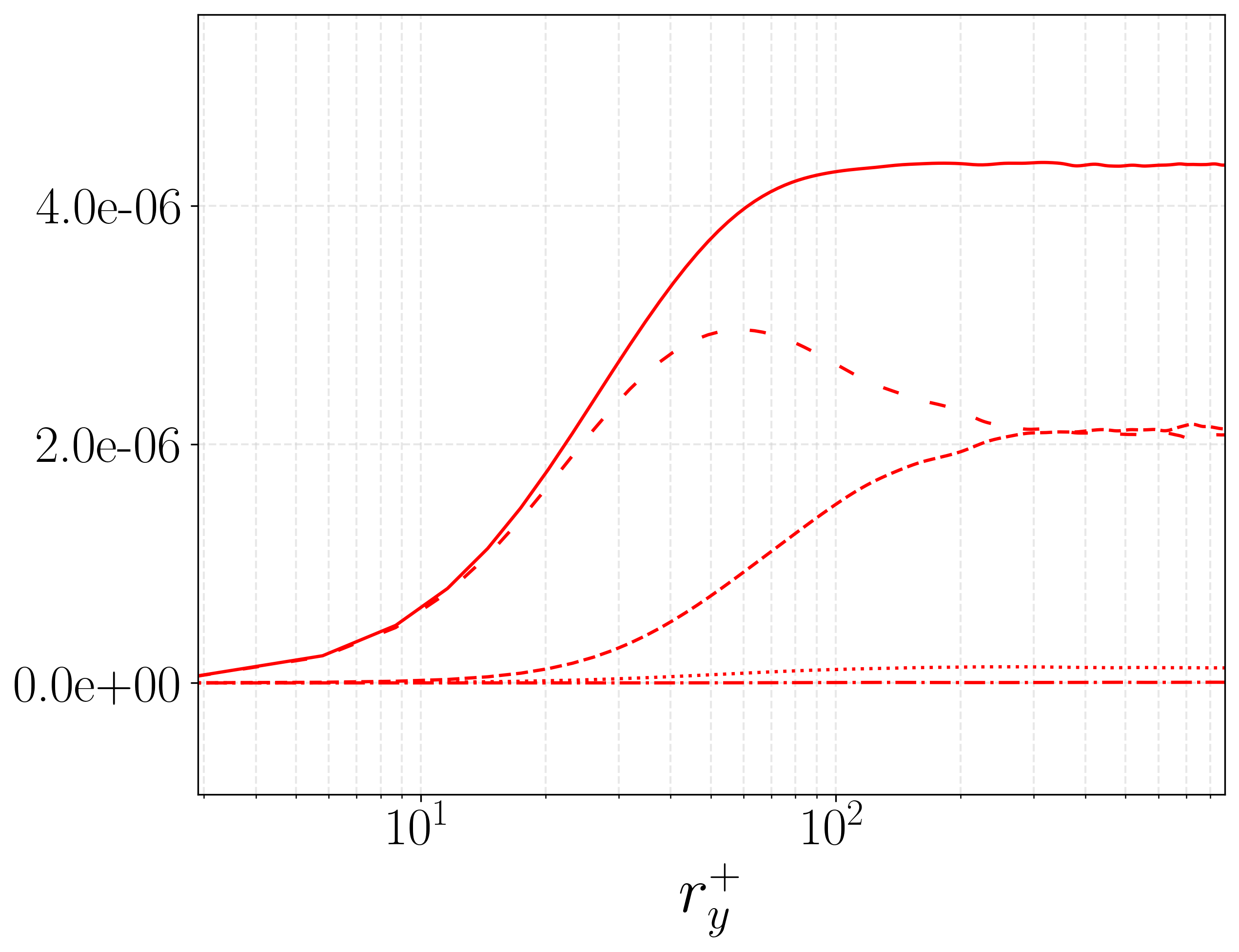}\mylab{-55mm}{46mm}{(\bbb)}
        \caption{Scale-by-scale enstrophy budget terms in the spanwise scales $r_y^+$ evaluated at $z_c^+ = Re_{\tau}$. (a) Complete budget, with $T_{r}^+$ (solid green line), $D_{r}^+$ (solid blue line), $T_{c}^+$ (solid cyan line), $\xi_{\delta \omega^2}^+$ (solid red line), $\Pi_{\delta \omega^2}^+$ (solid orange line), $2 \chi^+$ (dashed black line). The turbulent transport $-(\partial \langle \delta \omega^2 \widetilde{w} \rangle / \partial z_c)^+$ is shown as a dash-dotted cyan line. (b) Vortex stretching contributions: total $\xi_{\delta \omega^2}^+$ (red solid line), $2 \Omega_y^+ \langle \delta \omega_i \delta \left( \partial u_i / \partial y \right) \rangle^+$ (red dash-dotted line), $2 \langle \delta \omega_x \delta \omega_z \rangle^+ \partial U^+ / \partial z^+$ (red dotted line), $2 \langle \delta \omega_i \widetilde{\omega}_j \delta \left( \partial u_i / \partial x_j \right) \rangle^+$ (red dashed line), $2 \langle \delta \omega_i \delta \omega_j \widetilde{\left( \partial u_i / \partial x_j \right)} \rangle^+$ (red loosely-dashed line).}
\label{fig:GKE_1D_z1500}
\end{figure}

The budget terms are shown in figure \ref{fig:GKE_1D_z1500}(a). Overall, the budget in this region retains the main characteristics observed in the outer layer and, more generally, in homogeneous isotropic turbulence. The main distinction is the presence of non-negligible spatial transport $T_c$, which is almost entirely accomplished by its turbulent contribution $\partial \langle \delta \omega^2 \widetilde{w} \rangle/ \partial z_c$. This term predominantly supplies enstrophy to large spanwise scales. Such turbulent transport in the physical (wall-normal) direction is directly associated with the temporal growth of the turbulent boundary layer, which is primarily driven by large-scale anisotropic motions \citep{cimarelli2021numerical, boga2024numerical}. Furthermore, the budget reported in figure \ref{fig:GKE_1D_z1500}(a) exhibits a net excess of enstrophy production through vortex stretching $\xi_{\delta \omega^2}$, together with wall-normal turbulent transport, over enstrophy dissipation $2 \chi$, implying a positive temporal growth rate of the two-point enstrophy, $\partial \langle \delta \omega^2 \rangle / \partial t > 0$ at this location. This excess enstrophy production through vortex stretching must necessarily result from the local amplification of the vorticity transported into the region by the transport in physical space $T_c$, since vortex stretching can only increase enstrophy by acting on pre-existing vorticity. The decomposition of the vortex-stretching term shown in figure \ref{fig:GKE_1D_z1500}(b) also closely resembles that found in homogeneous isotropic turbulence. The contributions involving the mean flow remain negligible, while the fully fluctuating contributions account for almost the totality of the enstrophy production, with a similar behaviour as the one described in the previous section dealing with the outer region.

Hence, from a global perspective, the scale enstrophy budget in the TNTI region can be approximated as an unsteady isotropic case with the addition of a spatial transport,
\begin{equation}
\begin{gathered}
       \frac{\partial \langle \delta \omega^2 \rangle}{\partial t} + \frac{\partial \langle \delta \omega^2 \delta u_j \rangle}{\partial r_j} + \frac{\partial \langle \delta \omega^2 \widetilde{w} \rangle}{\partial z_c} - 2 \nu \frac{\partial^2 \langle \delta \omega^2 \rangle}{\partial r_j \partial r_j} \approx \\
       2 \langle \delta \omega_i \widetilde{\omega}_j \delta \left( \frac{\partial u_i}{\partial x_j} \right) \rangle + 2 \langle \delta \omega_i \delta \omega_j \widetilde{\left( \frac{\partial u_i}{\partial x_j} \right)} \rangle - 2 \chi.
\label{eq:enstr_TP_TBL_z1500}
\end{gathered}
\end{equation}
From a physical perspective, this indicates that the bulges intersected by the average over the wall-parallel plane, contain a fully turbulent flow which does not show major effects given by the spatial inhomogeneity, except for the spatial transport term.

Despite the global picture outlined above, the present analysis also reveals some signatures of a confinement effect in the TNTI region, identified here by the plane $z_c^+ = Re_{\tau}$. As discussed in the previous sections, at this location, both the single-point and two-point enstrophy statistics (figures \ref{fig:enstr_contr} and \ref{fig:Sii_z1500}) showed anisotropy in the vorticity field, while the scale enstrophy fluxes (figure \ref{fig:GKE_2D}(a)) revealed a slight reverse transfer towards large spanwise scales. A more detailed understanding of this behaviour can be obtained by examining the inertial interscale transport, $T_r$. Although this term is not among the dominant contributions to the scale-by-scale budget, it directly characterizes the interscale transport of enstrophy across scales and therefore provides valuable insight into the underlying dynamics.

\begin{figure}[t!]
\centering
\includegraphics[width=0.49 \linewidth]{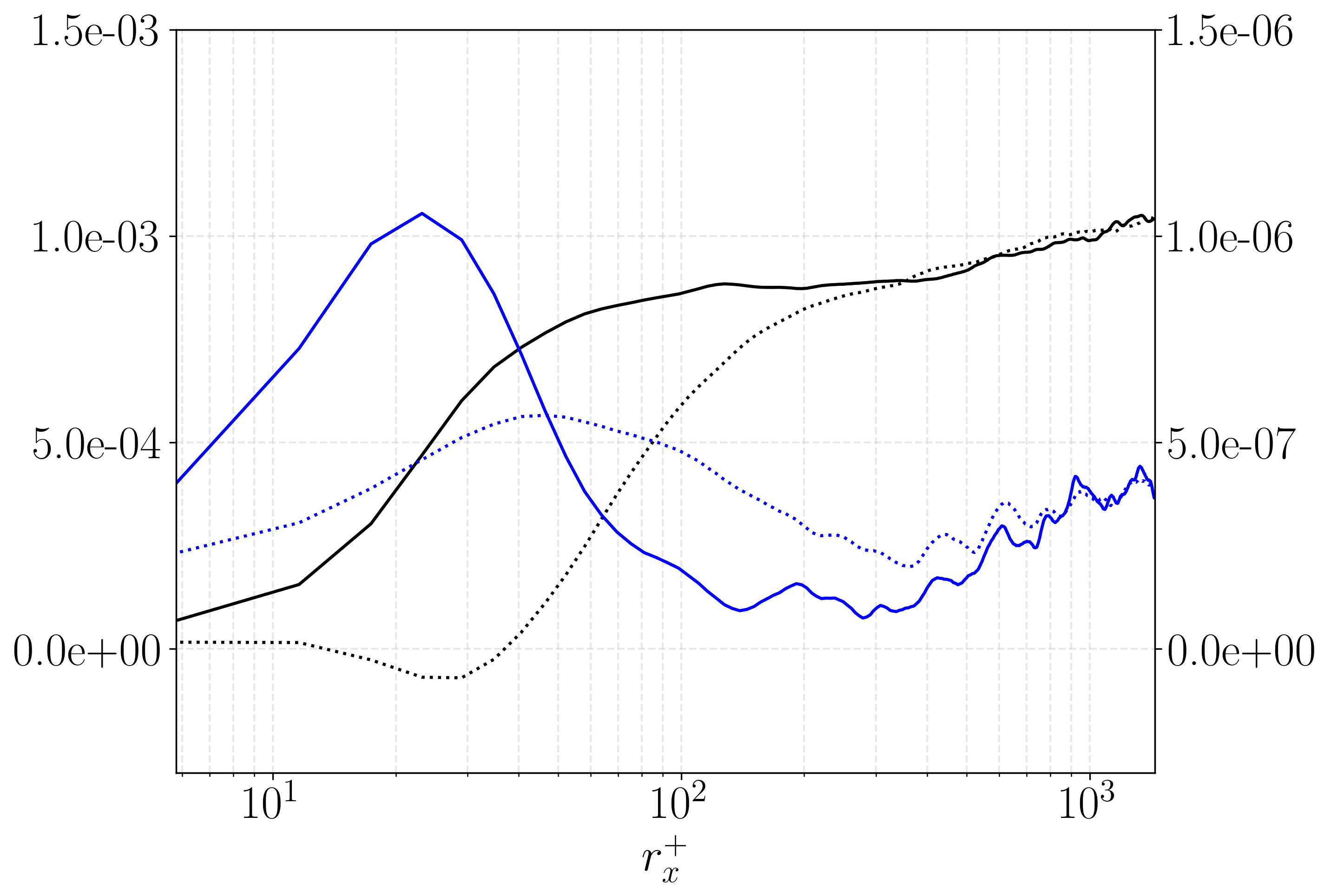}\mylab{-56mm}{40mm}{(\aaa)}
\includegraphics[width=0.49 \linewidth]{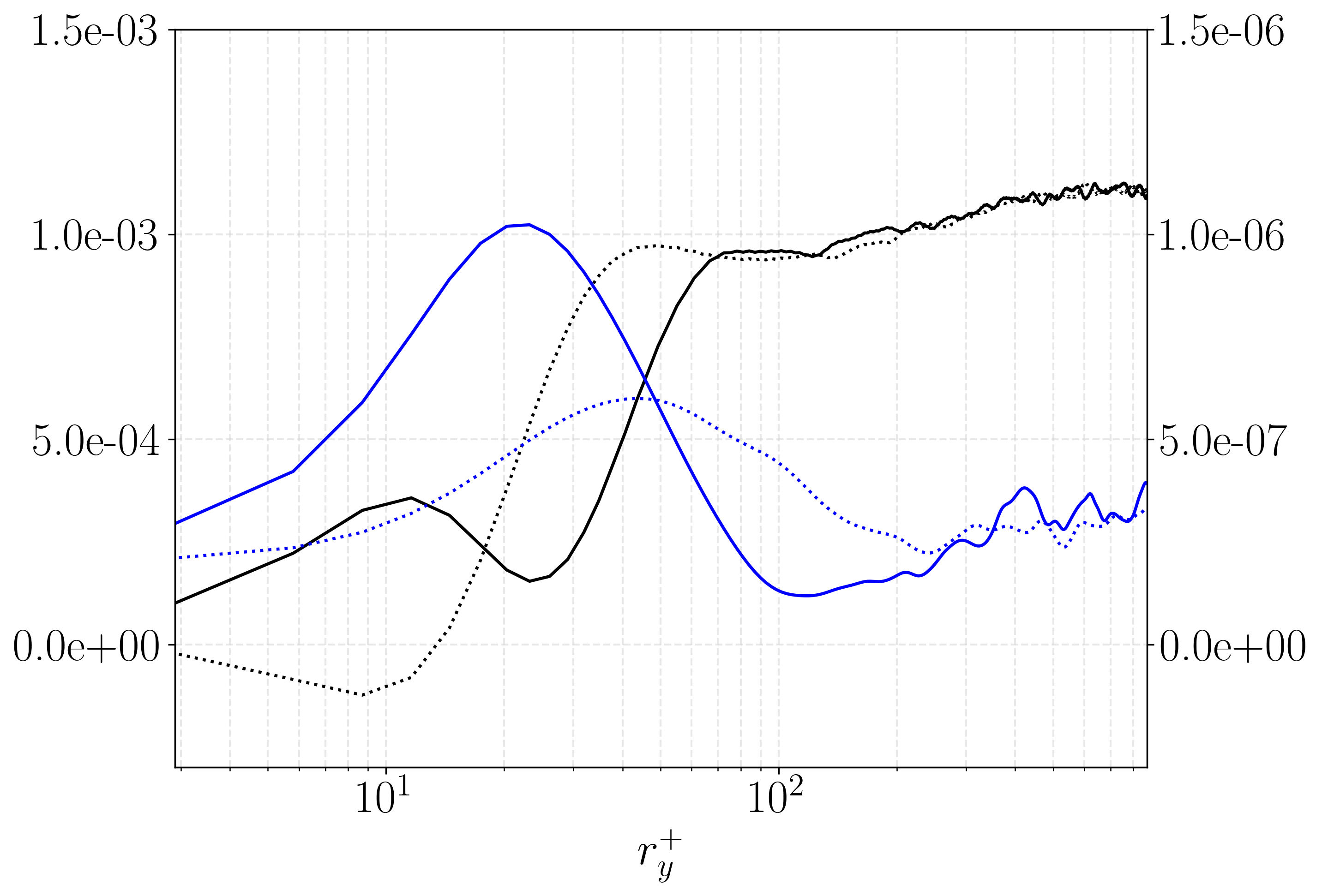}\mylab{-56mm}{40mm}{(\bbb)}
        \caption{Comparison between the inertial interscale enstrophy transport $-(\partial \langle \delta \omega^2 \delta u_j \rangle / \partial r_j)^+$ at $z_c^+ = 0.5$ (solid black line, left axis), and at $z_c^+ = Re_{\tau}$ (solid blue line, right axis) as a function of (a) the streamwise separation scale $r_x^+$ and of (b) the spanwise separation scale $r_y^+$. In both figures the wall-normal inertial interscale transport $-(\partial \langle \delta \omega^2 \delta w \rangle / \partial r_z)^+$ is reported as dotted black lines at $z_c^+=0.5$ and as dotted blue lines at $z_c^+=Re_{\tau}$.}
\label{fig:GKE_1D_inert_flux_comparison}
\end{figure}

Figure \ref{fig:GKE_1D_inert_flux_comparison}(b) compares $T_r$ in the TNTI region ($z_c^+=Re_{\tau}$) with its counterpart in the very near-wall region ($z_c^+=0.5$). Despite the different behaviour exhibited at the small scales, the two distributions share a common feature: both exhibit a positive contribution at large spanwise separations which is entirely sustained by the component associated with wall-normal contribution $\partial \langle \delta \omega^2 \delta w \rangle / \partial r_z$. From a physical perspective, this suggests that enstrophy at such locations is undergoing a direct transfer in the wall-normal scales, while feeding larger spanwise scales $r_y$. An analogous behaviour is observed when the inertial interscale transport $T_r$ is analysed in the streamwise separation scale $r_x$, in figure \ref{fig:GKE_1D_inert_flux_comparison}(a). This suggests that the process is not restricted to the spanwise direction but more generally promotes the growth of structures elongated within the wall- or interface-parallel plane.

These observations support the view that the near-wall and the TNTI regions share a conceptually similar confined dynamics. In both locations, the inertial interscale transfer tends to contract enstrophy-containing motions in the direction normal to the confining interface while transporting enstrophy towards larger scales in the directions parallel to that interface. By considering the relation between enstrophy and dissipation (\ref{eq_ens_diss}), we can equally infer that turbulent dissipation in the near-wall and  TNTI regions is a strongly anisotropic phenomenon performed in a very thin vertical layer over a variety of small and large horizontal scales.

\section{Conclusions}\label{sec:concl}

The present work investigates turbulent enstrophy dynamics in a temporally evolving turbulent boundary layer in order to provide an alternative perspective on the multiscale dynamics of regions affected by confinement effects, namely the near-wall region and the Turbulent/Non-Turbulent Interface region. In both regions, the classical picture of turbulence is significantly modified. Energy is transferred towards smaller scales in the wall-normal direction while, simultaneously, an inverse transfer towards larger wall-parallel scales takes place. As a consequence, the energy is dissipated through a non-classical process, occurring at scales that are small in the wall-/interface-normal direction and large in the wall-/interface-parallel directions. Since enstrophy is intrinsically linked to energy dissipation through its mean value, and to the energy cascade process through vortex stretching, it provides an alternative perspective in regions where both of these processes are fundamentally altered.

The first part of the analysis employed single-point statistics to provide a macroscopic picture of enstrophy dynamics and to characterize confinement effects through the decomposition of enstrophy and of the main budget terms into their streamwise, spanwise, and wall-normal contributions. The resulting picture is as follows. Enstrophy is generated at the wall and injected into the domain through viscous diffusion. In the viscous sublayer, the vorticity field is essentially two-dimensional, being composed almost exclusively of streamwise and spanwise components. Entering the buffer layer, the vortex stretching and tilting term reaches its maximum intensity, amplifying enstrophy while redistributing it among all three vorticity components and progressively establishing a fully three-dimensional vorticity field. Further away from the wall, in the outer region, a local equilibrium is established between enstrophy production by the fully fluctuating contributions to vortex stretching and viscous dissipation through vorticity gradients. In this region, enstrophy is approximately equally distributed among the three vorticity components, exhibiting an almost isotropic state. This balance is progressively lost approaching the TNTI region, where the spanwise vorticity component again becomes dominant.

While the single-point analysis provides a clear picture of some essential aspects of enstrophy dynamics, it does not capture its inherently multiscale nature. To address this aspect, the analysis has been extended to the space of scales by means of the generalized Kolmogorov equation framework applied to enstrophy. The two-point analysis provided a deeper insight into these mechanisms. The near-wall enstrophy field exhibits the clear footprint of structures that are located farther away from the wall. In particular, the streamwise vorticity field showed the wall-footprint of the quasi-streamwise vortices, which induce alternate spanwise velocity fluctuations near the wall, inducing a secondary streamwise vorticity. On the other hand, the spanwise vorticity field showed the footprint of the  high- and low-speed streaks, which induce large-scale streamwise velocity fluctuations near the wall, inducing spanwise vorticity fluctuations. In this sense, enstrophy is generated at the wall by mechanisms that are not sustained locally, but rather are just the footprint at the wall of the dynamics of the buffer and outer layers. In this very near-wall region, the dynamics is dominated by diffusive transports associated with wall-normal gradients. The vortex stretching term is already non-negligible in this region (e.g. at $z_c^+=0.5$), and is mainly composed of its purely fluctuating contributions and the ``pure stretching'' of the mean spanwise vorticity $2 \Omega_y \langle \delta \omega_y \delta (\partial v / \partial y) \rangle$. The prevalence of sheet-like structures in this region is reflected in the cancellation of the remaining mean-flow-related contributions: $2 \Omega_y \langle \delta \omega_x \delta (\partial u / \partial y) \rangle \approx - 2 \langle \delta \omega_x \delta \omega_z \rangle \partial U / \partial z$ and $2 \Omega_y \langle \delta \omega_z \delta (\partial w / \partial y) \rangle \approx 0$.

As enstrophy is transported away from the wall, its evolution exhibits two distinct behaviours depending on its spanwise scale. Enstrophy generated at the wall at scales $r_y^+ \lesssim 55$ remains confined within the viscous sublayer, where it undergoes a direct interscale transfer towards smaller scales before being dissipated. In contrast, enstrophy generated at larger scales $r_y^+ \gtrsim 55$ is effectively transported through the buffer layer into the outer region. This behaviour may suggest that only enstrophy at scales comparable to those of the structures responsible for turbulent transport effectively correlates with them, allowing it to pass through the buffer layer despite the repulsive effect of the local enstrophy production by vortex stretching.

In addition to enstrophy production, the buffer layer is also the region where enstrophy is redistributed among the different vorticity components, recovering the full 3-dimensionality of the vorticity field. In particular, the wall-normal vorticity, which is negligible in the viscous sublayer, is generated primarily by the tilting of the mean spanwise vorticity $\Omega_y$ by non-uniform wall-normal velocity fluctuations $w$. The newly generated wall-normal vorticity is tilted in the streamwise direction by the mean velocity gradient $\partial U / \partial z$, consistently with the hairpin-vortex regeneration cycle described in the literature.

The transport of enstrophy continues through a split transport regime. As enstrophy ascends from the wall, one fraction undergoes a direct interscale transfer towards smaller scales, while the remaining fraction is transferred towards larger scales. The crossover scale separating these two behaviours, denoted by $\ell_{\delta \omega^2}$, scales linearly with the wall distance up to a few hundred viscous units and nearly coincides with the corresponding energy crossover scale $\ell_{\delta q^2}$. Remarkably, both crossover scales are located close to the scale of maximum energy production $\ell_{\Pi}$, consistent with the attached-eddy interpretation. Within this framework, the break-up of attached eddies is associated with a direct transfer towards smaller scales, whereas their growth and reorganization is associated with a reverse transfer towards larger scales that scale linearly with the wall distance. The analogous scaling of the enstrophy crossover suggests that the mechanisms governing the scale organization of energy also leave an imprint in the enstrophy transport.

Finally, the two-point statistics show that the outer region is characterized by an almost isotropic distribution and dynamics of enstrophy, in agreement with the single-point analysis. As the TNTI region is approached, however, confinement effects become visible again. In this region, a slight predominance of the spanwise vorticity component, which is also distributed at larger scales, is observed, partially recalling the near-wall region. Similarly to the near-wall region, the wall-normal interscale transport showed a positive contribution at large spanwise scales, indicating that in both regions (even if with different intensities), the confinement effects induce an interscale transfer toward smaller wall-normal scales that feeds larger wall-parallel scales.

\begin{acknowledgements}
We wish to acknowledge the Italian supercomputing centre CINECA that, under the ISCRA B project
ESSPRO (Direct numerical simulation of Entrainment and Self-Sustaining PROcesses of turbulence in boundary layers) and the ISCRA C project TURBO (Direct numerical simulation of TURbulent temporal BOundary layers), provided the HPC resources for the simulations of the temporal boundary layer. Financial support was provided by the Italian Ministry of University and Research (MUR), through the PRIN 2022 PNRR Project 20224NY2YK - ADMIRE, CUP E53C24002710006, funded by the National Recovery and Resilience Plan (PNRR), Italy, Mission 4 Component C2 Investment 1.1.

Declaration of Interests. The authors report no conflict of interest.
\end{acknowledgements}

\bibliographystyle{jfm}
\bibliography{jfm}

\end{document}